\documentclass[useAMS,usenatbib,usegraphicx]{mn2e}
\usepackage{color}
\usepackage{epsf}
\usepackage{ulem}
\usepackage{epsfig}
\usepackage{amsmath}
\usepackage{amssymb}
\newcommand{\jbz}{\mathfrak{j}_0}
\newcommand{\jbo}{\mathfrak{j}_1}
\newcommand{\mps}{\mathfrak{P}^2}
\newcommand{\djbo}{\Delta\mathfrak{j}_1}

\newcommand{\hMpc}{h^{-1}{\rm\;Mpc}}

\newcommand{\snl}{\Sigma_{\rm nl}}

\newcommand{\sxi}{\xi_s(r)}
\newcommand{\e}{\mathrm{e}}

\newcommand{\ifrac}{\frac{3}{r_{i2}^3-r_{i1}^3}}
\newcommand{\jfrac}{\frac{3}{r_{j2}^3-r_{j1}^3}}
\newcommand{\tnii}{\int^{r_{i2}}_{r_{i1}}}
\newcommand{\jint}{\int^{r_{j2}}_{r_{j1}}}
\newcommand{\dom}{\frac{d\Omega}{4\pi}}
\newcommand{\rp}{r'}

\newcommand{\vol}{\frac{2}{V}}
\newcommand{\oms}{\sqrt{\Omega_m(1+z)^3 + \Omega_{\Lambda}}}

\title[A 2\% Distance to $z=0.35$: Fitting Techniques]
{A 2\% Distance to $z=0.35$ by Reconstructing Baryon Acoustic
Oscillations - II: Fitting Techniques}

\author[X. Xu et al.]
{Xiaoying Xu$^{1}$, Nikhil Padmanabhan$^{2}$, Daniel J. Eisenstein$^{3}$, 
Kushal T. Mehta$^{1}$, \newauthor Antonio J. Cuesta$^{2}$\\
$^{1}$ Steward Observatory, University of Arizona, 933 N. Cherry Ave., 
Tucson, AZ 85721; xxu@as.arizona.edu\\
$^{2}$ Dept. of Physics, Yale University, 
260 Whitney Ave., New Haven, CT 06511\\
$^{3}$ Harvard Smithsonian Center for Astrophysics, Harvard University, 
60 Garden St., Cambridge, MA 02138}

\begin{document}
\maketitle
\label{firstpage}
\begin{abstract}
We present results from fitting the baryon acoustic oscillation (BAO)
signal in the correlation function obtained from the first application of
reconstruction to a galaxy redshift survey, namely, the Sloan Digital Sky
Survey (SDSS) Data Release 7 (DR7) luminous red galaxy (LRG) catalogue. We
also introduce more careful approaches for deriving a suitable covariance
matrix and fitting model for galaxy correlation functions. These all
aid in obtaining a more accurate measurement of the acoustic scale
and its error. We validate our reconstruction, covariance matrix and
fitting techniques on 160 mock catalogues derived from the LasDamas
simulations in real and redshift space. We then apply these techniques
to the DR7 LRG sample and find that the error on the acoustic scale
decreases from $\sim3.5\%$ before reconstruction to $\sim1.9\%$ after
reconstruction. This factor of 1.8 reduction in the error is equivalent
to the effect of tripling the survey volume. We also see an increase
in our BAO detection confidence from $\sim3\sigma$ to $\sim4\sigma$
after reconstruction with our confidence level in measuring the correct
acoustic scale increasing from $\sim3\sigma$ to $\sim5\sigma$. Using
the mean of the acoustic scale probability distributions produced from
our fits, we find $D_v/r_s = 8.89 \pm 0.31$ before reconstruction and
$8.88 \pm 0.17$ after reconstruction.
\end{abstract}

\begin{keywords}
distance scale
-- cosmological parameters
-- large-scale structure of universe
-- cosmology: theory, observations
\end{keywords}


\section{Introduction}\label{sec:intro}

The interaction between matter and radiation prior to cosmological
recombination leaves an imprint on the present day distribution of matter
known as the baryon acoustic oscillations (BAO). As matter accreted in
overdensities under the influence of gravity, the resulting compression of
the primordial plasma caused temperatures and hence radiation pressure
to increase. When the radiation pressure became sufficiently high,
the photons pushed out from the overdensities in spherical sound waves
carrying the baryons along with them. The subsequent competition between
gravity and radiation pressure set up a system of standing sound waves
within the primordial plasma \citep{PY70, SZ70, BE84, H89, HS96, HW96,
EH98}. The frequency of these waves corresponds to a characteristic
spatial scale known as the acoustic scale or the sound horizon. This is
the distance traveled by the sound wave in the plasma before recombination
and is $\sim150$ comoving Mpc. After the photons stream off, the baryons
are deposited at these characteristic separations which can still be
seen in the galaxy distribution today. Therefore, this acoustic scale
can be used as a very accurate standard ruler for measuring cosmological
distances at large spatial separations \citep{EH98, EHT98, BG03, E03,
HH03, L03, SE03, M04, AR05, AQG05, Angulo05, GB05, DJT06}. This has
been demonstrated using galaxy redshift surveys such as described in
\citet{Cea05, Eea05, H06, Tea06, Pad07, Pea07a, Pea07b, Sanchez09, Kea10,
Pea10, Beutler11, Bea11a, Bea11b, Hea12, Sea12}. Forecasts have also
been made for BAO studies in future galaxy surveys (e.g. \citealt{Wea09,
ZKT09}) and neutral hydrogen surveys (e.g. \citealt{MW08, WLG08}).

In practice, the acoustic scale may appear slightly shifted from its
predicted linear theory position due to non-linear structure growth
\citep{MWP99, SE05, JK06, ESW07, Hea07, GBS07, Ma07, Aea08, CS08,
SBA08, Sea08, Smith08, PdW09}. Non-linear evolution may also smear the
acoustic peak making it more difficult to centroid, thereby resulting in
a poorer measurement of the acoustic scale. However, these uncertainties
can be largely removed by reconstruction \citep{Eea07, Sea08, Nea09,
PWC09, Sea10, Mea11}, which is the process by which galaxies are moved
back along the first-order displacements that arise due to non-linear
growth. In addition to partially undoing non-linear structure growth,
our reconstruction technique also includes a prescription to remove
the large-scale redshift-space distortion (known as Kaiser squashing;
\citealt{Kaiser87}) that further distorts the BAO signal along the
line-of-sight. This paper along with its companion papers, \citet{Peaipa}
and \citet{KMeaip}, presents the first application of reconstruction to
a galaxy redshift survey.

Since the goal of all BAO galaxy surveys is to measure the acoustic
scale to high precision, we intend for this paper to present a viable
procedure for attaining this goal. We include a discussion of the
necessary statistical tools, such as a new method for deriving a reliable,
smooth covariance matrix, and a robust fitting framework for measuring
the acoustic scale. We use our mock catalogues to demonstrate that making
slight adjustments to our fiducial model parameters such as $\snl$ (used
to model the degradation of the BAO signal due to non-linear structure
growth), the fitting range, the number of marginalization terms and the
input cosmology, do not alter the measured acoustic scale. This indicates
the robustness of our techniques.

Using these tools and the DR7 LRG sample, we measure the acoustic scale
to $3.5\%$ before reconstruction and $1.9\%$ after reconstruction. Our
post-reconstruction result is the highest precision measurement of
the acoustic scale at $z=0.35$ obtained through galaxy surveys to
date. Without reconstruction, we would need to increase the survey volume
by nearly a factor of 3 to achieve this same factor of 1.8 reduction
in the error. We also find that both measures of BAO significance we
consider improve by at least 1$\sigma$ after reconstruction.

In \S\ref{sec:mockc} we discuss the mock catalogues we use for our
analyses and our reconstruction parameters. In \S\ref{sec:techs} we
describe some of the covariance and fitting techniques used in previous
studies followed by an outline of the covariance matrix and fitting models
we employ for this study. \S\ref{sec:red_fit} describes the fitting
results on our redshift-space mocks. The analogous discussion for real
space is found in \S\ref{sec:real_fit}. We apply our techniques to the DR7
LRG sample in \S\ref{sec:drs} and conclude in \S\ref{sec:theend}. For more
details of the reconstruction method and the SDSS data set, we refer the
interested reader to the companion paper \citet{Peaipa} (Paper I). The DR7
cosmology results can be found in the other companion paper \citet{KMeaip}
(Paper III).

\section{Mock Catalogues and Reconstruction}\label{sec:mockc}

A variety of statistics such as the correlation function, the power
spectrum, and more recently $\omega_\ell(r_s)$ \citep{Xea10} are
available for measuring clustering and the BAO scale. As mentioned
in \S\ref{sec:intro}, the measurement of this scale is affected by
non-linear structure growth. Hence, in order to obtain an accurate
measurement of the acoustic scale through fitting the observational data,
we must first employ an algorithm known as reconstruction to partially
undo the effects of this non-linear evolution. We also need to develop a
technique that returns reliable error estimates for our chosen clustering
statistic (i.e. the covariances between different scales) and a method
to marginalize out the broadband (non-BAO) information from the statistic.

To this end, we compute correlation functions in both real and redshift
space with and without reconstruction from SDSS DR7 LRG mock catalogues
created using the LasDamas simulations \citep{Meaip}. The simulation
cosmology is $\Omega_m=0.25$, $\Omega_b=0.04$, $h=0.7$, $n_s=1$ and
$\sigma_{8, \rm matter}=0.8$ at $z=0$. There are a total of 160 mock
catalogues corresponding to our area of interest, the DR7 Northern
galactic cap, which has a sky coverage of 7189 deg$^2$. The redshift
range covered by the mocks is $0.16<z<0.44$ (note that this is slightly
different to the redshift range of the DR7 data, $0.16<z<0.47$).

The process of reconstruction is conceptually equivalent to running
gravity backwards \citep{Eea07}. This procedure helps remove some of the
smearing and shifting of the acoustic peak caused by non-linear structure
growth. We also include a prescription for removing the redshift-space
distortion caused by Kaiser squashing which can further broaden the
acoustic peak. Many past studies have tested the basic reconstruction
algorithm using simulations \citep{Nea09, Sea10, Mea11}, however, this
study marks its first application to a real galaxy redshift survey. The
reconstruction algorithm can be simply described as follows. We estimate
the matter density field from the observed galaxies using a simple bias
scaling from the measured galaxy density field. We then solve the linear
continuity equation $\nabla \cdot \vec{q} = -\delta$ where $\vec{q}$
is the displacement field and $\delta$ is the density field. This gives
us the first order displacements that arise from non-linear structure
growth. Finally, we shift the galaxies back along these displacement
vectors. For our reconstruction, we apply a Gaussian smoothing to the
matter density field using a smoothing scale of 15$\hMpc$ to reduce
sensitivity to small scale clustering which is poorly constrained in
large galaxy surveys.

For details of the computation and reconstruction, please see Paper
I. All correlation functions were computed in $3\hMpc$ bins from
$2.5-197.5\hMpc$. 

\section{Covariance Matrix and Fitting Techniques}\label{sec:techs}

\subsection{Overview of Past Approaches}
In past studies involving observational data, the most common method for
deriving the covariance matrix was to construct it from mock catalogues
generated from either simulations \citep{HRS06, Tea09} or perturbation
theory approaches \citep{SS02}. Perturbative methods are less accurate
than we would like and as we will show, the covariance matrices calculated
from mocks can still be noisy, even if the number of mocks used is
large. One can also assume the smooth Gaussian covariance matrix from
linear theory, however this neglects the non-linear contribution to the
noise. Hence, it is necessary to devise a scheme for approximating the
mock covariances with a smooth function or find alternate methods to
regularize the matrix. In this paper, we present a robust approximation
scheme, which we will show produces a faithful representation of the
expected covariances.

The acoustic scale can be measured from galaxy clustering statistics by
fitting the data with a template based on linear theory. The location
of the acoustic peak in this template must depend on a parameter that
specifies the magnitude of the acoustic scale relative to the fiducial
value. Typically in Fourier space, this requires a fitting model of
the form
\begin{equation}
P(k) = B(k)P_m(k/\alpha) + A(k).
\end{equation}
Here, $P_m(k)$ is the template power spectrum based on linear theory
and $\alpha$ is the scale dilation parameter that is used to adjust
the location of the acoustic peak. $A(k)$ and $B(k)$ are functions
involving nuisance parameters that can be used to marginalize out the
broadband shape of the power spectrum (i.e. scale-dependent bias and
redshift-space distortions). The broadband shape does not contain BAO
information but may bias the measurement of the BAO scale if not accounted
for properly. These terms can also help mitigate the effects of using
the wrong model cosmology. Analogously, in configuration space we have
\begin{equation}
\xi(r) = B(r)\xi_m(\alpha r) + A(r).
\end{equation}

In order to obtain an accurate measure of the acoustic scale, we require
this fitting model to be robust. This simply means that if we slightly
change the parameters that go into the model, the measured value of
$\alpha$ should always be consistent. For a fitting form where this is
true, even if we use model parameters that are not optimal, we will still
measure the correct acoustic scale. This is necessary since we use this
fitting form to derive the acoustic scale in the SDSS DR7 data and in
practice we are not certain of the exact model parameters to use.

In Fourier space, Pad\'{e} approximates and basis functions based on cubic
splines work well for both $A(k)$ and $B(k)$, while high order polynomials
may also be used for $A(k)$. This has been demonstrated for simulated
data (e.g. \citet{Sea08, PdW09, Sea10, Mea11}) as well as SDSS-II
observational data (e.g. \citet{Tea06, Pea07b, Pea10}). However, high
order polynomials and cubic spline forms do not transform particularly
nicely to configuration space due to poor numerical convergence of
the integration.

In configuration space, there have been attempts to model the
scale-dependent bias associated with the $B(r)$ term such as in
\citep{Bea11a}. As for $A(r)$, an array of forms have been used. In
theoretical works \citep{CS08, SBA08} and the DR6 motivated
observational work \citep{Sanchez09}, $A(r)$ was motivated by perturbation
theory and contained derivatives and integrations of the linear theory
correlation function. Other works based in simulations (e.g. \citet{CG11})
and SDSS observations (e.g. \citet{Eea05, Kea10}) did not use an $A(r)$
term at all. However, as we will show in this work, having a non-zero
$A(r)$ term aids greatly in removing unwanted broadband information and
ameliorating errors in the assumed model cosmology. This is especially
true if one is to take $B(r)=B$ and delegate the marginalization of
scale-dependent bias to the $A(r)$ term, as is done in most correlation
function analyses. We note here though, that the form for $A(r)$ does
not need to be complicated as we show in \S\ref{sec:red_norec_ff}.

\subsection{Covariance Matrices}\label{sec:red_norec_cm}

\begin{figure}
\vspace{0.4cm}
\centering
\begin{tabular}{c}
\epsfig{file=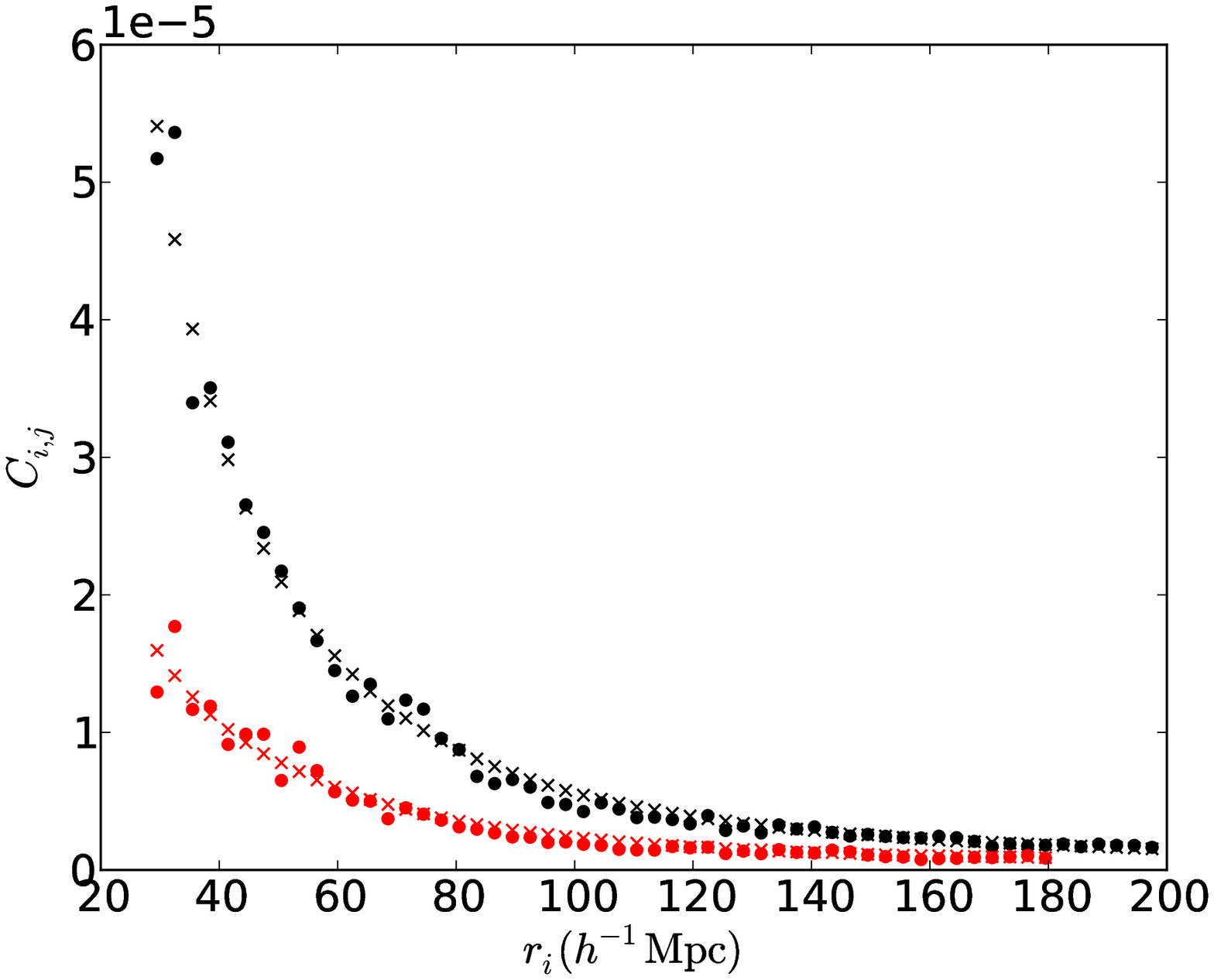, width=0.8\linewidth, clip=} \\
\epsfig{file=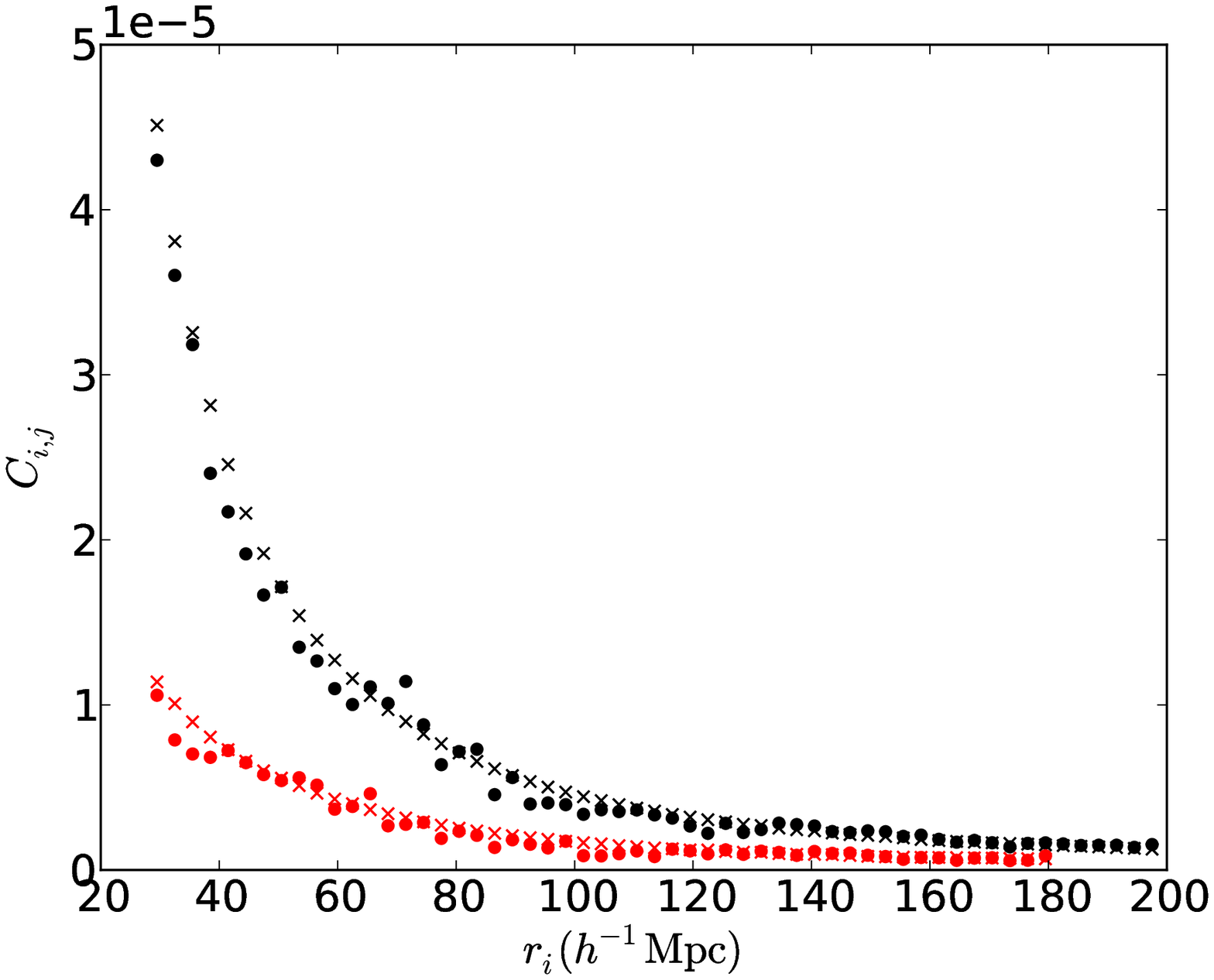, width=0.8\linewidth, clip=}
\end{tabular}
\caption{The diagonal (black) and 6th off-diagonal (red) of the mock
(circles) and modified Gaussian (crosses) covariance matrices in redshift
space before reconstruction (top) and after reconstruction (bottom). The
mock covariance matrix shows clear signs of noise. The modified Gaussian
covariance matrices take on the form given in Equations (\ref{eqn:modc}
\& \ref{eqn:noise}) with $\sigma_s=4\hMpc$.  Before reconstruction,
$c_0=0.89$, $c_1=0.46$, $c_2=1.34$, $c_3=2.32 \times 10^{-7}$ and after
reconstruction $c_0=0.89$, $c_1=0.30$, $c_2=1.45$, $c_3=1.87 \times
10^{-7}$. One can see that the modified Gaussian covariance matrices
are good smoothed approximations to the mock covariance values.
\label{fig:cov}} 
\end{figure}

We perform the analyses in this paper using the correlation function
statistic and hence, we require an estimate of the correlation function
covariances. As mentioned previously, the most obvious choice is to use
the covariance matrix calculated directly from the mock catalogues. The
value of the $i$th row and $j$th column of such a covariance matrix is
\begin{equation}
C_{ij} =
\frac{1}{N-1}\sum^{N}_{n=1}[\xi_n(r_i)-\bar{\xi}(r_i)][\xi_n(r_j)-\bar{\xi}(r_j)],
\label{eqn:mcov}
\end{equation}
where $N$ is the total number of mocks, $\xi_n(r)$ is the correlation
function calculated from the $n$th mock and $\bar{\xi}(r)$ is the average
of the mock correlation functions. However, we find that the covariances
calculated from 160 mocks are still noisy (see Figure \ref{fig:cov}). To
obtain a smooth approximation to the mock covariances, we introduce a
new technique which involves fitting a modified form of the Gaussian
covariance matrix to the data using a maximum likelihood approach.

The analytic Gaussian covariance matrix can be calculated as
\begin{equation}
C_{ij} = \vol\int
\frac{k^2dk}{2\pi^2}\jbz(kr_i)\jbz(kr_j)[P_c(k)+\aleph]^2
\label{eqn:tgcov}
\end{equation}
where $V$ is the volume of each mock, $\aleph$ is the shot-noise and
\begin{equation} 
\jbz(kr) = \frac{\sin(kr)}{kr}
\end{equation} 
is the $0$th order spherical Bessel function. $\aleph$ has 2 basic
components, linear shot-noise and non-linear shot-noise. In the standard
Gaussian covariance matrix, the linear shot-noise is assumed to be
Poisson, which implies $\aleph_{lin} = \bar{n}^{-1}$. Realistically
however, surveys span a range of redshifts, so $\bar{n}$ is dependent
on $z$. In addition, we must also consider the non-linear shot-noise
which arises due to non-linear structure growth at small scales. This
is typically not included in the calculation of the standard Gaussian
covariance matrix. We will address these issues in more detail shortly.

Due to the binning of data in our correlation function calculations, we
must also adjust our Gaussian covariance matrix calculation to reflect
this. Theoretically, the value of the binned correlation function at
the bin center $r_i$ is
\begin{eqnarray}
\xi(\bar{r_i}) &=& \frac{\int_{\Omega}\tnii d^3r \xi(r)}
{\int_{\Omega}\tnii d^3r} \\
&=& \ifrac \int_{\Omega}\tnii r^2dr \dom \int \frac{k^2dk}{2\pi^2}P(k)\jbz(kr)
\label{eqn:binxi}
\end{eqnarray}
where the bin limits are $(r_{i1},r_{i2})$ and $\xi(r)$ is the true
unbinned correlation function. Analogously, we may write the expression
for the binned covariance matrix as
\begin{eqnarray}
C_{ij} &=& \vol \ifrac\jfrac \nonumber \\
&& \centerdot \int_{\Omega}\tnii r^2dr \dom \int_{\Omega'}\jint \rp^2 dr' 
\dom' \nonumber \\
&& \centerdot \int \frac{k^2dk}{2\pi^2}\jbz(kr_i)\jbz(kr_j)[P_c(k)+\aleph]^2.
\end{eqnarray}
This can be shown to give 
\begin{equation}
C_{ij} =\vol \int \frac{k^2dk}{2\pi^2} \djbo(kr_i) \djbo(kr_j) [P_c(k)+\aleph]^2
\label{eqn:gcov}
\end{equation}
where 
\begin{equation}
\djbo(kr) = \frac{3}{r_2^3 - r_1^3}
\frac{[r_{2}^2\jbo(kr_{2}) - r_{1}^2\jbo(kr_{1})]}{k},
\end{equation}
\begin{equation}
\jbo(kr) = \frac{\sin(kr)}{(kr)^2} - \frac{\cos(kr)}{kr}
\end{equation}
is the $1$st order spherical Bessel function. Here, we have intentionally
written Equation (\ref{eqn:gcov}) to resemble Equation (\ref{eqn:tgcov}).

The input power spectrum $P_c(k)$ determines the sample variance of
the signal. In redshift space before reconstruction, we take $P_c(k)$
to have the form,
\begin{equation}
P_c(k) = b_0^2\int^{1}_{-1} (1+\beta\mu^2)^2F(\mu,k)P_t(k)d\mu
\label{eqn:pc}
\end{equation}
where $(1+\beta\mu^2)^2$ is the standard Kaiser squashing term
\citep{Kaiser87} with $\beta=f/b_0$, $f\sim \Omega_m^{0.6}$ and $b_0$
equal to the large-scale bias. $F(\mu,k)$ is a streaming model term used
to account for the Finger of God (FoG) effect. We take this term to be
exponential in configuration space and hence
\begin{equation}
F(\mu,k) = \frac{1}{(1+k^2\mu^2\sigma_s^2)^2}
\label{eqn:stream}
\end{equation}
in Fourier space, where $\sigma_s$ is the dispersion within a cluster and
is typically around $3-4\hMpc$. A Gaussian form for $F(\mu,k)$ can also
be used; however, we find little difference between the results. Kaiser
squashing and FoG are known as redshift-space distortions and arise from
observational biases in measured redshifts due to motions of galaxies
along the line-of-sight direction.

We determine $b_0^2$ by matching the configuration space transform
of $P_c(k)$ to the average of the mock correlation functions at
$r=50\hMpc$. This ensures that the amplitude of $P_c(k)$ matches the
average clustering amplitude in the mocks.

Our template power spectrum, $P_t(k)$, takes on the form
\begin{equation}
P_t(k) = [P_{\rm lin}(k)-P_{\rm smooth}(k)]\e^{-k^2\snl^2/2}+P_{smooth}(k),
\label{eqn:template}
\end{equation}
where $P_{\rm lin}(k)$ is the linear power spectrum at $z=0$. $P_{\rm
smooth}(k)$ is the dewiggled power spectrum described in \citet{EH98} and
$\snl$ is a smoothing parameter that is used to model the degradation in
the acoustic peak due to non-linear evolution \citep{CS06, ESW07, CS08,
M08, Sea08}. Before reconstruction, the overall shape of the acoustic
peak in the template matches the data best when $\snl\sim8\hMpc$; hence
we fix $\snl=8\hMpc$. We will show that varying this value has little
affect on the resulting covariance matrix later in this section.

In order to address the $z$ dependence of $\bar{n}$, we use the fact that
Equation (\ref{eqn:tgcov}) is really just the transform of the variance
in Fourier space, $[P_c(k)+\aleph]^2/V$, to the expected covariance in
configuration space. One can then imagine building up the inverse of this
variance, $I^2(k)$, as an integral over volume,
\begin{eqnarray}
I^2(k) &=& \int \frac{dV}{[P_c(k)+\aleph]^2} \nonumber \\
&=& \frac{c\Omega}{H_0}\int_{z_l}^{z_u} 
\left[P_c(k)+\frac{1}{\bar{n}(z)}\right]^{-2} \nonumber \\
&& \centerdot \frac{r^2(z)}{\oms}dz,
\label{eqn:Ik}
\end{eqnarray}
where we use 
\begin{equation}
dV = \frac{c}{H_0} \frac{r^2(z)}{\oms} dz d\Omega
\end{equation}
for a flat universe and assume $\bar{n}(z)$ has no angular
dependence. $z_u$ and $z_l$ are the upper and lower redshift limits
of the survey respectively. Now we can redefine the binned Gaussian
covariance matrix, Equation (\ref{eqn:gcov}), as
\begin{equation}
C_{ij} = 2 \int \frac{k^2dk}{2\pi^2} \djbo(kr_i) \djbo(kr_j) 
\mps(k)
\end{equation}
where $\mathfrak{P}^2(k) = [I^2(k)]^{-1}$. We calculate a model for
$\bar{n}(z)$that suits the DR7 data from the LasDamas random catalogue
and scale this to other cosmologies when necessary using the appropriate
volume ratios.

Since our binned Gaussian covariance matrix does not include non-linear
shot-noise, it underpredicts the mock covariance matrix. However, one
can imagine applying some modifications to the Gaussian covariance matrix
so that its shape better emulates that of the mock covariance matrix. We
assume a modification to the Gaussian covariance matrix of
\begin{equation}
C^m_{ij} = 2 \int
\frac{k^2dk}{2\pi^2}\djbo(kr_i)\djbo(kr_j)\mps(k;c_0,c_1,c_2) + c_3
\label{eqn:modc}
\end{equation}
where $\mps(k;c_0,c_1,c_2)$ corresponds to an $I^2(k)$, Equation
(\ref{eqn:Ik}), in which we make the substitution
\begin{align}
\label{eqn:noise}
P_c(k) + \frac{1}{\bar{n}(z)} \rightarrow&
c_0P_c(k) + \frac{c_1}{\bar{n}(z)}\int^{1}_{-1} 
(1+\beta\mu^2)^2F(\mu,k) d\mu \nonumber \\
& + \frac{c_2}{\bar{n}(z)} \\
=& \left[c_0b_0^2P_t(k) + \frac{c_1}{\bar{n}(z)}\right] \nonumber \\
& \centerdot \int^{1}_{-1} (1+\beta\mu^2)^2F(\mu,k) d\mu + 
\frac{c_2}{\bar{n}(z)}.
\end{align}
The $c_0$ term accounts for any remaining large-scale bias discrepancies
between $P_c(k)$ and the mock data. The $c_1$ term is used to represent
any effects streaming or Kaiser squashing may have on shot-noise. This
is associated with non-linear shot-noise. The $c_2$ term corresponds
to the standard Poisson shot-noise from linear theory. The $c_0$, $c_1$
and $c_2$ are parameters we use to scale the amplitudes of the various
components that go into the Gaussian covariance matrix in order to modify
its shape and $c_3$ can be associated with the integral constraint which
manifests itself as an additive offset in the correlation function.

The likelihood of any such $C^m(c_0,c_1,c_2,c_3)$ given a set of mock
catalogues is
\begin{eqnarray}
\mathcal{L} &=& \prod^{N}_{i=0} \mathcal{L}_i \\
&=& \prod^{N}_{i=0} (2\pi)^{-q/2} (\det C^m)^{-1/2} \e^{-\chi_i^2/2}
\label{eqn:like}
\end{eqnarray} 
where $N$ is the total number of mocks and $q$ is the number of points
to fit. $\chi^2_i = \vec{x}_i(C^m)^{-1}\vec{x}_i^T$ where $\vec{x}_i =
\xi_i(r) - \bar{\xi}(r)$ is a vector of dimension $q$. $\xi_i(r)$ is the
correlation function calculated from the $i$th mock and $\bar{\xi}(r)$ is
the average of the mock correlation functions. Equation (\ref{eqn:like})
can be re-written as
\begin{equation}
L = -2\log \mathcal{L} = Nq\log(2\pi)+N\log(\det C)+\sum^{N}_{i=0}\chi_i^2.
\end{equation}
We would like to find $C^m$ corresponding to the maximum of the 
likelihood function. This is equivalent to finding the $C^m$ that 
corresponds to the minimum of $L$.

Using a downhill simplex minimization scheme and fixing $\sigma_s=4\hMpc$,
we arrive at $c_0=0.89$, $c_1=0.46$, $c_2=1.34$ and $c_3=2.32 \times
10^{-7}$ for redshift space before reconstruction. Here, we have
fixed the value of $\sigma_s$ to reduce computation time, however,
it is possible to include it as a parameter in the maximum likelihood
fit. Allowing $\sigma_s$ to vary gives $\sigma_s=3.9\hMpc$ with most
modification parameters changing by less than $1\%$. Only $c_1$ changes by
$\sim3\%$ due to its partial degeneracy with $\sigma_s$ (when $\sigma_s$
is increased, a larger damping effect is placed on the power spectrum
term which can be compensated for by making $c_1$ larger). The fact that
the likelihood of the fixed $\sigma_s$ case is $0.99$ of the unfixed
case also suggests that fixing $\sigma_s$ is reasonable.

We also investigate the outcome of fixing $c_0=1$, i.e. assuming that
the sample variance given by our model power spectrum suits the data
perfectly. This does not change the log likelihood significantly and we
find that the acoustic scales and errors measured from the mocks as well
as the DR7 data are consistent with the case where $c_0$ is allowed to
vary. In addition, we find that changing the value of $\snl$ that goes
into $P_t(k)$ makes very little difference to the resulting covariance
matrix. Using $\snl=9\hMpc$ instead of $\snl=8\hMpc$ only changes all
the modification parameters and the maximum likelihood by $<1\%$.


The black dots in the top panel of Figure \ref{fig:cov} show the diagonal
(i.e. the $j=i$ elements) of the mock covariance matrix in redshift
space before reconstruction and the black crosses show the corresponding
diagonal of the modified Gaussian covariance matrix. Likewise, the 6th
off-diagonal (i.e. the $j=i+6$ elements) is overplotted in red. The
noise in the mock covariance matrix is obvious. It is evident from
the plot that the modified Gaussian covariance matrix is a good smooth
approximation to the mock covariance values. Hence, we use the modified
Gaussian covariance matrix derived from this maximum likelihood technique
with fixed $\sigma_s=4\hMpc$ as our estimate of the expected errors
on the mock correlation functions. The fitting technique described in
\S\ref{sec:red_norec_ff} utilizes this covariance matrix.

In redshift space after reconstruction, we take the input power spectrum
$P_c(k)$ to be
\begin{equation}
P_c(k) = b_0^2\int^{1}_{-1} F(\mu,k)P_t(k)d\mu
\label{eqn:pc_rec}
\end{equation}
which is just Equation (\ref{eqn:pc}) without the Kaiser term since
our reconstruction algorithm is designed to undo Kaiser squashing. We
assume $\snl=4\hMpc$ and retain $\sigma_s = 4\hMpc$ since we did not
apply any FoG compression. Fitting for the parameters of the modified
Gaussian covariance matrix using the maximum likelihood prescription, we
find $c_0=0.89$, $c_1=0.30$, $c_2=1.45$, $c_3=1.87 \times 10^{-7}$. The
diagonals and 6th off-diagonals of the post-reconstruction mock and
modified Gaussian covariance matrices are plotted in the bottom panel
of Figure \ref{fig:cov}. One can see that, as in the pre-reconstruction
case, our modified Gaussian approximation fits the mock covariances well.

\begin{figure}
\vspace{0.4cm}
\centering
\epsfig{file=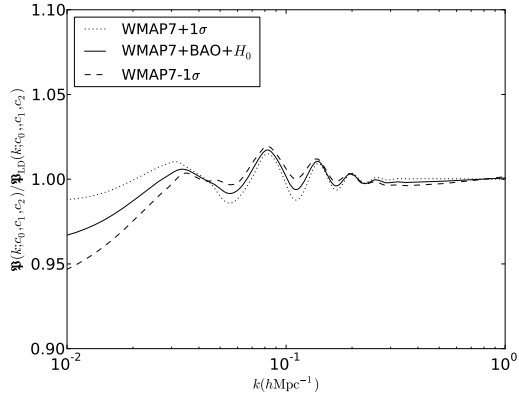, width=0.8\linewidth, clip=} 
\caption{The ratio of $\mathfrak{P}(k;c_0,c_1,c_2)$ terms (see Equation
(\ref{eqn:modc}) and surrounding text) found in the definition of the
modified Gaussian covariance matrix (MGCM). These MGCMs were all fit to
the covariances calculated from the LasDamas mocks in redshift space after
reconstruction. The numerator corresponds to MGCMs constructed using 3
non-LasDamas cosmologies. The denominator corresponds to the MGCM in the
LasDamas cosmology. The 3 non-LasDamas cosmologies are WMAP7+BAO+$H_0$
(solid line) and the 1$\sigma$ limits of this cosmology (+1$\sigma$ is
shown as the dotted line and -1$\sigma$ is shown as the dashed line). It
is seen that the 3 lines are all $\sim1$ to within $\sim5\%$. This
indicates that if we input a power spectrum with cosmology different to
LasDamas, our modification parameters can balance this input and the
noise terms to recover a covariance matrix that matches the expected
LasDamas covariances fairly well.
\label{fig:comppow}}
\end{figure}

In the post-reconstruction case, we also test that by using a different
cosmology from LasDamas to derive $P_t(k)$, it is still possible to
obtain a modified Gaussian covariance matrix that suits the mock data
using our maximum likelihood method. In Figure \ref{fig:comppow}, we
show $\mathfrak{P}(k;c_0,c_1,c_2)$ for the WMAP7+BAO+$H_0$ cosmology
\citep{Komatsu10} divided by the corresponding LasDamas values (solid
line). For reference, the WMAP7 cosmological parameters of relevance are
$H_0 = 70.2 \pm 1.4$, $100\Omega_b h^2 = 2.255 \pm 0.054$, $\Omega_c
h^2=0.1126 \pm 0.0036$, $n_s = 0.968 \pm 0.012$ and $\sigma_8 = 0.816
\pm 0.024$. The dotted (dashed) lines are for cosmologies derived by
adding (subtracting) the 1$\sigma$ errors from the WMAP7 values quoted
above. One can see that the 3 lines are all $\sim1$ to within $\sim5\%$
near the acoustic scale indicating that the modification parameters
are capable of adjusting the power spectrum and the noise terms in
the Gaussian covariance matrix to match the LasDamas covariances. In
\S\ref{sec:red_rec_fr}, we show that using these different covariance
matrices yield consistent acoustic scale measurements and errors to
those obtained using the correct LasDamas cosmology.

\subsection{Fitting Forms}\label{sec:red_norec_ff}

We fit the mock redshift-space correlation functions $\sxi$ over the
range $30<r<200\hMpc$ using the fiducial form (justification to follow)
\begin{equation}
\xi^{fit}(r) = B^2\xi_m(\alpha r)+A(r)
\label{eqn:fform}
\end{equation}
where
\begin{equation}
A(r) = \frac{a_1}{r^2} + \frac{a_2}{r} + a_3.
\label{eqn:aform}
\end{equation}
The parameters of the fit are $B^2$, $\alpha$, $a_1$, $a_2$ and $a_3$. The
latter are linear nuisance parameters.

The scale dilation parameter $\alpha$ represents how much the acoustic
peak in the data is shifted relative to that in the model. Therefore,
it \textit{is} our measurement of the acoustic scale and the parameter we
are most interested in extracting robustly from our fits. An $\alpha>1$
indicates a shift towards smaller scales and an $\alpha<1$ indicates a
shift towards larger scales.

The template correlation function, $\xi_m(r)$, takes on the form
\begin{equation} 
\xi_m(r) = \int \frac{k^2dk}{2\pi^2}P_m(k)\jbz(kr)\e^{-k^2a^2},
\end{equation}
where $P_m(k) = b^2P_t(k)$ and $P_t(k)$ is defined as in Equation
(\ref{eqn:template}). We perform the transformation from Fourier space to
configuration space using an additional Gaussian term to provide high-$k$
damping for the oscillatory transform kernel $\jbz(kr)$. This is conducive
to better numerical convergence in the integration. We pick $a=1\hMpc$,
a scale small enough such that the effects of the damping will not be
significant within our fitting range. 

\begin{figure}
\vspace{0.4cm}
\centering
\epsfig{file=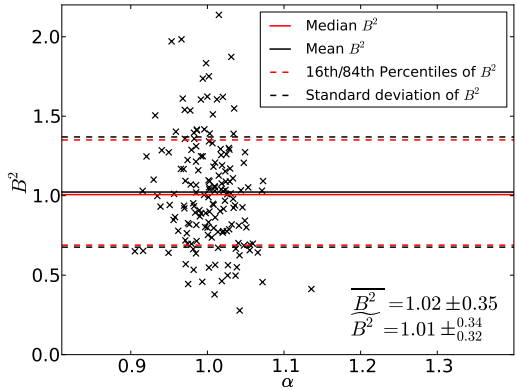, width=0.8\linewidth, clip=}
\caption{The values of $B^2$ versus $\alpha$ fit from the mocks
in redshift space before reconstruction. To ensure that $B^2$ is
non-negative, these values were obtained through fitting the 160 mock
redshift-space correlation functions using the non-linear parameter
$\log(B^2)$ instead of $B^2$. The solid red line indicates the
median $B^2$ value and the solid black line indicates the mean. The
dashed red lines indicate the 16th and 84th percentiles of $B^2$
(quoted with the median $\widetilde{B^2}$). The dashed black lines
correspond to the 1$\sigma$ deviations from the mean (quoted with the
mean $\overline{B^2}$). One can see that $B^2$ can reach values as
high as $\sim2.1$ and as low as $\sim0.3$. This substantial variation
is possible because the $A(r)$ term can compensate, and is therefore
not physically motivated. Hence to disfavour these extreme values, we
place a weak Gaussian prior on $\log(B^2)$ that has mean equal to 0 and
standard deviation equal to 0.4.
\label{fig:avb}} 
\end{figure}

The $b^2$ term is a constant normalization factor that we obtain by
taking the ratio of the mock correlation function being fit and the
configuration space transform of $P_t(k)$ at $r=50\hMpc$. This ensures
that the fitting normalization $B^2$ is of order unity. The normalization
must be positive, so we perform our fits with the non-linear parameter
$\log(B^2)$. Note that $B^2$ can vary substantially as long as the $A(r)$
function can compensate. This creates large variation in the amplitude
of the acoustic peak which is not physically motivated. We find that the
scatter in $B^2$ can be large with values being as high as $\sim2.1$
and as low as $\sim0.3$, especially in the mocks where the acoustic
signal does not appear to be as strong. This is summarized in Figure
\ref{fig:avb} where we have plotted $B^2$ versus best-fit $\alpha$
obtained through fitting the 160 mock correlation functions in redshift
space. For a careful description of the information plotted, please see
the figure caption. To disfavour extreme values of $B^2$, we place a weak
Gaussian prior on $\log(B^2)$ with a mean of 0 and standard deviation
of 0.4. For simplicity, we also apply this prior to redshift space with
reconstruction and real space with and without reconstruction.

We pick the form for $A(r)$ in Equation (\ref{eqn:aform}) due to
its simplicity in Fourier space. Since the transform of $r^{n}$ is
proportional to $k^{-3-n}$, in Fourier space $A(r)$ takes on the form
\begin{equation}
A'(k) = \frac{a'_1}{k} + \frac{a'_2}{k^2} + \frac{a'_3}{k^3}.
\label{eqn:akform}
\end{equation}
In addition to the fiducial $A(r)$ form in Equation (\ref{eqn:aform}),
we will also be analyzing various other forms of $A(r)$ throughout this
paper. We will refer to $A(r)=A'(k)=0$ as $poly0$, $A(r) = a_1/r^2$
(first order inverse polynomial in $k$) as $poly1$, $A(r) = a_1/r^2 +
a_2/r$ (second order inverse polynomial in $k$) as $poly2$ and $A(r) =
a_1/r^2 + a_2/r + a_3 + a_4r$ (fourth order inverse polynomial in $k$)
as $poly4$. Note that the fiducial form corresponds to $poly3$.

\begin{figure*}
\vspace{0.4cm}
\centering
\epsfig{file=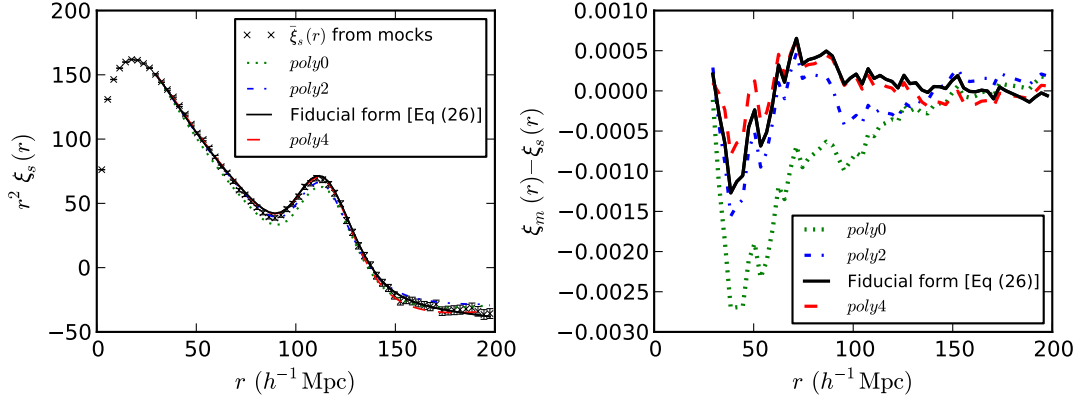, width=0.8\linewidth, clip=}
\caption{(left) Fits to the average redshift-space correlation function of
the mocks (black crosses) using Equation (\ref{eqn:fform}) with $A(r)$
being $poly0$ (dotted green line), $poly2$ (dash-dotted blue line),
fiducial form (Equation (\ref{eqn:aform})) (solid black line) and $poly4$
(dashed red line). (right) The corresponding residuals of the fits
(note that the fitting range is $30<r<2000\hMpc$). One can see that the
fit using the fiducial form matches the data better than the fits with
$poly0$ and $poly2$. However, the improvement between the fiducial form
and $poly4$ is negligible as reflected by the similar shapes of the solid
and dashed curves. These results motivate our choice of $A(r)$ given in
Equation (\ref{eqn:aform}). We have also allowed $\Sigma_{nl}$ to vary
in these fits. Using the fiducial form, we find $\Sigma_{nl}=8.1\hMpc$,
which is close to the value we assumed in deriving the covariance matrix.
\label{fig:akfit}}
\end{figure*}

We find that going up to the constant term in $A(r)$ as in the
fiducial form gives a good fit to the average of the mock correlation
functions. This is shown in Figure \ref{fig:akfit} where in the left
panel we have plotted the fits to the average mock, redshift-space
correlation function (black crosses) using Equation (\ref{eqn:fform})
and various forms for $A(r)$. The $poly0$, $poly2$, fiducial form and
$poly4$ cases are shown as the dotted green, dash-dotted blue, solid
black and the dashed red lines respectively. The corresponding residuals
are shown in the right panel.

We have also allowed $\Sigma_{nl}$ to vary in these fits and find
that for the fiducial form, $\Sigma_{nl}=8.1\hMpc$. This is close to
the value of $8\hMpc$ we assumed in the estimation of the covariance
matrix. The results from fits to the mean mock correlation functions
using the fiducial form are summarized in Table \ref{tab:mres}.

\begin{table}
\caption{\label{tab:mres} Fit results to average mock correlation functions}

\begin{tabular}{lcc}
\hline
&$\alpha$&$\Sigma_{nl}$ \\
&&($\hMpc$) \\
\hline

Redshift space w/o reconstruction&
$1.003$& 
$8.1$\\
Real space w/o reconstruction&
$1.002$&
$6.6$\\
Redshift space w/ reconstruction&
$1.003$&
$4.4$\\
Real space w/ reconstruction&
$0.999$&
$3.0$\\ 

\hline
\end{tabular}
\end{table}

The $\chi^2$ per degree-of-freedom (dof) goes down from 2.7 for $poly0$
to 1.4 for the fiducial form. The decrease from the fiducial form to
$poly4$ is much smaller (only $\sim0.2$) as evidenced by the similarity
in shape between the solid curve and the dashed curve. Although the
value of $\chi^2$ per dof is still large for the fiducial form, we note
that the error bars expected when fitting each individual mock will be
much larger and thus result in reasonable values of $\chi^2$ as will be
shown in \S\ref{sec:red_norec_fr}. In principle we could further lower
$\chi^2$ by taking $A(r)$ out to higher orders of $r$, however we then
run the risk of having the nuisance parameters fit the noise in the data.

Recall that our ultimate goal is to measure the acoustic scale, $\alpha$,
from the data. This can be done by finding the value of $\alpha$ that
gives rise to the best-fit model to the data. Our models are non-linear
in $\alpha$ and the normalization factor $\log(B^2)$, so we can nest a
linear least-squares fitter inside a non-linear fitting routine, which
in our case is a downhill simplex. The former calculates $a_1$, $a_2$
and $a_3$ for each value of $\alpha$ and $B^2$ the latter steps to. Then,
to find the best-fit $\alpha$, we use the non-linear fitter to minimize
the $\chi^2$ goodness-of-fit indicator
\begin{equation}
\chi^2(\alpha,B^2) = [\vec{d}-\vec{m}(\alpha,B^2)]^TC^{-1}[\vec{d}-\vec{m}(\alpha,B^2)]
\end{equation}
where $\vec{d}$ is the correlation function measured from the mocks
and $\vec{m}(\alpha,B^2)$ is the best-fit model at each $\alpha$ and
$B^2$. $C^{-1}$ is the inverse of the covariance matrix. Recall that we
use the modified Gaussian covariance matrix (MGCM) described in Equation
(\ref{eqn:modc}) of \S\ref{sec:red_norec_cm} here.

Based on our fiducial form defined in Equations (\ref{eqn:fform} \&
\ref{eqn:aform}), we define a fiducial model for redshift space over a
fitting range of $30<r<200\hMpc$. $\xi_m(k)$ is derived from the LasDamas
cosmology using $\snl=8\hMpc$. We denote the fiducial model with subscript
$[f]$ throughout this paper unless otherwise stated. We perform the
above prescribed fitting algorithm on all 160 of our mock catalogues
using the fiducial model to obtain a best-fit value of $\alpha$ for each.

\begin{figure}
\vspace{0.4cm}
\centering
\epsfig{file=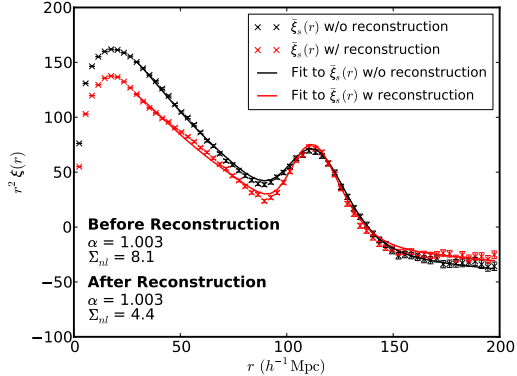, width=0.8\linewidth, clip=}
\caption{Fits to the average of the mock redshift-space correlation
functions before and after reconstruction. The black crosses are the mock
data before reconstruction and the black line is its best-fit model. The
red crosses are the mock data after reconstruction and the red line is
its best-fit model. We have allowed $\snl$ to vary in these fits, the
results are summarized on the plot. We find that before reconstruction,
the shift in the acoustic peak is already very small ($\alpha\sim1$),
so we do not expect reconstruction to shift the peak much closer to
its predicted linear theory position. However, we find that $\snl$
was reduced by a factor of 1.8 after reconstruction, indicating that
reconstruction was able to mitigate the acoustic peak smearing due to
non-linear structure growth.
\label{fig:comprec}}
\end{figure}

For redshift space with reconstruction, we use the same fiducial fitting
form defined by Equations (\ref{eqn:fform} \& \ref{eqn:aform}). The mean
of the mock redshift-space correlation functions before (black) and after
(red) reconstruction are shown in Figure \ref{fig:comprec}. The data
are represented by crosses and the fits to the data using the fiducial
fitting form are shown as solid lines. The results from these fits are
also summarized in Table \ref{tab:mres}.

Since before reconstruction, $\alpha$ is already very close to 1, we
would not expect reconstruction to have a large affect on the measured
acoustic scale, which is exactly what we see. However, we find that after
reconstruction, $\snl=4.4\hMpc$, which is a factor of 1.8 reduction from
its pre-reconstruction value of $\snl=8.1\hMpc$. This decrease in $\snl$
indicates that reconstruction was successful at reducing the smearing
(large $\snl$) of the acoustic peak caused by non-linear structure
growth. Visually, this can be seen as the sharpening of the acoustic
feature in the average of the mocks after reconstruction. A sharpened peak
is easier to centroid and should result in a more accurate measurement
of the acoustic scale.

We define the fiducial model in redshift space after reconstruction to be
identical to the pre-reconstruction model except with $\snl=4\hMpc$. This
is the same value as that used to derive the MGCM for post-reconstruction
redshift space. This is not a bad approximation as we have just shown the
fit to the average of the mock correlation functions has $\snl=4.4\hMpc$.

\section{LasDamas Redshift Space Results}\label{sec:red_fit}

\subsection{Without Reconstruction}\label{sec:red_norec_fr}
We begin by studying the LasDamas mocks in redshift space without
reconstruction. We perform our fits on the mocks using the fiducial
model and fitting techniques outlined in the previous section and find
that a few of the mocks do not give compelling measurements of $\alpha$
due to their relatively weak acoustic features. We attempt to identify
which mocks have poorly constrained values of $\alpha$ by performing
our fits at different test values of $\alpha_i$ using our fiducial model
and measuring the resulting $\chi^2$. This allows us to calculate
\begin{equation}
p(\alpha_i) = \frac{\e^{-\chi^2(\alpha_i)/2}}{\sum_{j}\e^{-\chi^2(\alpha_j)/2}
\Delta \alpha},
\label{eqn:pra}
\end{equation}
the probability of measuring the acoustic scale to be $\alpha = \alpha_i$
from a particular mock. Here, the denominator is a normalization factor
equivalent to integrating over all test values of $\alpha$ where $\Delta
\alpha$ is the difference between the test values. We calculate a mean
and a standard deviation for our $p(\alpha)$ distributions as
\begin{eqnarray}
\langle\alpha\rangle &=& \sum_{i}\alpha_i p(\alpha_i) \Delta \alpha \\
\label{eqn:pralph}
\sigma_\alpha &=& \sqrt{\sum_{i}{[\alpha_i-\langle\alpha\rangle]^2 
p(\alpha_i)}\Delta \alpha}.
\label{eqn:stddeva}
\end{eqnarray}
A small standard deviation indicates that the best-fit $\alpha$ measured
from the mock is well constrained. Conversely, a large standard deviation
indicates that it is difficult to measure an accurate value of $\alpha$
from the mock.

\begin{figure*}
\vspace{0.4cm}
\begin{tabular}{c}
\epsfig{file=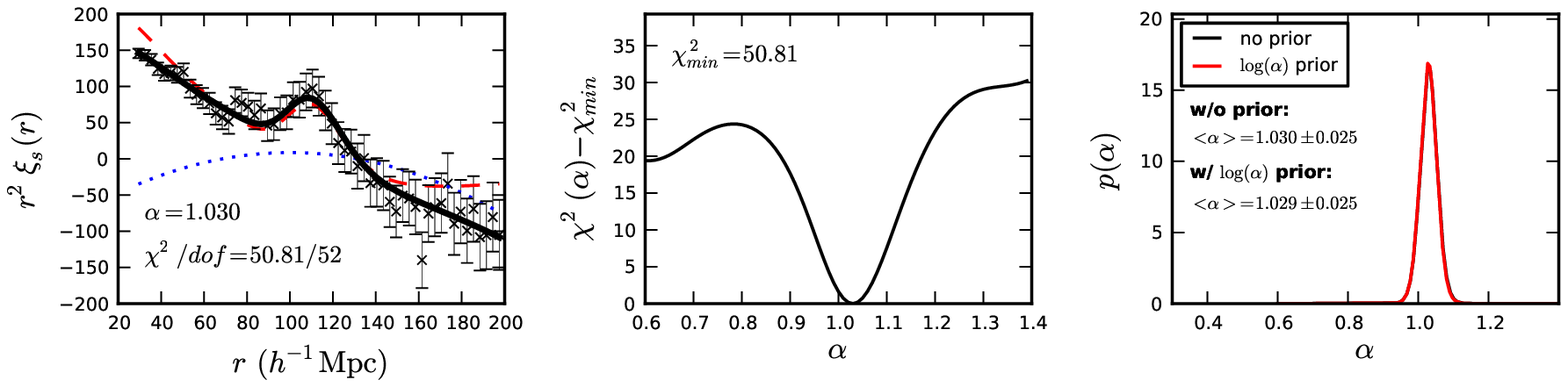, height=0.22\linewidth, width=0.95\linewidth, clip=} \\ 
\epsfig{file=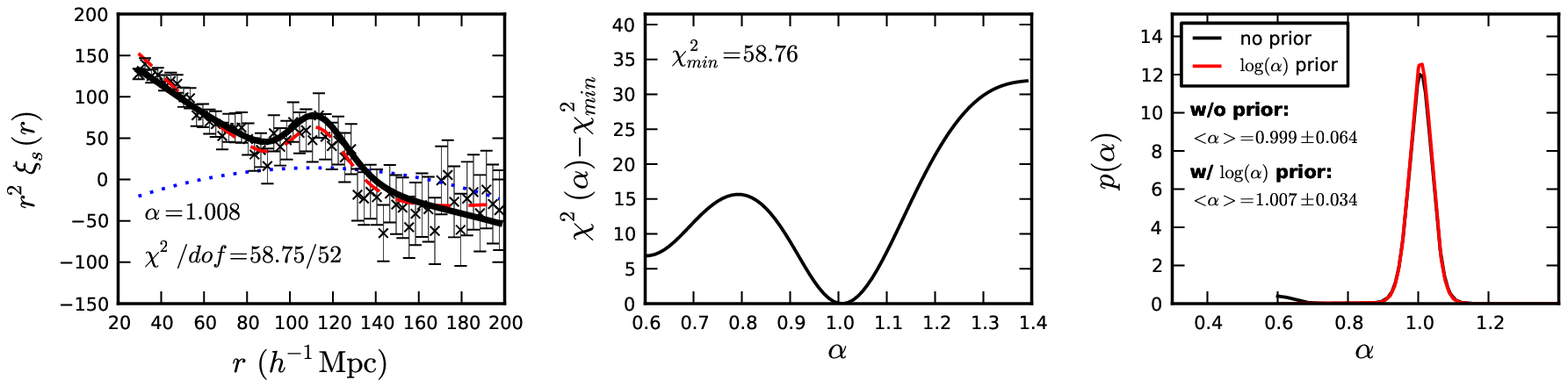, height=0.22\linewidth, width=0.95\linewidth, clip=} \\
\epsfig{file=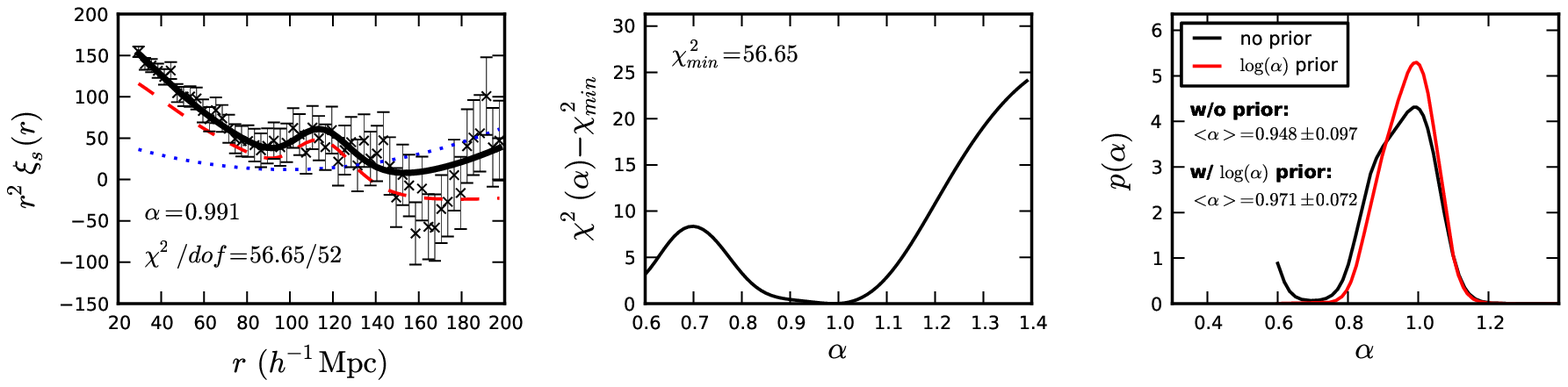, height=0.22\linewidth, width=0.95\linewidth, clip=} \\
\epsfig{file=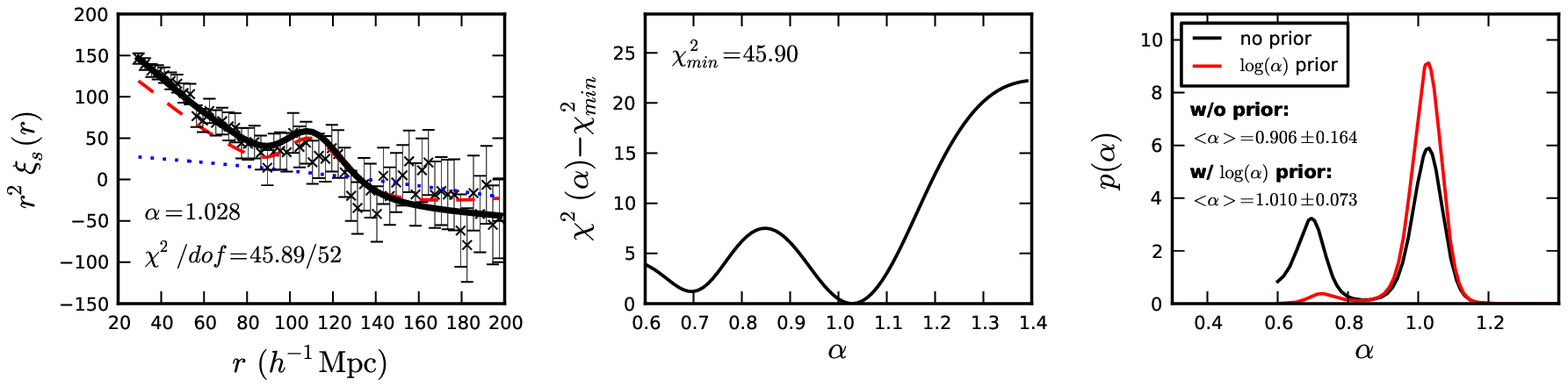, height=0.22\linewidth, width=0.95\linewidth, clip=} \\
\end{tabular}
\caption{Sample fit results from the fiducial model for our redshift-space
mocks, $\sxi$. (rows 1+2) Results from mocks that have well constrained
measures of $\alpha$. (rows 3+4) Results from mocks that have poorly
constrained measures of $\alpha$. (left column) Actual fits using the
fiducial model. The model (black line) is overplotted on the mock data
(black crosses with error bars). The dotted blue line corresponds to
the $A(r)$ term in the model and the dashed red line corresponds to
the $B^2\xi_m(\alpha r)$ term. Comparing rows 1 \& 2 with rows 3 \& 4
suggests that there must be a fairly prominent acoustic peak in order
to obtain a well constrained measurement of $\alpha$. (middle column)
The $\Delta\chi^2 = \chi^2(\alpha) - \chi^2_{min}$ curve. The large
differences in $\chi^2$ between the minimum and the plateaus of the
well constrained cases indicate that we have robust detections of the
$\chi^2$ minimum and hence the best-fit $\alpha$ which corresponds to this
minimum. In the poorly constrained mocks, the difference is much smaller
and there may be double minima at small $\Delta\chi^2$ from each other,
indicating a poor detection of the best-fit $\alpha$. (right column)
The $p(\alpha)$ distribution versus $\alpha$ (black line) calculated
from $\chi^2(\alpha)$, Equation (\ref{eqn:pra}). The red line is the same
curve but with a 15\% Gaussian prior on $\log(\alpha)$. We say best-fit
$\alpha$ is well constrained in a mock, when the standard deviation of
the $p(\alpha)$ distribution is small, and not well constrained when
the standard deviation is large, even after the prior is applied. In some
mocks, we see significant $\chi^2$ differences between the minimum and the
plateau, however, the $\sigma_\alpha$ measured may still be large. This
is due to a downturn in the $\chi^2(\alpha)$ curve at $\alpha\sim0.7$
(see the second row). Such a downturn is not physically motivated
because it is caused by the model attempting to hide the acoustic peak
in the larger errors at large $r$. Hence, we introduce the prior on
$\log(\alpha)$ to suppress this effect.
\label{fig:exfigs}}
\end{figure*}

In Figure \ref{fig:exfigs}, we have plotted the fit results using the
fiducial model for 2 of our mock redshift-space correlation functions,
$\sxi$, that appear to have well constrained values of $\alpha$ (upper 2
panels) and 2 that do not (lower 2 panels). These are representative of
the other well and poorly constrained mocks in our set. For a detailed
description of the information plotted, please see the figure caption.

The left column in each set shows the actual fit to the mock correlation
function using the fiducial model. The best-fit values of $\alpha$
and their corresponding minimum $\chi^2/dof$ are given on the plots. In
comparing the well constrained mocks to the poorly constrained mocks,
we can see that in order to obtain a fairly certain measurement of
best-fit $\alpha$, the mock must have a prominent acoustic peak. If
one ignores the best-fit models which can be used to guide the eye,
the acoustic features in both of the poorly constrained mocks are much
weaker than in the well constrained mocks.

The middle column in each set shows the $\Delta\chi^2 = \chi^2(\alpha)
- \chi^2_{min}$ curve for each mock. The $\chi^2(\alpha)$ here is the
same as that which appears in Equation (\ref{eqn:pra}) and $\chi^2_{min}$
is the minimum of $\chi^2(\alpha)$, i.e. $\chi^2$ at the best-fit value
of $\alpha$. One can see that for the well constrained mocks, the curve
is nearly parabolic around the minimum (expected if $\alpha$ is Gaussian
distributed) and then plateaus at extreme values of $\alpha$. The height
in $\Delta\chi^2$ of these plateaus can be used as a proxy for the
significance of the $\chi^2$ minimum. In the poorly constrained mocks,
the plateau occurs at much smaller $\Delta\chi^2$ values. In addition,
there may be double minima at small differences in $\chi^2$. These
indicate that we are not detecting the $\chi^2$ minimum (and hence
best-fit $\alpha$) robustly.


In the right panels, we use these $\chi^2(\alpha)$ curves
to calculate their corresponding $p(\alpha)$ distributions
using Equation (\ref{eqn:pra}). These are plotted as the black
lines. The red lines include an additional 15\% Gaussian prior on
$\log(\alpha)$, i.e. $\chi^2(\alpha) \rightarrow \chi^2(\alpha) +
\left( \frac{\log(\alpha)}{0.15} \right)^2$. We apply this weak prior
because in some of the cases where the best-fit $\alpha$ should be well
constrained, i.e. in the second row where the $\Delta\chi^2$ curve is
nicely parabolic around a minimum that is at a significant $\Delta\chi^2
\sim 15$ away from the plateau, we still measure a large $\sigma_\alpha$
from the $p(\alpha)$ distribution. This is due to a slight downturn
in the $\chi^2$ versus $\alpha$ curve (and hence an upturn in the
$p(\alpha)$ distribution) at $\alpha\sim0.7$. At these small $\alpha$,
the acoustic peak in the model is getting pushed out to large $r$. Here,
the error bars are larger so the fitter is having an easier time hiding
the acoustic peak in the errors. Since this downturn in $\chi^2$ is not
physically motivated, we apply this prior to downweight the $\chi^2$
values at extreme $\alpha$. One can see the effectiveness of the prior
by noticing that the upturn in $p(\alpha)$ disappears after the prior
is applied to the mock in the second row.

As mentioned previously, the acoustic scale is well constrained in the
mocks that have very small standard deviations in $\alpha$. In these
cases, the inferred standard deviation can become even smaller after the
prior is applied, not due to any dramatic change in the general shape of
the curve but rather because the tails become suppressed by the prior. The
mocks where $\alpha$ is not well constrained, however, have very broad
distributions with large standard deviations even after a prior is
applied. This suggests that we may segregate the well constrained mocks
from the poorly constrained mocks by setting a cutoff in the standard
deviation after applying the prior on $\log(\alpha)$. We also note here
that, after applying the prior, the mean $\alpha$ of the $p(\alpha)$
distribution should be fairly close to the best-fit $\alpha$ from the
fiducial model for the well constrained mocks. This is indeed what we
observe. Any discrepancy is likely due to the fact that the $p(\alpha)$
distribution is not exactly Gaussian.

\begin{figure}
\vspace{0.4cm}
\centering
\epsfig{file=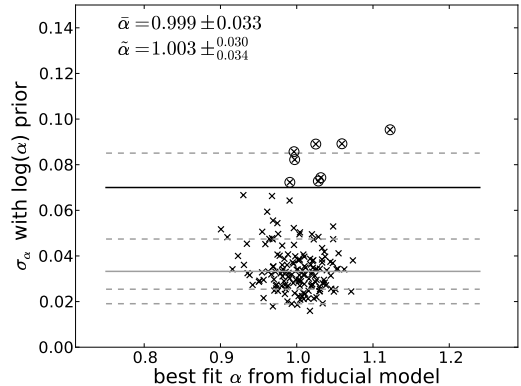, width=0.8\linewidth, clip=}
\caption{The standard deviations of $p(\alpha)$ for the redshift-space
mocks plotted against their best-fit $\alpha$ values measured using
the fiducial model. A large standard deviation indicates that $\alpha$
is poorly constrained in its corresponding mock. The solid grey line
indicates the median of the standard deviations while the dashed grey
lines indicate the 98th, 84th, 16th and 2nd percentiles. We see that most
of these poorly constrained mocks fall above a standard deviation of
7\%. Hence we impose a 7\% cutoff (black horizontal line) in standard
deviation and remove all the mocks with standard deviations above
this cutoff from our fitting sample. The mocks with uncertain $\alpha$
measurements based on this metric are circled. There are 8 of such mocks,
which is $\sim4\%$ of our sample. The mean and median values of best-fit
$\alpha$ measured using the fiducial model after removing the poorly
constrained mocks are listed on the plot.
\label{fig:sdva_rednr}}
\end{figure}

A plot of the standard deviations versus the best-fit $\alpha$ values from
the fiducial model are shown in Figure \ref{fig:sdva_rednr}. The median of
the standard deviations is indicated by the solid grey line and the 98th,
84th, 16th and 2nd percentile levels are indicated by the dashed grey
lines. We see that the poorly constrained mocks mostly lie at standard
deviations larger than 7\% (indicated by the black horizontal line in the
plot). Hence, we make a cutoff in standard deviation at 7\% and take all
mocks that lie above this cutoff to have poorly constrained measurements
of $\alpha$ (circled in black). Both of the poorly constrained mocks
shown in Figure \ref{fig:exfigs} fall into this category. For our
redshift-space mocks before reconstruction, we find that 8 ($\sim4\%$)
have fairly poor measurements of $\alpha$. The mean and median values
of the best-fit $\alpha$ from the fiducial model are given in the plot
after removing the poorly constrained mocks. We use this procedure to
remove these poorly constrained mocks from our $\alpha$-fitting sample
before proceeding. Note that they are still included in our covariance
matrix derivation.

To verify the robustness of our covariance modeling and the fiducial
model, we compare the $\alpha$ values we measure from the fiducial model
to those we measure when the fiducial model parameters are slightly
changed or if we fit using a different covariance matrix. A summary of
the results of these fits, after removing the poorly constrained mocks
as just described, can be found in Table \ref{tab:alphasred}. We quote the
mean of any quantity $x$ and its standard deviation as $\bar{x}$ and we
quote the median with the 84th/16th percentiles as $\tilde{x}$ throughout
this paper. For the fiducial model, we find $\bar{\alpha}=0.999\pm0.033$
and $\tilde{\alpha} = 1.003 \pm^{0.030}_{0.034}$. This means that we
can measure the shift in the acoustic scale to about 3-3.5\% accuracy
from our mocks before reconstruction.

\begin{table*}
\caption{Redshift space fitting results for various models}
\label{tab:alphasred}

\begin{tabular}{@{}lccccc}

\hline
Model&
$\bar{\alpha}$&
$\tilde{\alpha}$&
$\overline{\Delta\alpha}$\footnotemark[1]&
$\widetilde{\Delta\alpha}$&
$\overline{\chi^2}/dof$\\

\hline
\multicolumn{6}{c}{Redshift Space without Reconstruction}\\
\hline

Fiducial $[f]$ &
$0.999 \pm 0.033$&
$1.003 \pm^{0.030}_{0.034}$&
--&
--&
52.96/52\\
\\[-1.5ex]
Fit with 15\% larger $\Omega_m$ using fiducial $A(r)$.\footnotemark[2] &
$0.998 \pm 0.034$&
$1.001 \pm^{0.029}_{0.035}$&
$-0.002 \pm 0.002$&
$-0.001 \pm^{0.001}_{0.002}$&
53.29/52\\
\\[-1.5ex]
Fit with $n_s=0.96$ using fiducial $A(r)$. &
$1.001 \pm 0.033$&
$1.004 \pm^{0.030}_{0.034}$&
$0.002 \pm 0.001$&
$0.001 \pm^{0.001}_{0.001}$&
52.92/52\\
\\[-1.5ex]
Fit with $N_{rel}=4$ using fiducial $A(r)$. &
$1.006 \pm 0.033$&
$1.008 \pm^{0.032}_{0.033}$&
$0.007 \pm 0.005$&
$0.006 \pm^{0.001}_{0.001}$&
52.85/52\\
\\[-1.5ex]
Fit with $\snl \rightarrow 0$. &
$0.996 \pm 0.036$&
$0.997 \pm^{0.032}_{0.032}$&
$-0.003 \pm 0.020$&
$-0.004 \pm^{0.013}_{0.013}$&
54.29/52\\
\\[-1.5ex]
Fit with $\snl \rightarrow \snl+2$. &
$1.001 \pm 0.034$&
$1.005 \pm^{0.028}_{0.034}$&
$0.002 \pm 0.005$&
$0.002 \pm^{0.004}_{0.005}$&
53.28/52\\
\\[-1.5ex]
Fit with $poly0$. &
$0.995 \pm 0.035$&
$0.996 \pm^{0.034}_{0.030}$&
$-0.004 \pm 0.012$&
$-0.003 \pm^{0.007}_{0.008}$&
56.03/55\\
\\[-1.5ex]
Fit with $poly2$. &
$0.997 \pm 0.033$&
$1.002 \pm^{0.030}_{0.035}$&
$-0.002 \pm 0.004$&
$-0.001 \pm^{0.002}_{0.003}$&
54.44/53\\
\\[-1.5ex]
Fit with $poly4$. &
$0.999 \pm 0.033$&
$1.002 \pm^{0.031}_{0.033}$&
$0.000 \pm 0.001$&
$0.000 \pm^{0.000}_{0.000}$&
51.81/51\\
\\[-1.5ex]
Fit with $50<r<200\hMpc$ fitting range. &
$1.000 \pm 0.033$&
$1.004 \pm^{0.030}_{0.033}$&
$0.001 \pm 0.005$&
$0.001 \pm^{0.003}_{0.003}$&
45.73/45\\
\\[-1.5ex]
Fit with $20<r<200\hMpc$ fitting range. &
$1.002 \pm 0.033$&
$1.004 \pm^{0.033}_{0.033}$&
$0.003 \pm 0.008$&
$0.003 \pm^{0.006}_{0.006}$&
59.45/57\\
\\[-1.5ex]
Fit with $70<r<150\hMpc$ fitting range. &
$0.999 \pm 0.033$&
$1.001 \pm^{0.033}_{0.031}$&
$0.000 \pm 0.010$&
$-0.000 \pm^{0.009}_{0.008}$&
21.83/22\\
\\[-1.5ex]
Fit using mock covariance matrix. &
$1.002 \pm 0.027$&
$1.003 \pm^{0.025}_{0.026}$&
$0.003 \pm 0.022$&
$0.003 \pm^{0.018}_{0.017}$&
52.80/52\\
\hline
\multicolumn{6}{c}{Redshift Space with Reconstruction}\\
\hline
Fiducial $[f]$ &
$1.001 \pm 0.021$&
$1.001 \pm^{0.020}_{0.022}$&
--&
--&
53.69/52\\
\\[-1.5ex]
Fit with 15\% larger $\Omega_m$ using fiducial $A(r)$.\footnotemark[2] &
$1.001 \pm 0.021$&
$1.001 \pm^{0.020}_{0.022}$&
$-0.000 \pm 0.001$&
$-0.000 \pm^{0.001}_{0.001}$&
51.86/52\\
\\[-1.5ex]
Fit with $n_s=0.96$ using fiducial $A(r)$. &
$1.002 \pm 0.021$&
$1.002 \pm^{0.020}_{0.022}$&
$0.001 \pm 0.000$&
$0.001 \pm^{0.000}_{0.000}$&
51.84/52\\
\\[-1.5ex]
Fit with $N_{rel}=4$ using fiducial $A(r)$. &
$1.006 \pm 0.021$&
$1.006 \pm^{0.020}_{0.022}$&
$0.005 \pm 0.001$&
$0.005 \pm^{0.001}_{0.001}$&
51.95/52\\
\\[-1.5ex]
Fit with $\snl \rightarrow 0$. &
$1.001 \pm 0.022$&
$1.001 \pm^{0.022}_{0.020}$&
$-0.000 \pm 0.004$&
$-0.001 \pm^{0.004}_{0.003}$&
53.83/52\\
\\[-1.5ex]
Fit with $\snl \rightarrow \snl+2$. &
$1.002 \pm 0.021$&
$1.001 \pm^{0.022}_{0.020}$&
$0.001 \pm 0.004$&
$0.001 \pm^{0.002}_{0.004}$&
53.99/52\\
\\[-1.5ex]
Fit with $poly0$. &
$1.000 \pm 0.021$&
$1.000 \pm^{0.019}_{0.020}$&
$-0.002 \pm 0.004$&
$-0.001 \pm^{0.004}_{0.004}$&
57.34/55\\
\\[-1.5ex]
Fit with $poly2$. &
$1.000 \pm 0.021$&
$1.001 \pm^{0.021}_{0.022}$&
$-0.001 \pm 0.002$&
$-0.001 \pm^{0.001}_{0.001}$&
55.41/53\\
\\[-1.5ex]
Fit with $poly4$. &
$1.001 \pm 0.021$&
$1.001 \pm^{0.021}_{0.022}$&
$-0.000 \pm 0.000$&
$-0.000 \pm^{0.000}_{0.000}$&
52.68/51\\
\\[-1.5ex]
Fit with $50<r<200\hMpc$ fitting range. &
$1.001 \pm 0.021$&
$1.000 \pm^{0.023}_{0.021}$&
$0.000 \pm 0.002$&
$0.000 \pm^{0.001}_{0.002}$&
46.58/45\\
\\[-1.5ex]
Fit with $20<r<200\hMpc$ fitting range. &
$1.005 \pm 0.021$&
$1.004 \pm^{0.022}_{0.020}$&
$0.004 \pm 0.003$&
$0.004 \pm^{0.003}_{0.003}$&
60.06/57\\
\\[-1.5ex]
Fit with $70<r<150\hMpc$ fitting range. &
$1.001 \pm 0.026$&
$1.003 \pm^{0.022}_{0.019}$&
$-0.000 \pm 0.012$&
$0.000 \pm^{0.004}_{0.005}$&
22.69/22\\
\\[-1.5ex]
Fit using mock covariance matrix. &
$1.004 \pm 0.017$&
$1.003 \pm^{0.017}_{0.014}$&
$0.003 \pm 0.015$&
$0.001 \pm^{0.015}_{0.010}$&
54.22/52\\
\hline
\end{tabular}

\medskip
$^{1}$ $\Delta \alpha = \alpha_{[i]} - \alpha_{[f]}$, where $i$ is the model number. \\
$^{2}$ We scale the measured sound horizons to the LasDamas cosmology where necessary.

\end{table*}

\begin{figure}
\vspace{0.4cm}
\begin{tabular}{c}
\epsfig{file=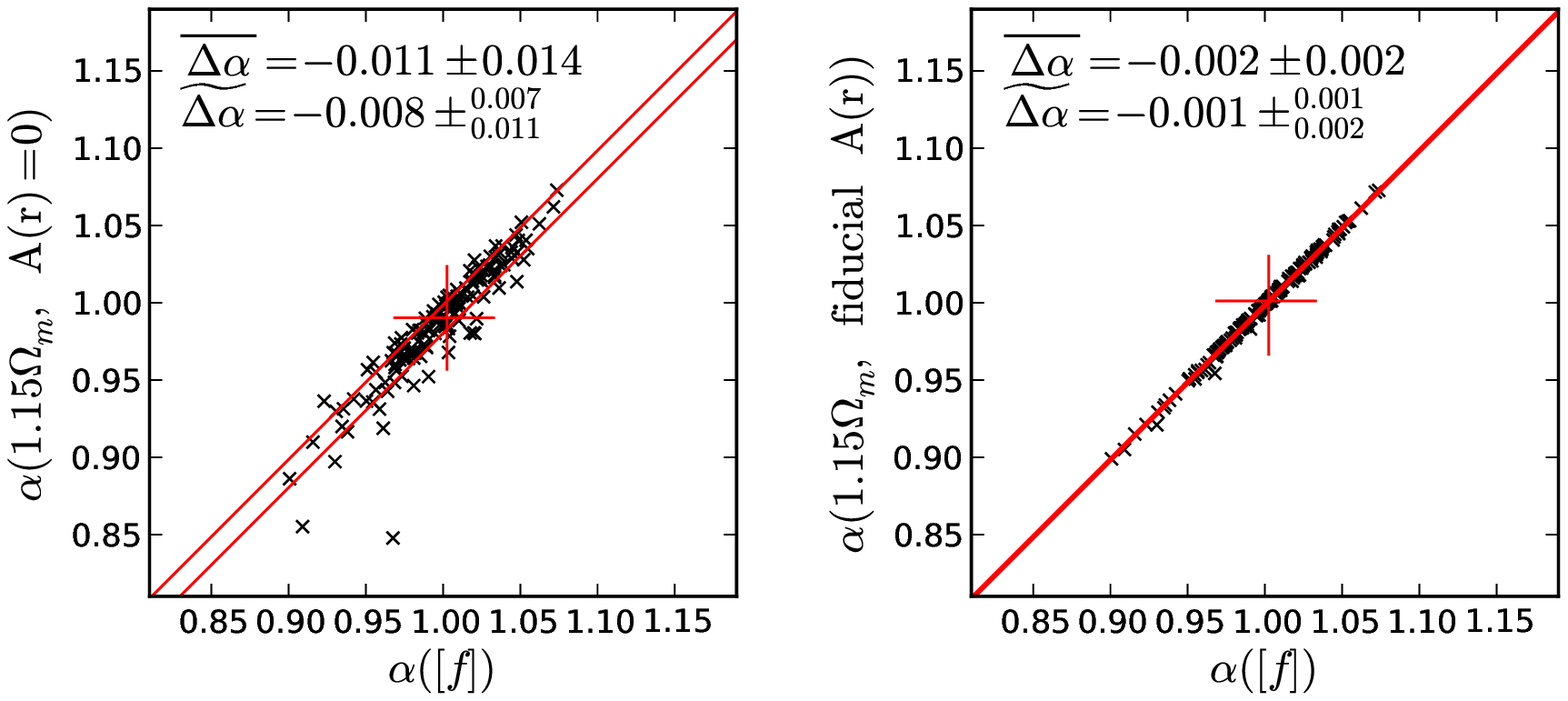, height=0.42\linewidth, width=0.95\linewidth, clip=} \\
\epsfig{file=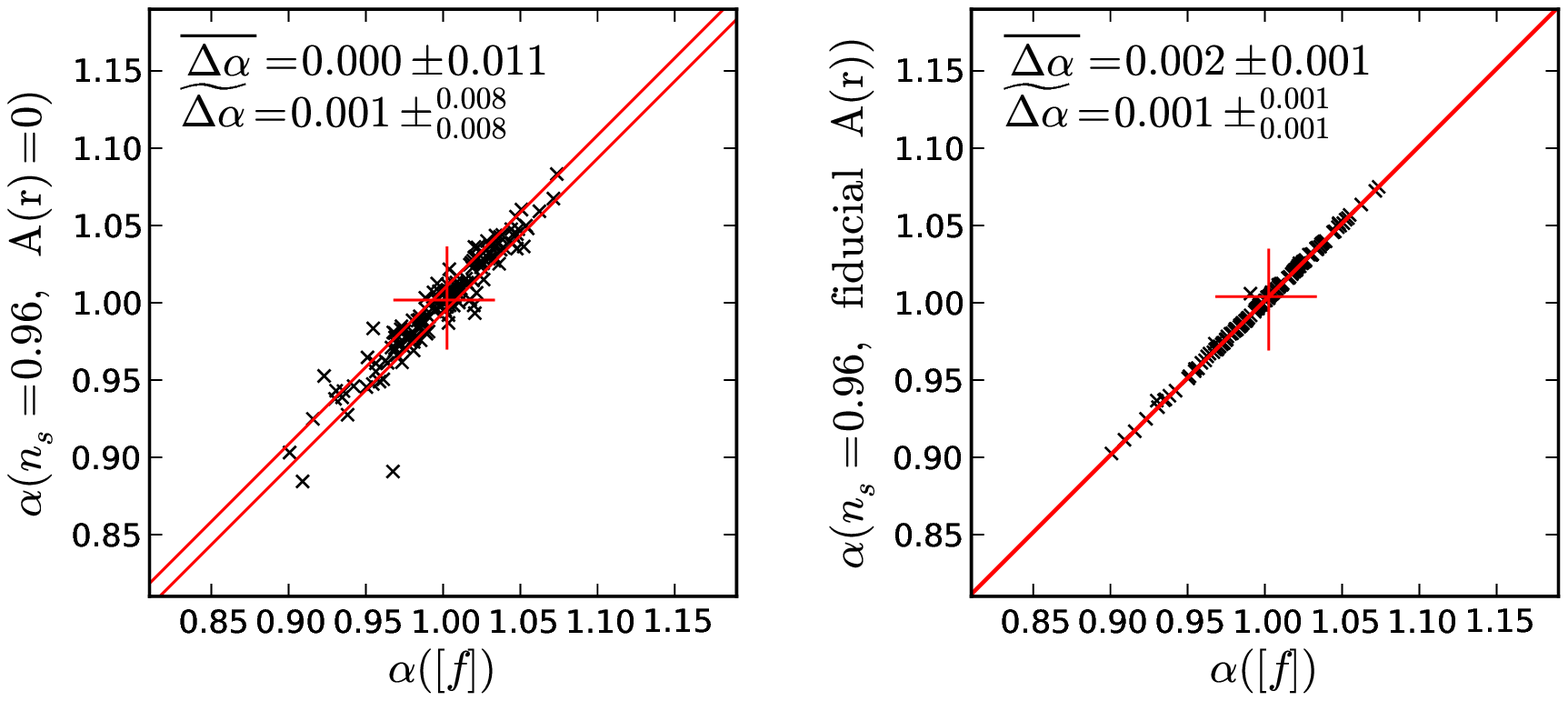, height=0.42\linewidth, width=0.95\linewidth, clip=} \\
\epsfig{file=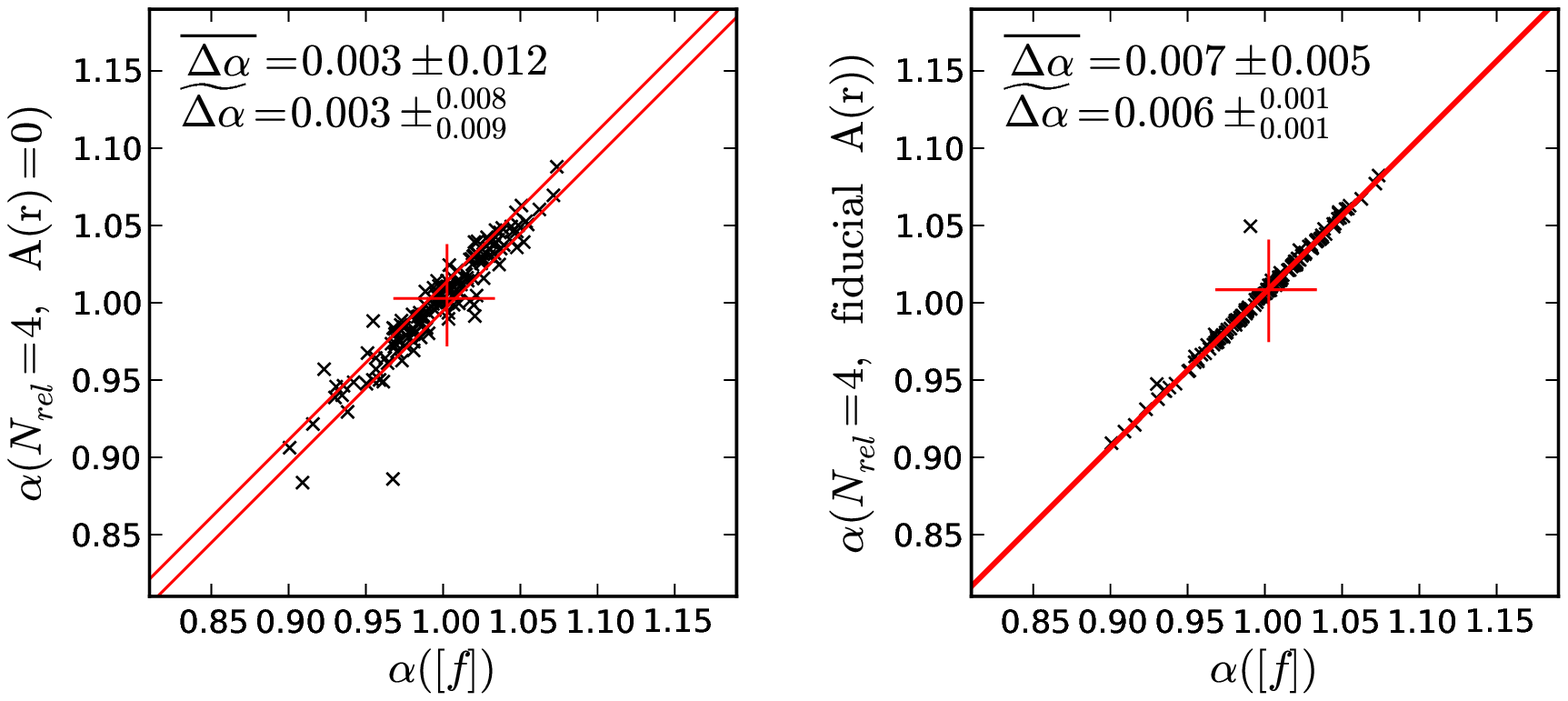, height=0.42\linewidth, width=0.95\linewidth, clip=} 
\end{tabular}
\caption{Validation of our fitting method using LasDamas mocks: varying
template cosmology. Here we have plotted the redshift-space $\alpha$
values measured using the fiducial model (i.e. using the true LasDamas
cosmology) on the $x$-axis versus the $\alpha$ values measured using
templates derived from slightly different cosmologies on the $y$-axis. For
the incorrect cosmology templates, we have performed the fits using
$poly0$ (left) and the fiducial $A(r)$ form (right). The $\alpha$
values from the incorrect cosmologies have been scaled to the correct
cosmology where necessary. The red cross indicates the median $\alpha$
values with their 16th and 84th percentiles. The red lines indicate the
16th and 84th percentiles of $\Delta \alpha = \alpha_{y\mathrm{-axis}} -
\alpha_{x\mathrm{-axis}}$. These values are given in the plots. Overall,
we see that the fiducial $A(r)$ form is better at recovering the correct
acoustic scale than $poly0$ and that our fiducial model is robust in
recovering the correct acoustic scale even when the template power
spectrum is derived from a slightly different cosmology. (top) Results
when we fit with a template cosmology where $\Omega_m$ is 15\% higher
than LasDamas. (middle) Results when we fit with a template cosmology
where $n_s=0.96$. (bottom) Results when we fit with a template where
there are 4 relativistic neutrino species. We see that the $\Delta\alpha$
offset is larger in this case, especially when fitting with the fiducial
$A(r)$ form. This $\sim0.6\%$ offset is likely a result of the template
BAO shape deviating slightly from that in the mock data.
\label{fig:cosmo}} 
\end{figure}

Our first test is to see whether we can recover the true acoustic
scale using our fiducial model but with $P_m(k)$ derived from slightly
different cosmologies to that used by LasDamas. Figure \ref{fig:cosmo}
shows the $\alpha$ values derived from these incorrect cosmologies versus
the $\alpha$ values obtained through fits using the fiducial model
(i.e. with the correct cosmology). The $\alpha$ values obtained from
the incorrect cosmologies have been scaled to the correct cosmology
where necessary by multiplying the ratio of the sound horizons,
$r_{s,\rm{lin}}(\rm{correct})/r_{s,\rm{lin}}(\rm{incorrect})$, where the
$r_{s,\rm{lin}}$ are calculated using Equation (6) in \citet{EH98}. For
a more detailed discussion of the sound horizon calculation, please
refer to Paper III. The figure caption gives an explicit description
of the items plotted. Note that we define $\Delta \alpha \equiv
\alpha_{y\mathrm{-axis}} - \alpha_{x\mathrm{-axis}}$, where we always
have $\alpha_{x\mathrm{-axis}}$ equal to the values of $\alpha$ measured
using the fiducial model.

The top left panel shows the $\alpha$ values from a fit using $A(r)=0$
(i.e. $poly0$) and a cosmology with a 15\% larger value of $\Omega_m$
(and hence $\Omega_b$). This difference should give rise to an acoustic
scale that is about 5\% smaller. The top right panel shows the results
from the same fit with the fiducial $A(r)$ form instead of $poly0$. We
expect the mean and median $\Delta \alpha$ values to be $\sim0$ if
we can recover the true acoustic scale using an incorrect cosmology
template (i.e. if the $\alpha$ values plotted on the 2 axes are perfectly
correlated). We see that this result is recovered with $\sim0.2\%$ scatter
when we fit using the fiducial $A(r)$ form. This is $\sim1$\% smaller
than the scatter found when fitting with $poly0$, another indication of
the advantages of fitting with a non-zero $A(r)$.

The middle left panel of Figure \ref{fig:cosmo} shows the $\alpha$
values from fits using $poly0$ and a cosmology with $n_s=0.96$ plotted
against the results from the fiducial model ($n_s=1.0$). The difference in
$n_s$ should not affect the position of the acoustic scale, but only the
shape of the model. The analogous results using the fiducial $A(r)$ form
instead of $poly0$ are shown in the middle right. The correct acoustic
scale is recovered with $\Delta \alpha$ very close to 0 and $\sim0.1$\%
scatter when the fiducial $A(r)$ form is used. The corresponding $poly0$
fit does a poorer job with a scatter in $\Delta \alpha \sim 0$ of about
$1\%$. Overall, the fiducial model seems to be able to recover the
true acoustic scale even if its power spectrum template has a slightly
different cosmology. This and the previous example show how important
it is to fit with a non-zero $A(r)$ term if we are not certain of the
true model cosmology to be used (i.e. in the case of actual observations).

The $y$-axis of the bottom panels in Figure \ref{fig:cosmo} correspond to
$\alpha$ values measured using $poly0$ (left) and fiducial $A(r)$ (right)
with a template cosmology consisting of 4 relativistic neutrino species
($N_{rel}=4$) instead of the standard 3. We use the same $\Omega_m$,
$\Omega_b h^2$, and epoch of matter-radiation equality as the $N_{rel}=3$
case, so that the rough shape of the power spectrum is preserved. This
requires $H_0=74.3$ km/s/Mpc. As in the previous cases, the scatter in
$\Delta\alpha$ is smaller if we employ the fiducial $A(r)$ rather than
$poly0$. However, we find a mean offset of 0.6\% when using an $N_{rel}=4$
template, after scaling by the appropriate sound horizon. We believe this
is because the shape of the template around the BAO feature is slightly
different from the $N_{rel}=3$ case, e.g., because the baryon fraction
in this model is different. While the 0.6\% offset is much smaller than
the statistical errors of the DR7 data set, larger surveys might need
to iterate their fits to converge to a sufficiently accurate template
when investigating variations in the number of relativistic species.

\begin{figure}
\vspace{0.4cm}
\begin{tabular}{c}
\epsfig{file=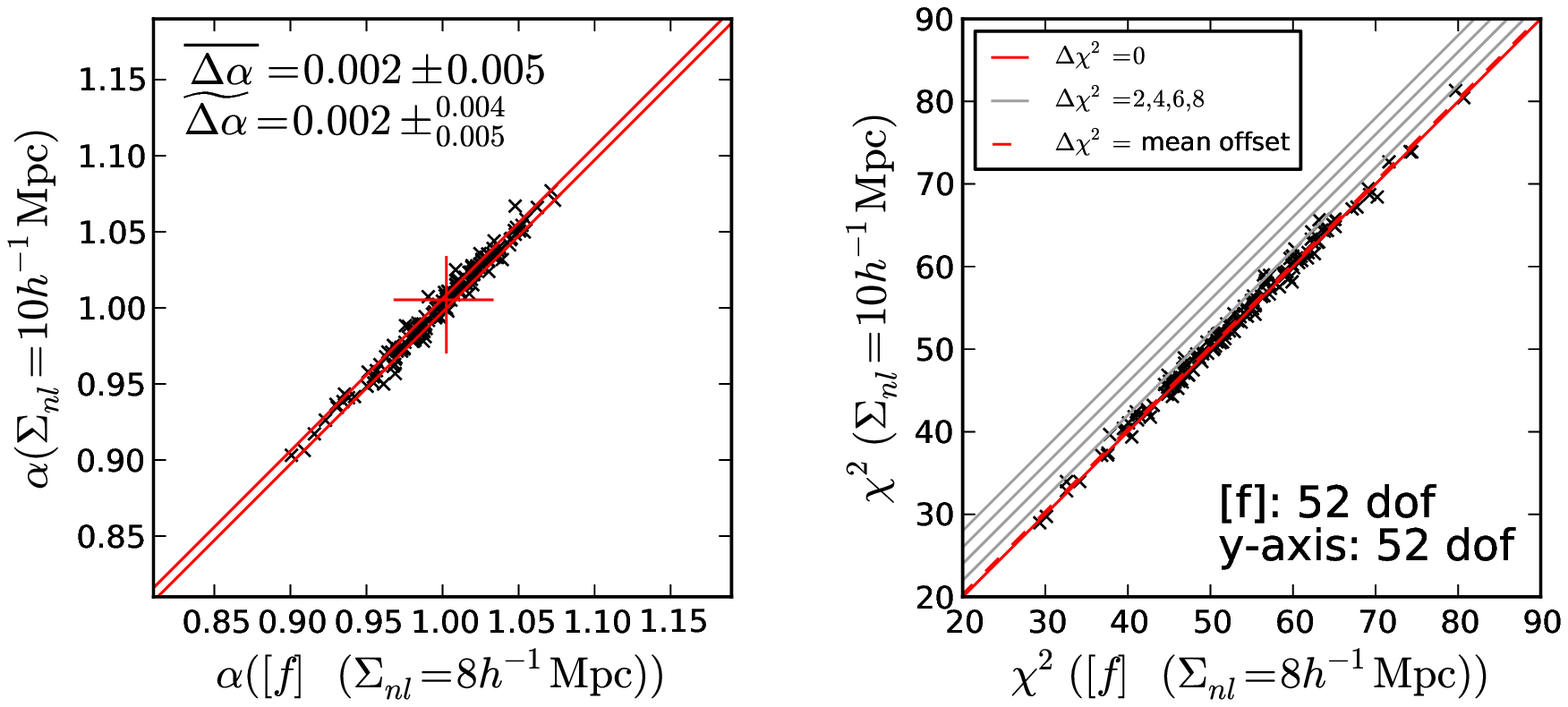, height=0.42\linewidth, width=0.95\linewidth, clip=} \\
\epsfig{file=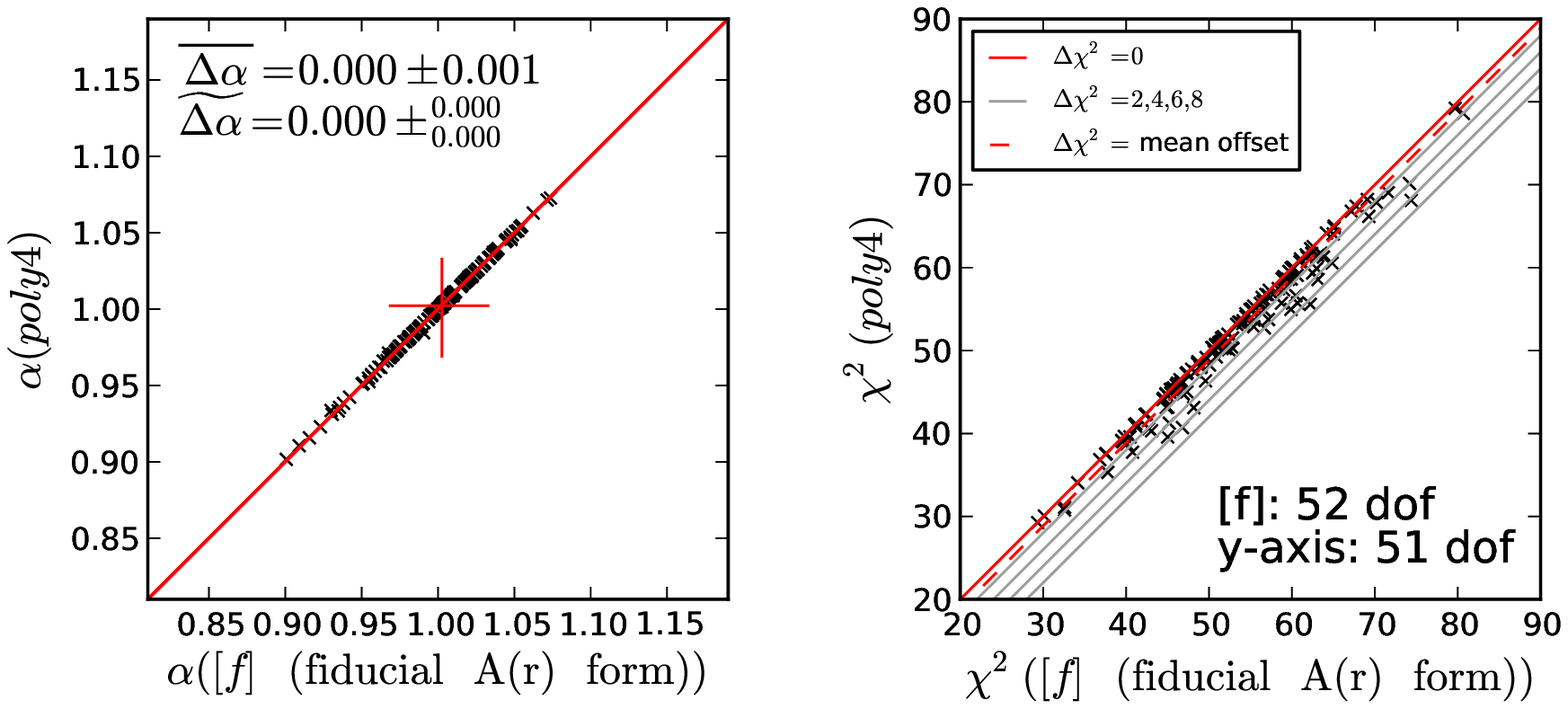, height=0.42\linewidth, width=0.95\linewidth, clip=} \\
\epsfig{file=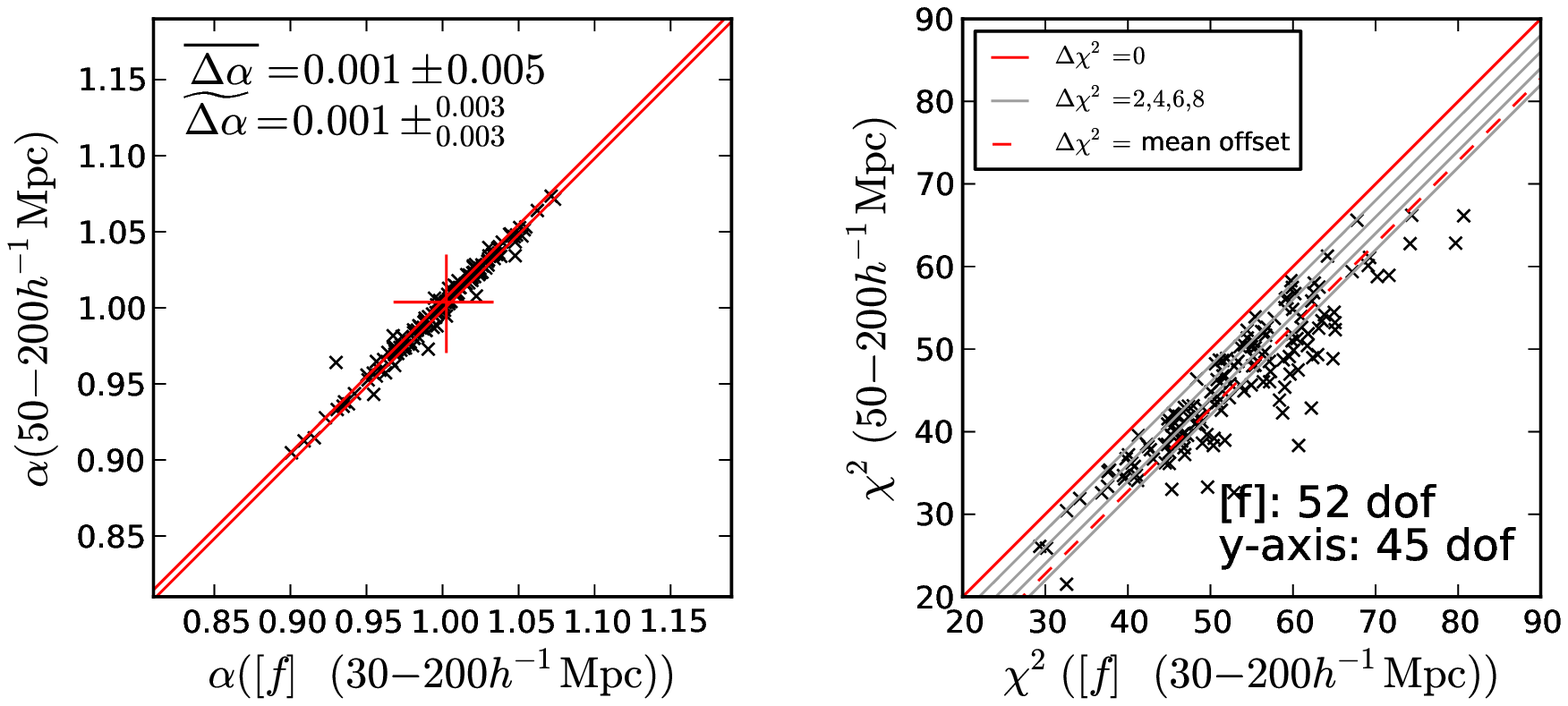, height=0.42\linewidth, width=0.95\linewidth, clip=} \\
\epsfig{file=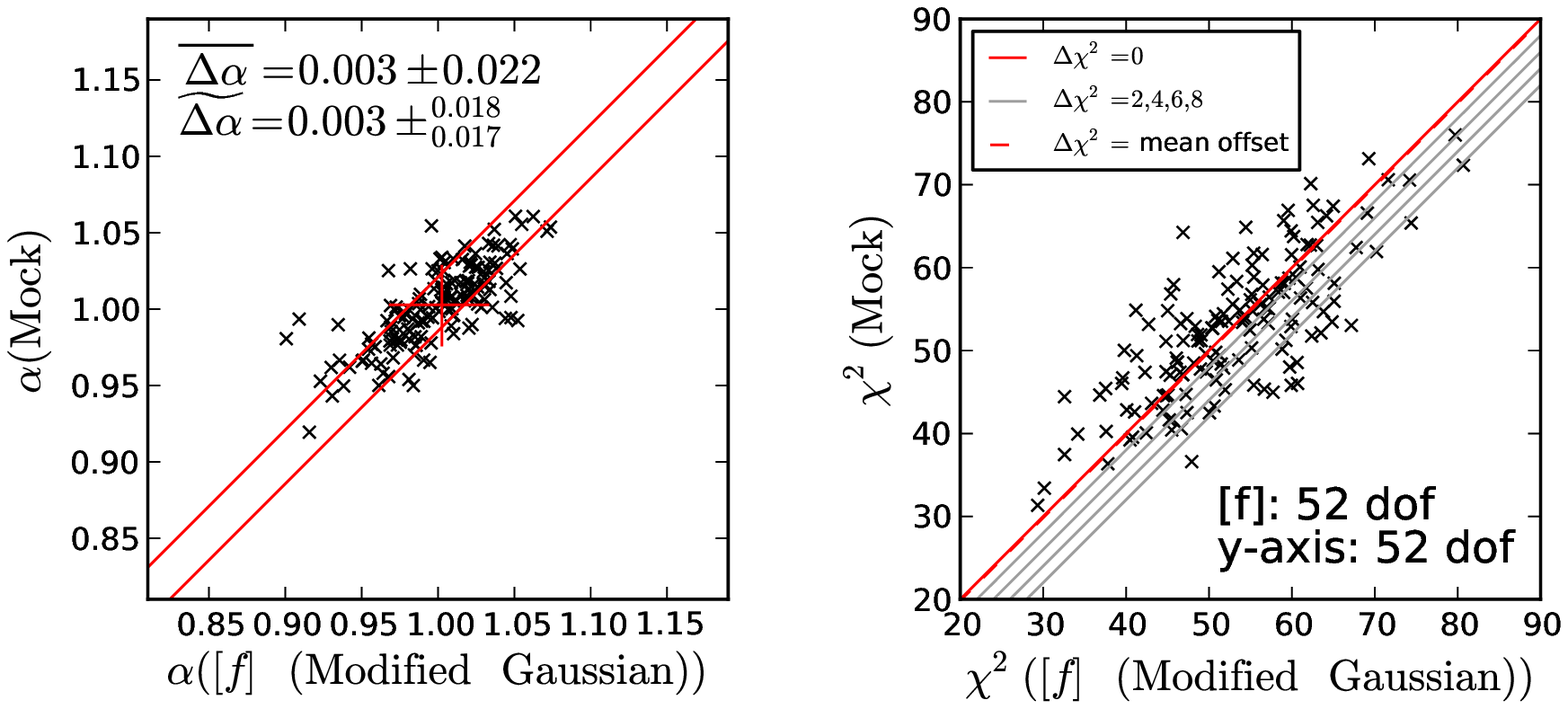, height=0.42\linewidth, width=0.95\linewidth, clip=} 
\end{tabular}
\caption{Validation of our fitting method using LasDamas mocks: varying
fitting/model parameters. This figure is akin to Figure \ref{fig:cosmo}
in that the left panels show similar plots of redshift-space $\alpha$
values measured using the fiducial model ($x$-axis) versus those
measured using models in which the fiducial parameters are slightly
tweaked ($y$-axis). However, instead of varying the template cosmology,
here we vary other fiducial model parameters such as $\snl$ (top),
the order of $A(r)$ ($2^{nd}$ row) and the fitting range ($3^{rd}$
row). The tight correlations shown in all of these plots indicate the
robustness of our covariance matrix estimators and the robustness of our
fiducial model to small changes in model parameters. The right panels show
corresponding plots of the measured best-fit $\chi^2$ values. In all cases
we see that the $\chi^2$ values shift by reasonable amounts given the
addition or subtraction of degrees-of-freedom as we change the fiducial
parameters. (top) Results when we fit using $\snl=10\hMpc$. ($2^{nd}$
row) Results when we fit using $poly4$. ($3^{rd}$ row) Results when
we use a fitting range of $50<r<200\hMpc$. (bottom) For completeness,
we show the comparison between fits using the mock covariance matrix
(Equation (\ref{eqn:mcov})) and fits using the MGCM. A correlation
between the 2 sets of $\alpha$ can be seen, but the noisiness of the
mock covariance matrix is responsible for the larger scatter. Similarly,
the corresponding $\chi^2$ plot shows a fair bit of scatter. However,
the average $\chi^2$ values obtained using these 2 different covariance
matrices match nicely.
\label{fig:comps}} 
\end{figure}

Next we test how changing the value of $\snl$ in $P_m(k)$ of the fiducial
model affects the measured acoustic peak position. In the top left panel
of Figure \ref{fig:comps}, we plot the $\alpha$ values measured using
fits with $\snl=10\hMpc$ versus those derived from the fiducial model
($\snl=8\hMpc$) in redshift space. One can see a tight correlation between
the 2 sets of $\alpha$ with consistent mean and median values. The mean
and median $\Delta \alpha$ values are consistent with 0 and only have
$\sim0.5\%$ scatter. The top right panel shows the corresponding $\chi^2$
values from the fits. The number of degrees of freedom is calculated by
subtracting the number of fitting parameters (5 in our fiducial form:
$B^2$, $a_1$, $a_2$, $a_3$ and $\alpha$) from the number of data points
being fit (57 for our fiducial fitting range of $30<r<200\hMpc$). The
tight correlation in $\alpha$ and the very small change in $\chi^2$/dof
between the 2 models suggest that the value of $\alpha$ is not sensitive
to small changes in $\snl$. However, if we use a less sensible value of
$\snl$ like $\snl=0\hMpc$ which corresponds to no acoustic peak smearing
(very unlikely, especially before reconstruction), $\Delta\alpha$
is still consistent with 0 but the scatter increases to $1-2\%$. This
suggests that the fiducial form defined by Equations (\ref{eqn:fform}
\& \ref{eqn:aform}) returns consistent values of $\alpha$ as long as
a reasonable value of $\snl$ is used.

The left panel of the $2^{nd}$ row in Figure \ref{fig:comps} shows the
$\alpha$ values measured from fits using an $A(r)$ that is an order
higher than the fiducial form (i.e. $poly4$) versus the $\alpha$ values
measured using the fiducial model. Again, a tight correlation exists
between the 2 sets of $\alpha$ with the mean and median values agreeing
nicely. The mean and median values of $\Delta \alpha$ are consistent with
0 and have negligible scatter. The right panel in the $2^{nd}$ row shows
the analogous plot for the $\chi^2$ values. One can see that the average
$\chi^2$ decreases by $\sim1$ as one expects when increasing the number
of nuisance parameters by 1. This suggests that continuing to increase
the order of $A(r)$ beyond that in the fiducial model offers little
improvement to the fits. However, as long as one does not increase the
order to a point where noise in the data is being fit, one should measure
consistent values of $\alpha$. When $A(r)$ is taken to be an order less
than fiducial (i.e. $poly2$), the scatter goes up slightly to $\sim0.3$\%
and when $poly0$ is used, the scatter increases to $\sim1\%$. Hence,
decreasing the order of $A(r)$ is feasible, but decreasing the order by
too much will give a less consistent measurement of $\alpha$.

Finally we test how adjusting the fitting range affects our measurements
of $\alpha$. Changing the minimum of the fitting range from 30$\hMpc$
(fiducial) to 50$\hMpc$ seems to have little affect on $\alpha$. The
$3^{rd}$ row of Figure \ref{fig:comps} shows the $\alpha$ and $\chi^2$
values obtained using these 2 fitting ranges. One can see that the mean
and median $\alpha$ values agree nicely and that the 2 sets of $\alpha$
values are obviously correlated. $\Delta\alpha$ is again consistent
with 0 and has very small scatter ($\sim0.4\%$). The $\chi^2$ values
decreased by about 7 on average, which is expected since the number
of data points fit decreased by 7. We perform similar experiments by
shifting the fitting range to $20<r<200\hMpc$ and $70<r<150\hMpc$. In
both cases, $\Delta \alpha=0$ lies within slightly larger scatter
($\sim0.7-1\%$). In the prior case, this is likely due to non-linear
effects at small scales coming into play. These effects are not well
modeled by our fitting template. In the latter case, the larger scatter
is likely caused by some of the acoustic information being cut out by
using such a small fitting range.

For completeness, we also show the $\alpha$ values obtained through
using the mock covariance matrix (Equation (\ref{eqn:mcov})) versus
those obtained using the MGCM and the fiducial model. Since the mock
covariance matrix is noisy, we expect there to be significant scatter in
the $\alpha$ versus $\alpha$ and $\chi^2$ versus $\chi^2$ plots. These
are shown in the bottom panels of Figure \ref{fig:comps}. A correlation
between the two $\alpha$ sets is still visible, but it is not as tight
as those in the upper panels. $\Delta \alpha$ is still consistent with 0
but the scatter is now $\sim2\%$. Note that the average $\chi^2$ values
of the 2 cases match well. This indicates that the MGCM is a reasonable
approximation to the covariances we expect in our mock data.

\subsection{With Reconstruction}\label{sec:red_rec_fr}
\begin{figure}
\vspace{0.4cm}
\centering
\epsfig{file=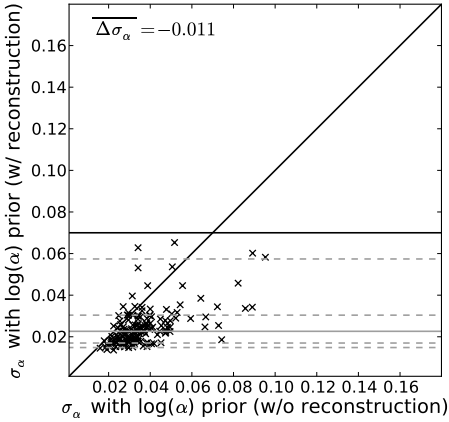, width=0.7\linewidth, clip=}  
\caption{The standard deviations of $p(\alpha)$ for each mock before
reconstruction versus those after reconstruction. The diagonal black line
is a 1-1 line to guide the eye. Only a few of the mocks have slightly
larger standard deviations after reconstruction, most of the mocks
lie very much below the diagonal line. The median change in standard
deviation is 1.1\% which implies that our ability to constrain $\alpha$
increases significantly after reconstruction. Note that also, after
reconstruction, there are no longer any poorly constrained mocks with
standard deviations larger than 7\% (black horizontal line), the cutoff
imposed in Figure \ref{fig:sdva_rednr}. The grey solid and dashed lines
are as in Figure \ref{fig:sdva_rednr}.
\label{fig:comperrs_redr}} 
\end{figure}

Next we study the LasDamas mocks in redshift space after
reconstruction. We find that after reconstruction, our ability to
constrain the acoustic scale in each individual mock as measured
by the standard deviation of $\alpha$ is greatly improved. We plot
$\sigma_\alpha$ before reconstruction against those after reconstruction
in Figure \ref{fig:comperrs_redr}. The black diagonal line is the 1-1
line. One can see that only a few of the mocks have larger standard
deviations after reconstruction but they are not much larger. Most of
the points lie significantly below the line with the median change
in $\sigma_\alpha$ equal to 1.1\% as indicated on the plot. Hence,
in general, reconstruction can significantly improve our ability to
constrain $\alpha$.

\begin{figure*}
\vspace{0.4cm}
\begin{tabular}{c}
\epsfig{file=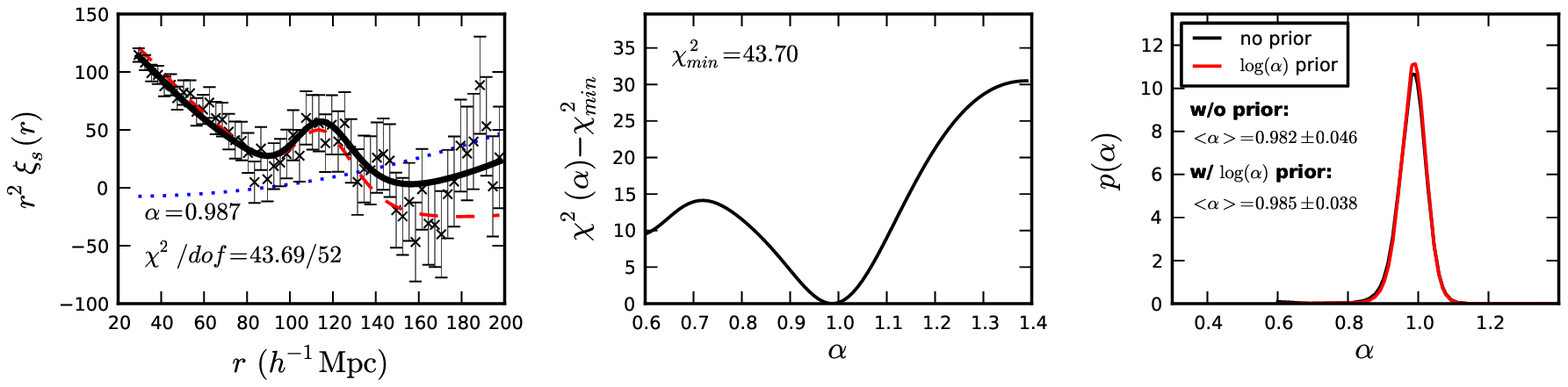, height=0.22\linewidth, width=0.95\linewidth, clip=}\\
\epsfig{file=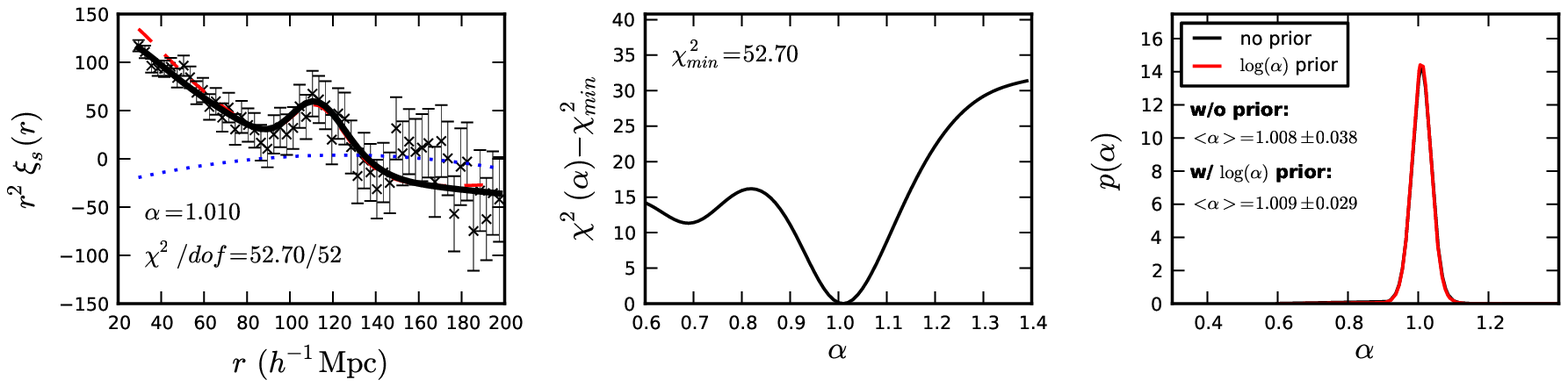, height=0.22\linewidth, width=0.95\linewidth, clip=} \\
\end{tabular}
\caption{The same poorly constrained mocks as in Figure \ref{fig:exfigs}
after reconstruction. One can see that reconstruction has improved our
ability to obtain a solid measurement of $\alpha$ in both cases. The
acoustic peaks are now clearly visible, there are significant differences
in $\chi^2$ between the minima of the $\Delta\chi^2$ curves and the
plateaus, and the $p(\alpha)$ distributions are now regular Gaussians
with standard deviations $\sim1.9$ and $\sim2.5$ times smaller than
before reconstruction. This type of improvement is characteristic of
the other previously poorly constrained mocks in our sample and again
emphasizes the utility of reconstruction.
\label{fig:exfigs_rec}} 
\end{figure*}

One can also see that after reconstruction, there are no longer any
poorly constrained mocks that lie above the 7\% cutoff (black horizontal
line) imposed in the unreconstructed case. The solid grey line indicates
the mean $\sigma_\alpha$ after reconstruction and the dashed grey lines
correspond to the 98th, 84th, 16th and 2nd percentiles, similar to Figure
\ref{fig:sdva_rednr} for redshift space without reconstruction.

As we saw in \S\ref{sec:red_norec_fr}, the mocks where $\alpha$ is well
constrained have strong acoustic features. Figure \ref{fig:comperrs_redr}
showed that in reconstructed redshift space our measurements of
best-fit $\alpha$ should be much more reliable. This implies that
the acoustic peak in the poorly constrained mocks from before should
be more prominent after reconstruction as we would expect. In Figure
\ref{fig:exfigs_rec}, we show the same 2 poorly constrained mocks as in
Figure \ref{fig:exfigs}. The fit results from both of these mocks clearly
demonstrate how effective reconstruction is. The acoustic peaks can be
clearly seen now and the $\chi^2$ minima corresponding to the best-fit
$\alpha$ values are significantly different from the plateau values. The
$p(\alpha)$ curves have also become more Gaussian in shape with standard
deviations much smaller than before (by factors of $\sim1.9$ and $\sim2.5$
respectively). These are characteristic of the improvements seen for
the other mocks which were poorly constrained before reconstruction.

\begin{figure}
\vspace{0.4cm}
\centering
\epsfig{file=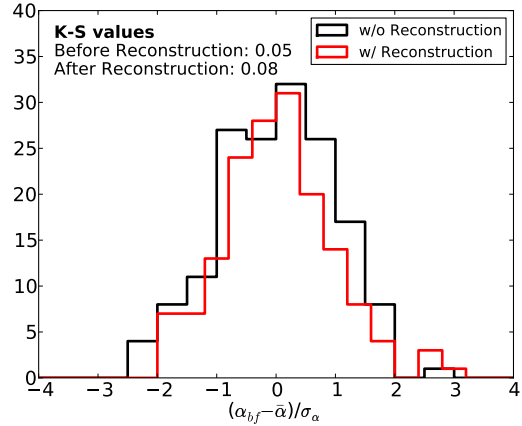, width=0.8\linewidth, clip=}  
\caption{The distributions of ($\alpha_{bf} -
\bar{\alpha}$)/$\sigma_\alpha$ before (black) and after (red)
reconstruction, where $\alpha_{bf}$ is the best-fit value of $\alpha$
for each mock and $\bar{\alpha}$ is the mean of the best-fit values. This
is a good measure of the signal-to-noise ratio of our best-fit $\alpha$
values. Both distributions are nearly Gaussian as indicated by the
K-S values shown in the plot. The Gaussianity of these distributions
implies that the $\sigma_\alpha$ values measured from $p(\alpha)$
are representative estimates of the error on $\alpha$ for each mock.
\label{fig:snva}}
\end{figure}

In Figure \ref{fig:snva}, we have plotted the distribution of
($\alpha_{bf} - \bar{\alpha}$)/$\sigma_\alpha$ which is a proxy for
the signal-to-noise of our $\alpha$ measurement. Here, $\alpha_{bf}$
is the best-fit value of $\alpha$ for each mock and $\bar{\alpha}$ is
the mean of the best-fit values. The distribution before reconstruction
is shown in black and the distribution after reconstruction is shown
in red. One can see that both distributions are roughly Gaussian. A
standard K-S test gives a value of $\sim0.05$ before reconstruction and
$\sim0.08$ after reconstruction in comparison to a Gaussian distribution
(recall that a value close to 0 indicates a better match to the
normal distribution). This demonstrates that the standard deviations of
$p(\alpha)$ are a representative estimate of the errors on the best-fit
values of $\alpha$ for each mock.

\begin{figure*}
\vspace{0.4cm}
\centering
\epsfig{file=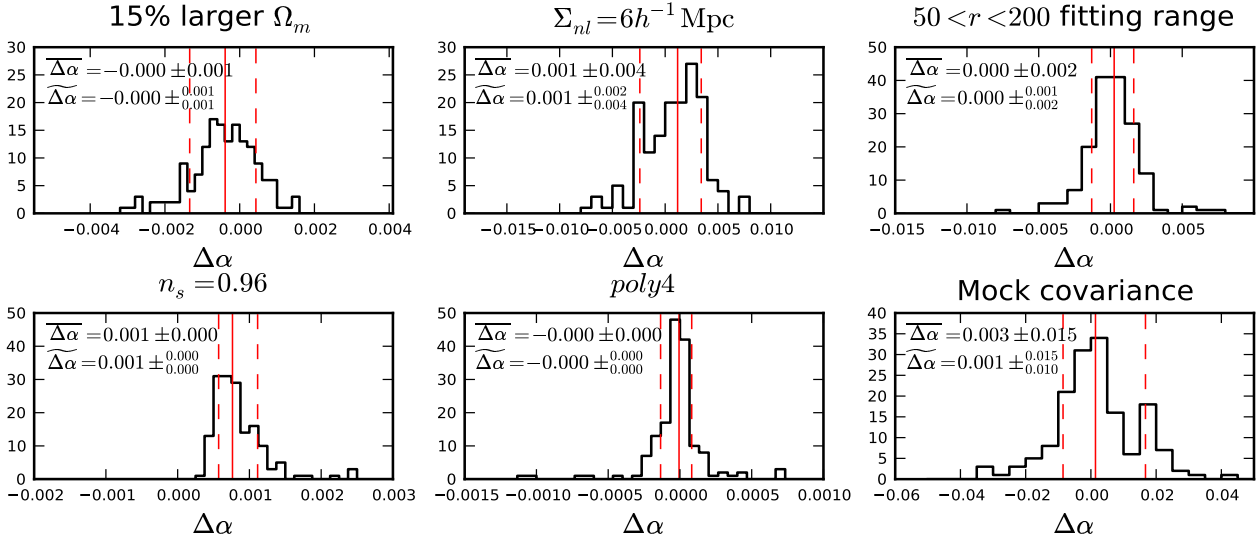, height=0.4\linewidth, width=0.94\linewidth, clip=} 
\caption{Validation of our fitting method in redshift space after
reconstruction using LasDamas mocks. The contents of this figure are
comparable to Figures \ref{fig:cosmo} \& \ref{fig:comps}, however,
the scatter plots have been replaced by histograms of $\Delta \alpha
= \alpha_{[i]} - \alpha_{[f]}$ here. The $\alpha_{[i]}$ are measured
from models that are derived by slightly changing the fiducial model
parameters. These are indicated above each plot. The solid red lines mark
the median $\Delta\alpha$. The dashed red lines indicate the 16th and
84th percentiles. One can see that $\Delta \alpha$ is very close to 0 with
small scatter in most of these cases. The slightly larger scatter in the
case where we fit using the mock covariance matrix is likely due to the
noisiness of that matrix. This indicates that the value of $\alpha$ is
insensitive to small changes in template cosmology, $\Sigma_{nl}$, order
of $A(r)$, fitting range and covariance matrix estimator used. Hence, our
basic fitting form and our covariance matrix estimators are robust. The
results shown in this figure are all consistent with those found in
unreconstructed redshift space.
\label{fig:comps_redr}}
\end{figure*}

Next, we again tweak the fiducial model parameters slightly and test
the robustness of our fitting form and our covariance matrix. The
results of the fits are summarized in Table \ref{tab:alphasred} and
in Figure \ref{fig:comps_redr}. This figure is essentially analogous
to Figure \ref{fig:cosmo} and Figure \ref{fig:comps}, however, we
have replaced the scatter plots with histograms of $\Delta \alpha =
\alpha_{[i]} - \alpha_{[f]}$. Here, $\alpha_{[i]}$ are the slightly
tweaked models as indicated by the titles. The fiducial model has mean
$\bar{\alpha}_{[f]} = 1.001 \pm 0.021$ and median $\tilde{\alpha}_{[f]}
= 1.001 \pm^{0.020}_{0.022}$ (recall that before reconstruction
these were $\bar{\alpha}=0.999\pm0.033$ and $\tilde{\alpha} = 1.003
\pm^{0.030}_{0.034}$). This indicates that the error on the acoustic
scale decreased by about a factor of 1.6 after reconstruction. We know
that $V \propto \sigma^{-2}$, where $V$ is the survey volume required
to achieve a variance $\sigma^2$. Therefore, we would have to increase
the survey volume by $\sim2.5$ times to achieve this same factor of 1.6
decrease in the error. Also, note that in general, the scatters in the
mean and median $\alpha$ and $\Delta \alpha$ values from the various fits
are smaller after reconstruction, another indication of its effectiveness.

The various panels of Figure \ref{fig:comps_redr} show $\Delta \alpha$
values for different tweaks to the fiducial model. The median values
are marked by the red lines (see caption for more details). Further
cases are summarized in Table 1. The only case that shows a relatively
large scatter in $\Delta\alpha$ is when we fit using the mock covariance
matrix instead of the MGCM; this can be attributed to the higher noise
in the mock covariance matrix. Also, as in the pre-reconstruction case,
we see that for $N_{rel}=4$, $\Delta\alpha\sim0.5\%$ which is slightly
larger than the other cases. However, in general, $\Delta \alpha \sim 0$
with very small scatter.

It should also be noted here that the cases which had noticeably larger
scatter in $\Delta\alpha$ before reconstruction ($\snl=0\hMpc$, $poly0$
and fitting ranges of $20<r<200\hMpc$ and $70<r<150\hMpc$), no longer
do post-reconstruction. This is because reconstruction undoes non-linear
structure growth and brings the correlation function closer to its linear
theory form (i.e. $\xi_s(r) \rightarrow \xi_m(r)$ and $\snl \rightarrow
0\hMpc$). The consistency in the measured values of $\alpha$ indicate
that after reconstruction, our fiducial model is even more robust against
changes in model parameters.

\begin{figure}
\vspace{0.5cm}
\epsfig{file=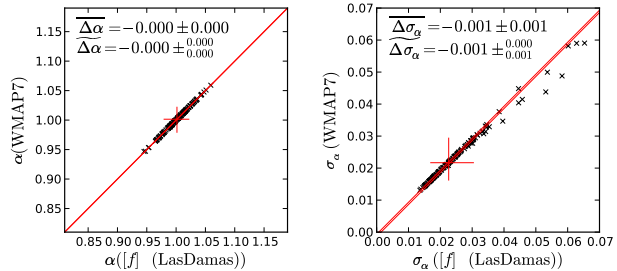, height=0.42\linewidth, width=0.95\linewidth, clip=} 
\caption{Testing the effects of using the wrong cosmology to derive the
covariance matrix and construct the model template. (left) $\alpha$ values
measured from the LasDamas mocks in redshift space after reconstruction
using a fitting template and MGCM (see Figure \ref{fig:comppow}) based
on the WMAP7 cosmology versus those measured using the fiducial model
(LasDamas fitting template and its corresponding MGCM). The $\alpha$
values from the WMAP7 cosmology have been rescaled to the LasDamas
cosmology. (right) The analogous plot for $\sigma_\alpha$ measured from
$p(\alpha)$. One can see that perfect correlations exist between the
axes of both plots. This indicates that our acoustic scale measurements
are not affected by deriving the MGCM using the wrong cosmology. Our
maximum likelihood method is capable of modifying the matrix from the
incorrect cosmology to match that expected from the correct cosmology.
\label{fig:compcov}}
\end{figure}

The left panel of Figure \ref{fig:compcov} shows the $\alpha$
values measured using the WMAP7 cosmology and its MGCM described in
\S\ref{sec:red_norec_cm} versus the $\alpha$ values measured using the
MGCM for the LasDamas cosmology. One can see that, after rescaling by
the ratio of the sound horizons, a perfect correlation exists between the
2 sets of $\alpha$ values. This is also true for the WMAP7$\pm1\sigma$
cosmologies shown in Figure \ref{fig:comppow}. The right panel of Figure
\ref{fig:compcov} shows the corresponding values of $\sigma_\alpha$
measured from $p(\alpha)$. Again, a strong correlation exists and
similar trends are observed for the 2 other WMAP7-like cosmologies. Hence,
the measurement of the acoustic scale is not affected by the cosmology
used for the covariance matrix or the fitting model. This demonstrates
the robustness of our fiducial model in dealing with a fitting template
constructed using the wrong cosmology as well as our maximum likelihood
approach for deriving a suitable covariance matrix for the data.

\begin{figure}
\vspace{0.5cm}
\epsfig{file=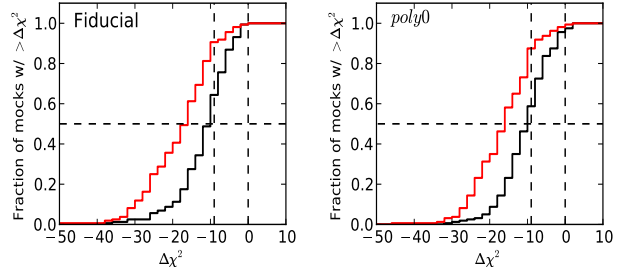, height=0.42\linewidth, width=0.95\linewidth, clip=}
\caption{The detectability of the BAO feature in
redshift space before and after reconstruction. (left) The
normalized cumulative distribution function of $\Delta\chi^2 =
\chi^2_{\mathrm{BAO}}-\chi^2_{\mathrm{no\;BAO}}$ from fits using the
fiducial $A(r)$ term. The solid black line shows the distribution before
reconstruction and the solid red line shows the distribution after
reconstruction. The horizontal dashed black line at fraction=50\%
indicates the value of $\Delta\chi^2$ that splits the mocks in
half (i.e. the median $\Delta\chi^2$ value). We can see that the
average $\Delta\chi^2$ is negative in both of these cases. The
vertical dashed black lines indicate where $\Delta\chi^2=0$ and -9
($3\sigma$). (right) The distribution of $\Delta \chi^2$ values from
fits using $poly0$. Again the average $\Delta\chi^2$ is negative both
before and after reconstruction. In all cases, the majority of mocks lie
beyond the $3\sigma$ line, especially in the reconstructed case. This
indicates that a detection of the BAO in our mock data is favoured over
a non-detection.
\label{fig:hibao}} 
\end{figure}

We conclude this section by demonstrating and comparing the
detectabilities of the BAO in the reconstructed (solid red line)
and unreconstructed (solid black line) mocks as shown in Figure
\ref{fig:hibao}. We have plotted the normalized cumulative distribution
of $\Delta\chi^2 = \chi^2_{\mathrm{BAO}}-\chi^2_{\mathrm{no\;BAO}}$
for fiducial $A(r)$ (left) and $poly0$ (right). Here, the $\chi^2$
values for each mock are calculated by marginalizing over the nuisance
parameters only while fixing $\alpha$ at the best-fit value from
the fiducial model or $poly0$ fits. $\chi^2_{\mathrm{BAO}}$ is the
$\chi^2$ obtained in this fashion using a template $\xi_m(r)$ containing
BAO. $\chi^2_{\mathrm{no\;BAO}}$ is the analogous value obtained using
a template that has no BAO feature. While it is true that in the no BAO
fits, the value of $\alpha$ we impose may not give the minimum $\chi^2$,
the lack of a BAO feature in the model eliminates its ability to constrain
$\alpha$ in these fits. Hence, comparing the BAO and no BAO $\chi^2$
values at the fiducial or $poly0$ best-fit $\alpha$ is a reasonable way
to circumvent this problem. We obtain the BAO-less model by setting
$\snl=1000\hMpc$ to completely damp out any acoustic signal. This
cumulative distribution indicates the fraction of mocks that lie more
negative of a given $\Delta\chi^2$ value. Note that we have plotted all
160 mocks here (i.e. we did not throw out any poorly constrained mocks).

If the data favours a model containing BAO, $\chi^2_{\mathrm{BAO}}$ should
be smaller than $\chi^2_{\mathrm{no\;BAO}}$ (i.e. $\Delta\chi^2$ should
be negative). The intersections of the dashed horizontal black lines and
the distributions correspond to values of $\Delta\chi^2$ that halve the
data and hence indicate the median $\Delta\chi^2$ values. One can see that
these medians are negative for all cases which indicates that on average
the data favours models containing BAO. The median $\Delta\chi^2$ before
reconstruction is $\sim-10$ and after reconstruction, it is $\sim-16$
for the fits performed using fiducial $A(r)$. The two vertical dashed
black lines indicate where $\Delta\chi^2=0$ and $\Delta\chi^2=-9$. The
latter corresponds to where a model containing BAO is favoured at
$3\sigma$ above a model without BAO. Before reconstruction, about 56\%
of the mocks lie above (more negative of) this 3$\sigma$ line. After
reconstruction, this number increases to 88\%. This again indicates that
our reconstruction algorithm is helping to restore acoustic information
back into the acoustic peak. Hence, the robustness of the BAO detection
is further improved by reconstruction.

Although there are some mocks that do not favour a model with BAO at very
high confidence and even a few mocks that do not favour a model with
BAO at all ($\Delta\chi^2>0$), this does not indicate that we are not
detecting the BAO. It is merely a statement that if we take observations
of many different regions of the universe, there is a finite chance
that the BAO signal will not be robustly detected in some of these
regions. This is in contrast to the conclusions drawn in \citet{CG11}.

The median value of $\Delta\chi^2$ is slightly more negative when the
fit is performed using the fiducial model versus when it is performed
using $poly0$ both before and after reconstruction. However, we see that
even the simple $poly0$ fits favour a model containing BAO over one
without BAO. Before reconstruction, about 53\% of the mocks lie above
the 3$\sigma$ line and after reconstruction, about 82\% lie above this
line. These numbers are very similar to those obtained in the fiducial
model case.

We have also performed this experiment for a few other fitting ranges
(50-200$\hMpc$ and 70-150$\hMpc$). The former yielded similar results,
however, the latter had slightly less dramatic $\Delta\chi^2$ values. This
is not unexpected because in these cases, the $A(r)$ terms are less
constrained and can therefore absorb some of the BAO signal.

\section{LasDamas Real Space Results}\label{sec:real_fit}

\subsection{Covariance Matrices}\label{sec:real_norec_cm}

As in redshift space, the covariance matrix derived from the mock
correlation functions through Equation (\ref{eqn:mcov}) is noisy. Hence,
we again use a modified Gaussian covariance matrix as a smooth
approximation to the mock covariance matrix. However, in real space,
we do not have any redshift-space observational effects such as Kaiser
squashing or FoG. Therefore, we take the input power spectrum to the
covariance matrix calculation to be
\begin{equation}
P_c(k) = b_0^2P_t(k)
\end{equation}
where the value of $b_0^2$ is determined as it was in redshift space.

We then introduce similar modification parameters to the redshift-space
case, namely, we assume the covariance matrix can be modeled by the form
\begin{equation}
C^m_{ij} = 2\int
\frac{k^2dk}{2\pi^2}\djbo(kr_i)\djbo(kr_j)\mps(k;c_0,c_2) + c_3.
\end{equation}
Here, $\mps(k;c_0,c_2)$ corresponds to an $I^2(k)$, Equation (\ref{eqn:Ik}),
in which we make the substitution
\begin{equation}
P_c(k) + \frac{1}{\bar{n}(z)} \rightarrow c_0P_c(k) + \frac{c_2}{\bar{n}(z)}.
\end{equation}
Note that this is the same as Equation (\ref{eqn:noise}) except with
$c_1=0$ and a different form for $P_c(k)$. Using the same maximum
likelihood prescription as that described in \S\ref{sec:red_norec_cm}, we
can derive values for the modification parameters $c_0$, $c_2$ and $c_3$.

We use $\snl=7\hMpc$ for calculating $P_c(k)$ in real space before
reconstruction. As in the redshift-space case, the value of $\snl$
used has negligible affect on the derived matrix. With this $P_c(k)$
we find $c_1=0.98$, $c_2=1.50$ and $c_3=5.57\times10^{-8}$.

After reconstruction, we take $\snl=3\hMpc$ in real space. The modified
Gaussian covariance matrix we obtain has the modification parameters
$c_0 = 0.89$, $c_2 = 1.57$ and $c_3 = 8.85 \times 10^{-8}$.

\subsection{Fitting Forms}\label{sec:real_norec_ff}

We use the same fiducial fitting form in both real space with
and without reconstruction as in redshift space for measuring the
shift in the acoustic scale, $\alpha$. This is described by Equations
(\ref{eqn:fform} \& \ref{eqn:aform}). In real space before reconstruction,
we define the fiducial model to make use of this fiducial form with
$\xi_m(r)$ derived from the LasDamas cosmology and $\snl=7\hMpc$
over a fitting range of $30<r<200\hMpc$. If we fit the average mock
real-space correlation function allowing $\snl$ to vary, we obtain
$\alpha=1.002$ and $\snl = 6.6\hMpc$, so our assumption for $\snl$
is not bad. In practice, like in redshift space, the measured $\alpha$
values for each individual mock are insensitive to our choice of $\snl$
as is shown in Table \ref{tab:alphasreal}. The error bars on our mock
data are approximated by the modified Gaussian covariance matrix (MGCM)
derived in the previous section.

\begin{table*}
\caption{Real space fitting results for various models}
\label{tab:alphasreal}

\begin{tabular}{@{}lccccc}

\hline
Model&
$\bar{\alpha}$&
$\tilde{\alpha}$&
$\overline{\Delta\alpha}$\footnotemark[1]&
$\widetilde{\Delta\alpha}$&
$\overline{\chi^2}/dof$\\

\hline
\multicolumn{6}{c}{Real Space without Reconstruction}\\
\hline

Fiducial $[f]$ &
$1.001 \pm 0.030$&
$1.000 \pm^{0.031}_{0.027}$&
--&
--&
53.34/52\\
\\[-1.5ex]
Fit with 15\% larger $\Omega_m$ using fiducial $A(r)$.\footnotemark[2] &
$1.000 \pm 0.030$&
$0.999 \pm^{0.031}_{0.028}$&
$-0.001 \pm 0.001$&
$-0.001 \pm^{0.001}_{0.001}$&
53.58/52\\
\\[-1.5ex]
Fit with $n_s=0.96$ using fiducial $A(r)$. &
$1.002 \pm 0.030$&
$1.001 \pm^{0.030}_{0.027}$&
$0.001 \pm 0.001$&
$0.001 \pm^{0.001}_{0.000}$&
53.39/52\\
\\[-1.5ex]
Fit with $N_{rel}=4$ using fiducial $A(r)$. &
$1.007 \pm 0.030$&
$1.005 \pm^{0.030}_{0.027}$&
$0.006 \pm 0.001$&
$0.005 \pm^{0.001}_{0.001}$&
53.36/52\\
\\[-1.5ex]
Fit with $\snl \rightarrow 0$. &
$0.998 \pm 0.032$&
$0.997 \pm^{0.032}_{0.029}$&
$-0.003 \pm 0.013$&
$-0.003 \pm^{0.009}_{0.012}$&
53.79/52\\
\\[-1.5ex]
Fit with $\snl \rightarrow \snl+2$. &
$1.003 \pm 0.030$&
$1.003 \pm^{0.031}_{0.031}$&
$0.002 \pm 0.005$&
$0.002 \pm^{0.004}_{0.005}$&
53.83/52\\
\\[-1.5ex]
Fit with $poly0$. &
$0.999 \pm 0.031$&
$1.001 \pm^{0.030}_{0.031}$&
$-0.002 \pm 0.008$&
$-0.001 \pm^{0.006}_{0.007}$&
56.19/55\\
\\[-1.5ex]
Fit with $poly2$. &
$1.000 \pm 0.030$&
$0.999 \pm^{0.031}_{0.028}$&
$-0.001 \pm 0.003$&
$-0.001 \pm^{0.002}_{0.002}$&
54.65/53\\
\\[-1.5ex]
Fit with $poly4$. &
$1.001 \pm 0.029$&
$1.000 \pm^{0.031}_{0.027}$&
$0.000 \pm 0.000$&
$0.000 \pm^{0.000}_{0.000}$&
52.03/51\\
\\[-1.5ex]
Fit with $50<r<200\hMpc$ fitting range. &
$1.002 \pm 0.029$&
$1.001 \pm^{0.031}_{0.025}$&
$0.001 \pm 0.003$&
$0.001 \pm^{0.003}_{0.003}$&
46.18/45\\
\\[-1.5ex]
Fit with $20<r<200\hMpc$ fitting range. &
$1.002 \pm 0.028$&
$1.000 \pm^{0.029}_{0.026}$&
$0.001 \pm 0.006$&
$0.002 \pm^{0.005}_{0.006}$&
59.33/57\\
\\[-1.5ex]
Fit with $70<r<150\hMpc$ fitting range. &
$1.002 \pm 0.031$&
$0.999 \pm^{0.031}_{0.020}$&
$0.001 \pm 0.011$&
$0.001 \pm^{0.008}_{0.008}$&
21.99/22\\
\\[-1.5ex]
Fit using mock covariance matrix. &
$1.002 \pm 0.023$&
$1.003 \pm^{0.021}_{0.025}$&
$0.001 \pm 0.016$&
$0.002 \pm^{0.013}_{0.016}$&
53.15/52\\
\hline
\multicolumn{6}{c}{Real Space with Reconstruction}\\
\hline
Fiducial $[f]$ &
$0.998 \pm 0.020$&
$0.999 \pm^{0.019}_{0.019}$&
--&
--&
53.44/52\\
\\[-1.5ex]
Fit with 15\% larger $\Omega_m$ using fiducial $A(r)$.\footnotemark[2] &
$0.998 \pm 0.020$&
$0.999 \pm^{0.019}_{0.020}$&
$-0.001 \pm 0.002$&
$-0.000 \pm^{0.001}_{0.001}$&
53.78/52\\
\\[-1.5ex]
Fit with $n_s=0.96$ using fiducial $A(r)$. &
$0.999 \pm 0.020$&
$1.000 \pm^{0.020}_{0.019}$&
$0.001 \pm 0.001$&
$0.001 \pm^{0.001}_{0.001}$&
53.48/52\\
\\[-1.5ex]
Fit with $N_{rel}=4$ using fiducial $A(r)$. &
$1.003 \pm 0.020$&
$1.003 \pm^{0.020}_{0.019}$&
$0.004 \pm 0.001$&
$0.004 \pm^{0.001}_{0.000}$&
53.68/52\\
\\[-1.5ex]
Fit with $\snl \rightarrow 0$. &
$0.998 \pm 0.020$&
$0.999 \pm^{0.021}_{0.020}$&
$-0.000 \pm 0.002$&
$-0.000 \pm^{0.002}_{0.002}$&
53.47/52\\
\\[-1.5ex]
Fit with $\snl \rightarrow \snl+2$. &
$0.999 \pm 0.020$&
$0.999 \pm^{0.020}_{0.019}$&
$0.000 \pm 0.003$&
$0.001 \pm^{0.002}_{0.003}$&
53.66/52\\
\\[-1.5ex]
Fit with $poly0$. &
$0.997 \pm 0.021$&
$0.999 \pm^{0.019}_{0.020}$&
$-0.001 \pm 0.005$&
$-0.001 \pm^{0.003}_{0.004}$&
56.44/55\\
\\[-1.5ex]
Fit with $poly2$. &
$0.998 \pm 0.020$&
$0.999 \pm^{0.019}_{0.021}$&
$-0.001 \pm 0.002$&
$-0.001 \pm^{0.001}_{0.001}$&
54.82/53\\
\\[-1.5ex]
Fit with $poly4$. &
$0.998 \pm 0.020$&
$0.999 \pm^{0.019}_{0.019}$&
$0.000 \pm 0.000$&
$-0.000 \pm^{0.000}_{0.000}$&
52.00/51\\
\\[-1.5ex]
Fit with $50<r<200\hMpc$ fitting range. &
$0.999 \pm 0.020$&
$1.001 \pm^{0.019}_{0.020}$&
$0.001 \pm 0.002$&
$0.001 \pm^{0.001}_{0.002}$&
46.80/45\\
\\[-1.5ex]
Fit with $20<r<200\hMpc$ fitting range. &
$0.996 \pm 0.020$&
$0.999 \pm^{0.018}_{0.021}$&
$-0.002 \pm 0.004$&
$-0.001 \pm^{0.002}_{0.004}$&
58.24/57\\
\\[-1.5ex]
Fit with $70<r<150\hMpc$ fitting range. &
$0.999 \pm 0.021$&
$1.000 \pm^{0.019}_{0.022}$&
$0.001 \pm 0.007$&
$0.001 \pm^{0.004}_{0.005}$&
21.89/22\\
\\[-1.5ex]
Fit using mock covariance matrix. &
$0.999 \pm 0.017$&
$1.001 \pm^{0.014}_{0.019}$&
$0.001 \pm 0.012$&
$-0.000 \pm^{0.010}_{0.010}$&
52.85/52\\
\hline
\end{tabular}

\medskip
$^{1}$ $\Delta \alpha = \alpha_{[i]} - \alpha_{[f]}$, where $i$ is the model number. \\
$^{2}$ We scale the measured sound horizons to the LasDamas cosmology where necessary.

\end{table*}

Fitting the average of the reconstructed real-space mock correlation
functions while allowing $\snl$ to vary gives $\alpha=0.999$ and
$\snl=3.0\hMpc$. As in redshift space, the value of $\alpha$ prior to
reconstruction is already very close to 1 and hence, we do not expect
reconstruction to shift the acoustic peak much closer to its predicted
linear theory position. However, $\snl$ decreased by a factor of $\sim2.2$
from its pre-reconstruction value, implying that reconstruction was
effective at removing the smearing of the acoustic peak caused by
non-linear structure growth. 

In our fiducial model for real space after reconstruction, we take
$\snl=3\hMpc$, as we did in the calculation of the modified Gaussian
covariance matrix. All other parameters of the fiducial model are
analogous to the unreconstructed case described above. The same fitting
algorithm as described in \S\ref{sec:red_norec_ff} is used.

\subsection{Without Reconstruction Fitting Results}\label{sec:real_norec_fr}

\begin{figure}
\vspace{0.4cm}
\centering
\epsfig{file=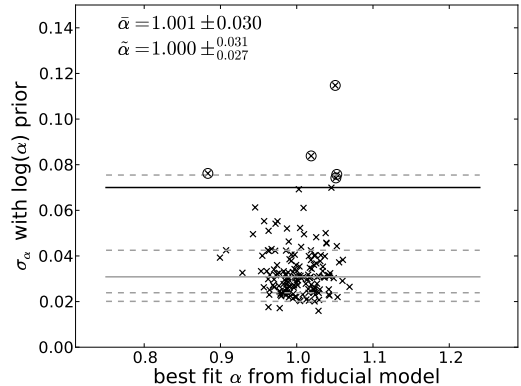, width=0.8\linewidth, clip=}
\caption{The standard deviations measured from $p(\alpha)$ versus the
best-fit values of $\alpha$ from the fiducial model for each mock in real
space. We impose a cutoff at a standard deviation of 7\% (marked by the
black horizontal line) as in redshift space. There are 5 mocks that lie
above this line (circled in black). We take these mocks to have poorly
measured values of $\alpha$ and discard them from our sample. The mean
and median value of $\alpha$ after discarding these poorly constrained
mocks are given on the plot.
\label{fig:sdva_realnr}}
\end{figure}

We use the same technique as that described in \S\ref{sec:red_norec_fr}
to identify and remove the mock correlation functions that do not provide
a well constrained measurement of $\alpha$ from our fitting sample. The
corresponding real-space plot to Figure \ref{fig:sdva_rednr} is shown
in Figure \ref{fig:sdva_realnr}. We use the same 7\% cutoff in standard
deviation ($\sigma_\alpha$) as in redshift space. This is marked by the
black horizontal line. There are 5 mocks ($\sim3$\%) that lie above this
cut off (circled in black) which we take to have poorly constrained
values of $\alpha$ and discard from our sample. The mean and median
values of $\alpha$ after removing these poorly constrained mocks are
indicated on the plot.

We test the robustness of our covariance matrix modeling and the fiducial
model as we did in redshift space. Namely, we compare the values of
$\alpha$ we measure using the fiducial model to those measured using
a model in which the fiducial parameters are slightly changed, or by
a fit in which we use the mock covariances rather than the MGCM. The
resultant $\alpha$, $\Delta \alpha$ and $\chi^2$ values are given in Table
\ref{tab:alphasreal}. Note that in general, the values of $\Delta\alpha$
and their scatters are slightly smaller in real space than in redshift
space. This is not unexpected since observational effects in redshift
space such as FoG and Kaiser squashing tend to broaden the acoustic peak
further, making it more difficult to obtain a precise measurement.

The results listed in Table \ref{tab:alphasreal} indicate that the
trends in real space are the same as those found in redshift space. In
particular, the values of $\alpha$ measured by slightly changing the
input cosmology, $\snl$, the order of $A(r)$ and the fitting range are
consistent with the values measured using the fiducial model, usually
with $\Delta\alpha < 0.2\%$. As in redshift space, the worst case is
when we change the template to use $N_{rel}=4$; this has a deviation of
0.6\%. Again, this is likely the result of the $N_{rel}=4$ correlation
function template having a BAO peak that does not quite match the mocks
well enough.  If we go to less sensible $\snl$ such as $\snl=0\hMpc$,
or decide to not use an $A(r)$ term, or fit using a less optimal fitting
range, the scatter in $\Delta\alpha$ increases as it did in redshift
space. Noise in the mock covariance matrix is again the likely culprit
causing the larger scatter in $\Delta \alpha$ between the MGCM fits and
the mock covariance fits. These results all imply that our covariance
modeling and our fiducial model are generally robust in real space
as well.

\subsection{With Reconstruction Fitting Results}\label{sec:real_rec_fr}

\begin{figure}
\vspace{0.4cm}
\centering
\epsfig{file=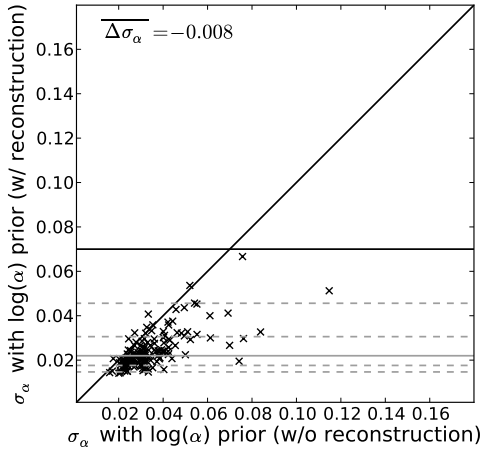, height=0.7\linewidth, clip=}  \\
\caption{The analogous plot in reconstructed real space to Figure
\ref{fig:comperrs_redr} for reconstructed redshift space. Once again,
most of the points lie significantly below the 1-1 line. The median
decrease in standard deviation is $\sim0.8\%$ as shown in the plot. Note
that there are no longer any poorly constrained mocks with standard
deviations above our 7\% cutoff. This once again illustrates how useful
and effective reconstruction is.
\label{fig:comperrs_realr}} 
\end{figure}

As in redshift space, we find that after reconstruction, we are able to
obtain much tighter constraints on the $\alpha$ values measured from
each individual mock in real space. Figure \ref{fig:comperrs_realr}
demonstrates this by plotting the standard deviation of $p(\alpha)$
for each mock before reconstruction against the value obtained
after reconstruction. Note that this is the analogue to Figure
\ref{fig:comperrs_redr} for redshift space. One can once again see that
most of the points lie significantly below the 1-1 line which indicates
that reconstruction effectively sharpened up the acoustic peak allowing
for more robust detections. The median decrease in standard deviation is
0.8\% and there are no longer any mocks that lie above our $\sigma_\alpha$
cutoff of 7\%.

Next, we once again test how slightly adjusting the fiducial model
parameters affects our measurements of $\alpha$. The results from these
fits are given in Table \ref{tab:alphasreal}. One can see that after
reconstruction, the scatters in $\alpha$ are very similar between real
space and redshift space.

We see that changing the fitting template cosmology, adjusting the value
of $\snl$, changing the order of $A(r)$ and altering the fitting range
mostly have little effect on the value of $\alpha$ measured. The only
case with $\Delta\alpha$ worse than 0.2\% is the $N_{rel}=4$ case, which
measures 0.4\%. In general, we still find that our fiducial model and
our prescription for deriving a suitable covariance matrix such as the
MGCM are robust. These results are all consistent with previous results.

Lastly, we investigate the detectability of the BAO in both
unreconstructed and reconstructed real space. We find that the median
$\Delta\chi^2 = \chi^2_{\mathrm{BAO}}-\chi^2_{\mathrm{no\;BAO}}$ values
are again negative and similar in magnitude to the redshift-space
cases. This suggests that the data is better fit by a model
containing BAO rather than a model without BAO. We also note that the
post-reconstruction real-space $\Delta\chi^2$ values are more negative
than before reconstruction. Hence, we conclude that we have a firm
detection of the acoustic signal in our mocks, with the detection being
even more robust after reconstruction.

\section{Measuring the BAO in SDSS DR7}\label{sec:drs}

\subsection{Covariance Matrices}
In this section, we apply the techniques described in \S\ref{sec:techs}
for redshift space to the DR7 LRG full sample. We use the form for the
modified Gaussian covariance matrix given in Equation (\ref{eqn:modc})
for redshift space with and without reconstruction. We adopt the
modification parameters ($c_0$, $c_1$, $c_2$ and $c_3$) derived for
the LasDamas mocks in both of these cases, assuming that the overall
shape of the covariance matrix should be modified in the same way for
both DR7 and LasDamas. However, we now switch to the WMAP7 cosmology in
constructing $P_c(k)$.

The $b_0^2$ coefficient in Equation (\ref{eqn:pc}) is chosen such that
$P_c(k)$ matches the DR7 correlation function at $r=50\hMpc$. This again
ensures that the amplitude of $P_c(k)$ matches the clustering amplitude
of DR7, an essential condition when reusing the modification parameters
to adjust the shape of the Gaussian covariance matrix.

In computing the pre-reconstruction covariance matrix, we retain
$\snl=8\hMpc$ and for post-reconstruction, we retain $\snl=4\hMpc$. We
also note that since the DR7 data goes out to $z=0.47$, we extend our
$\bar{n}(z)$ model derived from the LasDamas random catalogue out to
$z=0.47$ as well, after scaling it to the WMAP7 cosmology.

\subsection{Fit Results}
We compute the DR7 correlation functions in the WMAP7 cosmology. For
details of the computation and reconstruction, please see Paper I. We
present only the fitting results here.

\begin{figure*}
\vspace{0.4cm}
\begin{tabular}{c}
\epsfig{file=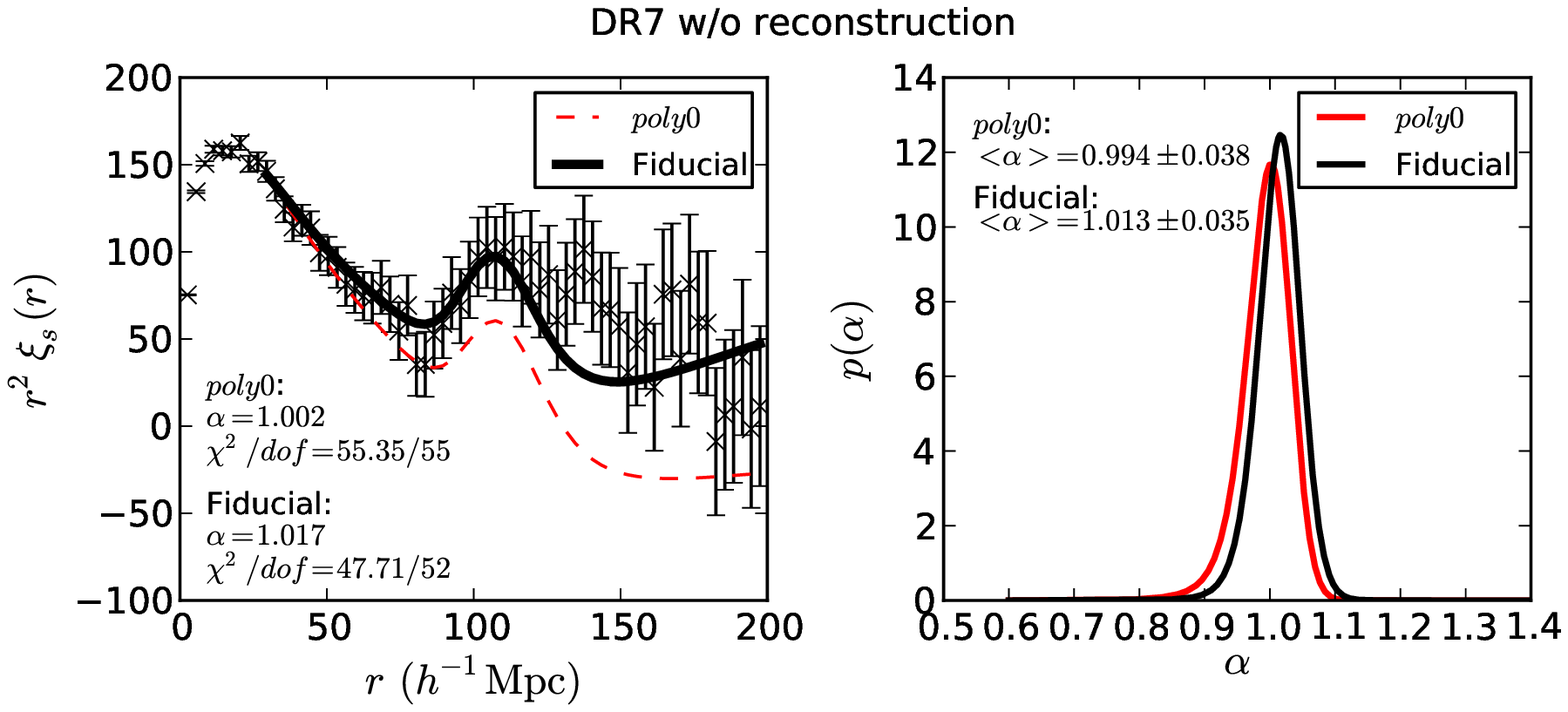, height=0.3\linewidth, width=0.73\linewidth, clip=}\\
\epsfig{file=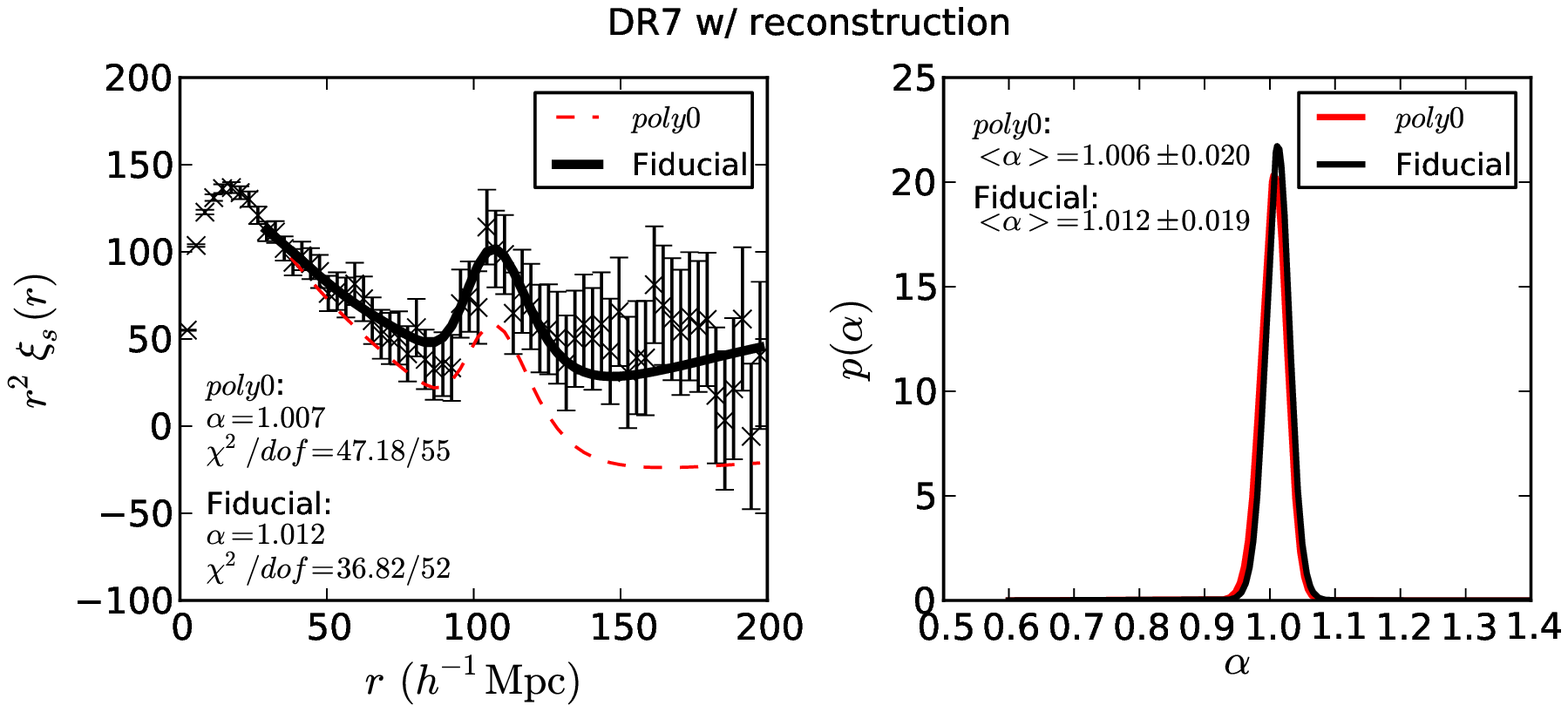, height=0.3\linewidth, width=0.73\linewidth, clip=} \\
\end{tabular}
\caption{DR7 fit results. (top) Before reconstruction. (bottom) After
reconstruction. The left column shows the fits to the DR7 data using the
fiducial model (solid black line) and $poly0$ (dashed red line). The right
column shows the $p(\alpha)$ distributions for fits using the fiducial
model (black line) and $poly0$ (red line). Here we have again applied the
15\% prior in $\log(\alpha)$ as described in \S\ref{sec:red_norec_fr}. As
with the LasDamas mocks, we use $\snl=8\hMpc$ in the fiducial model before
reconstruction and $\snl=4\hMpc$ after reconstruction. The similarities in
$\chi^2$, $\alpha$ and $\sigma_\alpha$ between the fiducial and $poly0$
cases indicate that the covariance matrix does not demand an $A(r)$ term
in the model. However, our mock correlation function analyses suggest
that having an $A(r)$ term is useful for marginalizing out errors due to
assuming the wrong cosmology and broadband effects that are not included
in our fitting model. The effectiveness of $A(r)$ in marginalizing over
the conspicuous excess large-scale power seen in these DR7 correlation
functions (left panels) also exemplifies its utility.
\label{fig:dr7fits}}
\end{figure*}

Figure \ref{fig:dr7fits} shows the results of our fits to the
DR7 data using the fiducial model and fitting algorithm outlined
in \S\ref{sec:techs}. These results are also summarized in Table
\ref{tab:dr7res} along with the fit results from varying fiducial model
parameters such as $\snl$ and fitting range.

\begin{table}
\caption{\label{tab:dr7res} DR7 fit results for various models}

\begin{tabular}{lcc}
\hline
Model&$\alpha$&$\chi^2$ \\
\hline

\multicolumn{3}{c}{Redshift Space without Reconstruction}\\
\hline
Fiducial $[f]$&
$1.017 \pm 0.035$\footnotemark[1]&
47.71/52\\
\\[-1.5ex]
$\snl=0$&
$1.025 \pm 0.029$&
49.89/52\\
\\[-1.5ex]
$\snl \rightarrow \snl+2$&
$1.011 \pm 0.039$&
47.86/52\\
\\[-1.5ex]
$poly0$&
$1.002 \pm 0.038$&
55.35/55\\
\\[-1.5ex]
$poly2$&
$1.016 \pm 0.034$&
47.72/53\\
\\[-1.5ex]
$poly4$&
$1.016 \pm 0.039$&
42.74/51\\
\\[-1.5ex]
$50-200\hMpc$ fitting range&
$1.011 \pm 0.040$&
40.44/45\\
\hline
\multicolumn{3}{c}{Redshift Space with Reconstruction}\\
\hline
Fiducial $[f]$&
$1.012 \pm 0.019$&
36.82/52\\
\\[-1.5ex]
$\snl=0$&
$1.012 \pm 0.017$&
35.99/52\\
\\[-1.5ex]
$\snl \rightarrow \snl+2$&
$1.012 \pm 0.021$&
38.12/52\\
\\[-1.5ex]
$poly0$&
$1.007 \pm 0.020$&
47.18/55\\
\\[-1.5ex]
$poly2$&
$1.012 \pm 0.019$&
37.14/53\\
\\[-1.5ex]
$poly4$&
$1.012 \pm 0.019$&
36.34/51\\
\\[-1.5ex]
$50-200\hMpc$ fitting range&
$1.012 \pm 0.019$&
33.21/45\\
\hline
\end{tabular}

\medskip
$^{1}$ Here, the quoted $\alpha$ is the best-fit value rather than the
mean of the probability distribution $p(\alpha)$ as in Paper I. These
2 values may be slightly different but are well within error of each
other.
\end{table}

The 2 panels at the top illustrate the pre-reconstruction results
and the 2 panels at the bottom illustrate the post-reconstruction
results. We fix $\snl$ in our model templates to the same values as in
the covariance matrices. The left column shows the data with the fiducial
model fit overplotted (black line). The dashed red line corresponds to
a fit using $poly0$ instead of fiducial $A(r)$. The best-fit $\alpha$
and $\chi^2$ values are quoted on the plot. The right column shows
the $p(\alpha)$ distributions for the fits using the fiducial model
(black line) and the fits using $poly0$ (red line). The means of the
distributions are quoted on the plot along with their standard deviations
$\sigma_\alpha$. Taking the best-fit $\alpha$ value from the fiducial
model fit and the $\sigma_\alpha$ from the $p(\alpha)$ distribution,
we measure the DR7 acoustic scale to correspond to $\alpha = 1.017
\pm 0.035$ before reconstruction and $\alpha = 1.012 \pm 0.019$ after
reconstruction. Using the mean of the $p(\alpha)$ probability distribution
instead gives $\alpha = 1.013 \pm 0.035$ before reconstruction and $\alpha
= 1.012 \pm 0.019$ after reconstruction. One can see that the two values
are the same after reconstruction, however, they are slightly different
before reconstruction due to the slight asymmetry of the $p(\alpha)$
distribution. The pre-reconstruction error is comparable to the 3.3\%
found by \citet{Pea10} for a similar sample.

This factor of 1.8 decrease in the error after applying reconstruction is
similar to what we saw for the mock catalogues. Since the survey volume
required to achieve a certain variance is inversely proportional to the
variance, we would have to increase the survey volume by about a factor
of 3 to achieve this same reduction in the error. This clearly shows how
effective reconstruction is at improving our measurement of the acoustic
scale. We can convert these $\alpha$ values into $D_v(z)/r_s$ measurements
at a median redshift of $z=0.35$ according to \citet{Eea05}, i.e.
\begin{equation}
\alpha = \frac{D_v(z)/r_s}{D_{v,f}(z)/r_{s,f}}
\end{equation}
where the subscript $f$ denotes the fiducial WMAP7 cosmology, $D_v(z)$
is the spherically averaged distance to redshift $z$ and $r_s$ is the
sound horizon. In the WMAP7 cosmology we have $r_{s,f}=152.76\rm{Mpc}$
and $D_{v,f}(z) = 1340.2\rm{Mpc}$. The best-fit $\alpha$ values then
give $D_v(z)/r_s = 8.92 \pm 0.31$ before reconstruction and $D_v(z)/r_s
= 8.88 \pm 0.17$ after reconstruction. The means of the $p(\alpha)$
distributions give $D_v(z)/r_s = 8.89 \pm 0.31$ before reconstruction
and $D_v(z)/r_s = 8.88 \pm 0.17$ after reconstruction.

From Table \ref{tab:dr7res}, we see that the $\alpha$ values obtained
by varying $\snl$, order of $A(r)$ and fitting range are all consistent
with each other within the errors. In particular, after reconstruction,
we see that all cases have very similar errors and all give an $\alpha$
value within $0.1\%$ of the others except the $poly0$ case. This is as
expected from our analysis of the mock catalogues.

The DR7 correlation function exceeds the linear theory prediction at
large $r$, suggesting extra large-scale power. This can be seen in
Figure \ref{fig:dr7fits} by comparing the data to the fit using the
$A(r)=0$ model (dashed red line). While this offset appears large to
the eye, we stress that the data points are correlated such that these
coherent offsets are only weakly constrained. This is demonstrated by
the fact that the fiducial $A(r)$ fit, which adds three marginalization
parameters and largely compensates the offset, does not decrease $\chi^2$
by a very significant amount. Hence, while such extra power could be a
sign of unaddressed systematic errors in the data set or some exotic
cosmology, the statistical significance of the extra power is weak. In
addition, the measured $\alpha$ and $\sigma_\alpha$ values are consistent,
which suggests that the data does not strongly demand a non-zero $A(r)$
in the model. However, we see that the fiducial $A(r)$ fit matches the
data much better. Also, our analysis of the mock correlation functions
indicates that we should err on the side of caution and marginalize over
a non-zero $A(r)$ term to remove any broadband affects not accounted
for in the model that could bias our measurement of the acoustic scale.

\begin{figure}
\vspace{0.4cm}
\centering
\epsfig{file=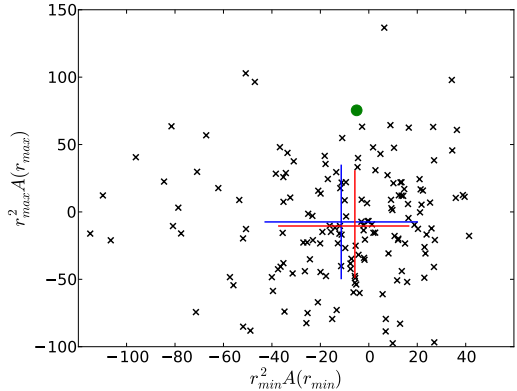, width=0.8\linewidth, clip=}  
\caption{The values of the fiducial $A(r)$ term at the edges of the
fitting range for the 160 LasDamas mocks (black crosses) in redshift
space before reconstruction. The large blue cross indicates the mean
and standard deviation. The large red cross indicates the median and
16th/84th percentile levels. The DR7 point is overplotted as the green
circle. One can see that this point falls within 2$\sigma$ of the
LasDamas average which implies that the shape of the DR7 $A(r)$ term
is not unexpected. Hence, even though $A(r)$ is providing a significant
amount of marginalization to account for the excess power at large scales
in the DR7 correlation function, it is not an inordinately large amount
in the context of LasDamas.
\label{fig:amm}}
\end{figure}

To further address the excess large-scale power, we study whether the
magnitude of the fiducial $A(r)$ term in the best-fit model to the DR7
data is unusual in the context of the LasDamas mocks. Figure \ref{fig:amm}
shows the values of the $A(r)$ term at the edges of the fitting range
(i.e. at $r\sim30\hMpc$ and $r\sim200\hMpc$) for the LasDamas mocks
before reconstruction (black crosses). The large blue cross indicates
the mean and standard deviation of the LasDamas values and the large
red cross indicates the median and the 16th/84th percentiles. The
DR7 point is overplotted as the green circle and clearly falls within
2$\sigma$ of the LasDamas average. A similar plot can be made for the
post-reconstruction fits.

\subsection{Comparison with LasDamas Cosmology}
We also compute the DR7 correlation functions with and without
reconstruction using the LasDamas cosmology. We apply the same fitting
algorithm, but change the cosmology of the covariance matrix and template
model to that of LasDamas. We again adopt the LasDamas modification
parameters to the Gaussian covariance matrix and the same $\snl$ values.

We find $\alpha = 1.053 \pm 0.034$ in redshift space before reconstruction
and $\alpha = 1.044 \pm 0.019$ after reconstruction. Converting these
$\alpha$ values to $D_v/r_s$ at a median redshift of $z=0.35$, we
find $D_v/r_s = 8.95 \pm 0.30$ before reconstruction and $D_v/r_s =
8.87 \pm 0.17$ after reconstruction. These values are consistent with
those obtained from the DR7 data in the WMAP7 cosmology when we factor
in errors. The values of $\sigma_\alpha$ are also consistent.

\subsection{Significance of the BAO Detection}
The BAO detection significance is an obvious question that must be
addressed. However, its characterization is non-trivial. There are
essentially 2 different tests which need to be evaluated. The first
considers the possibility that we have not detected the BAO signal in
our data, either because it does not actually exist or we just have not
observed it. The second assumes the BAO peak does exist and asks how
robustly we have measured its location.

\begin{figure*}
\vspace{0.4cm}
\begin{tabular}{cc}
\epsfig{file=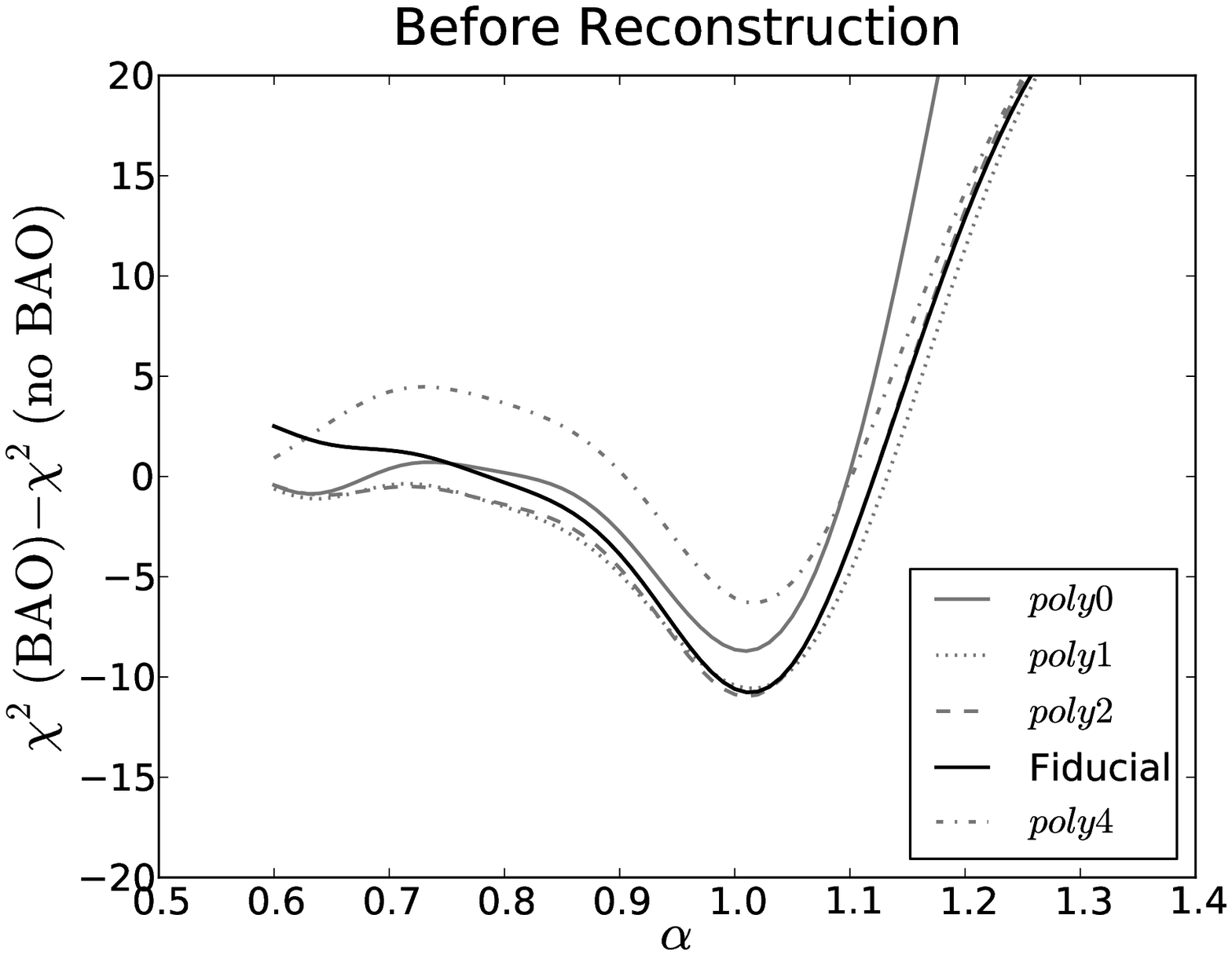, width=0.4\linewidth, clip=}
\epsfig{file=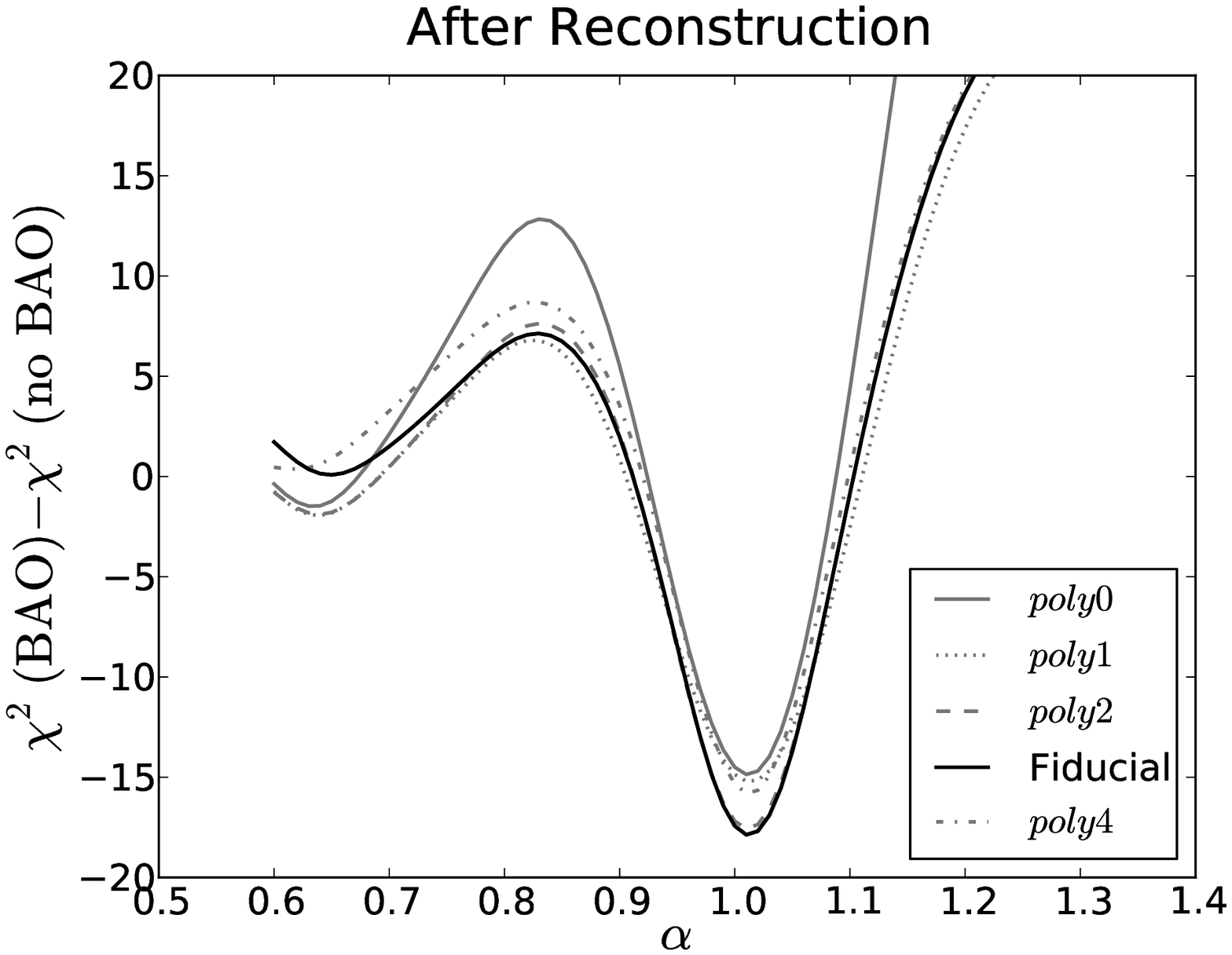, width=0.4\linewidth, clip=} \\ 
\epsfig{file=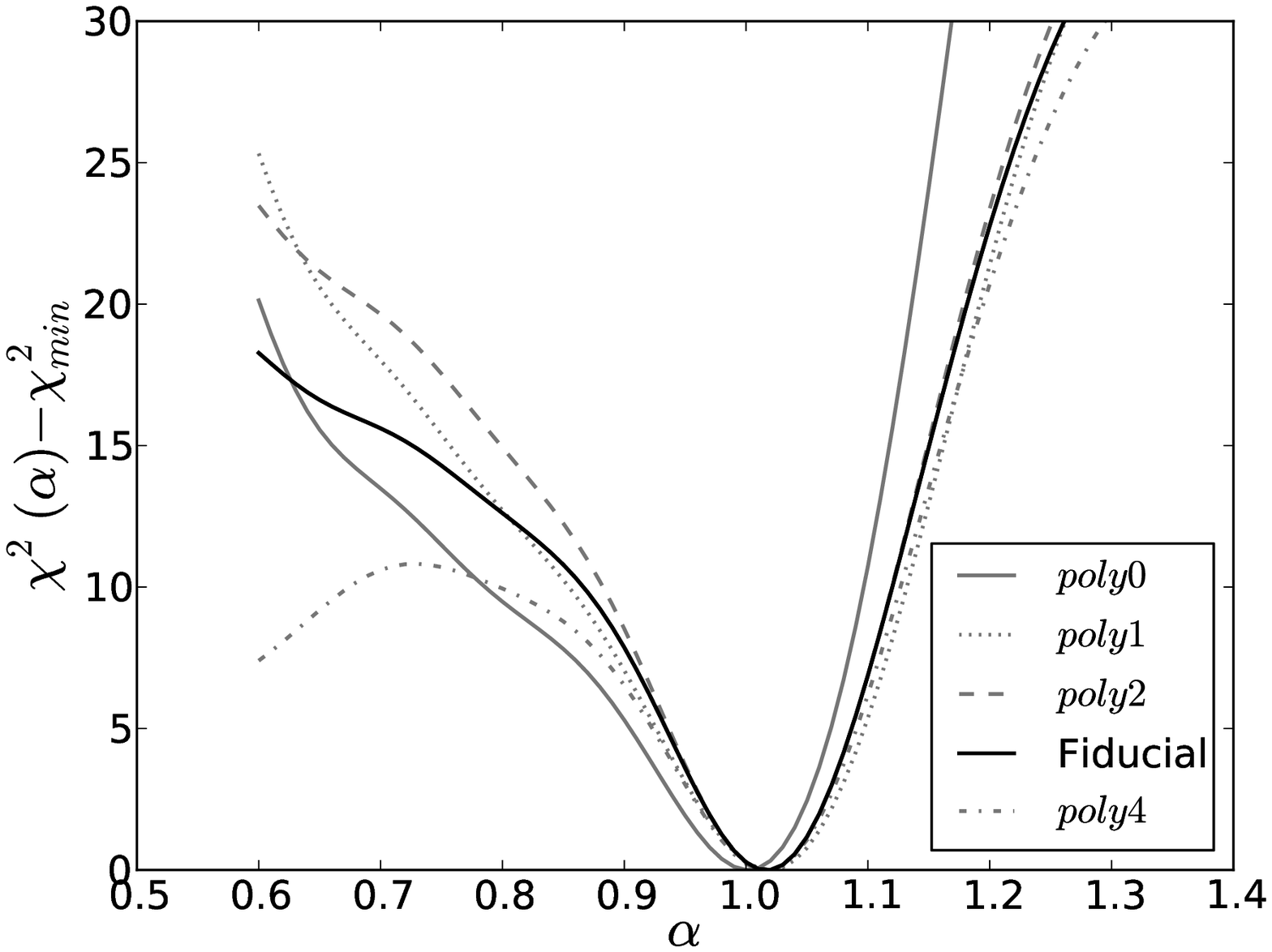, width=0.4\linewidth, clip=}
\epsfig{file=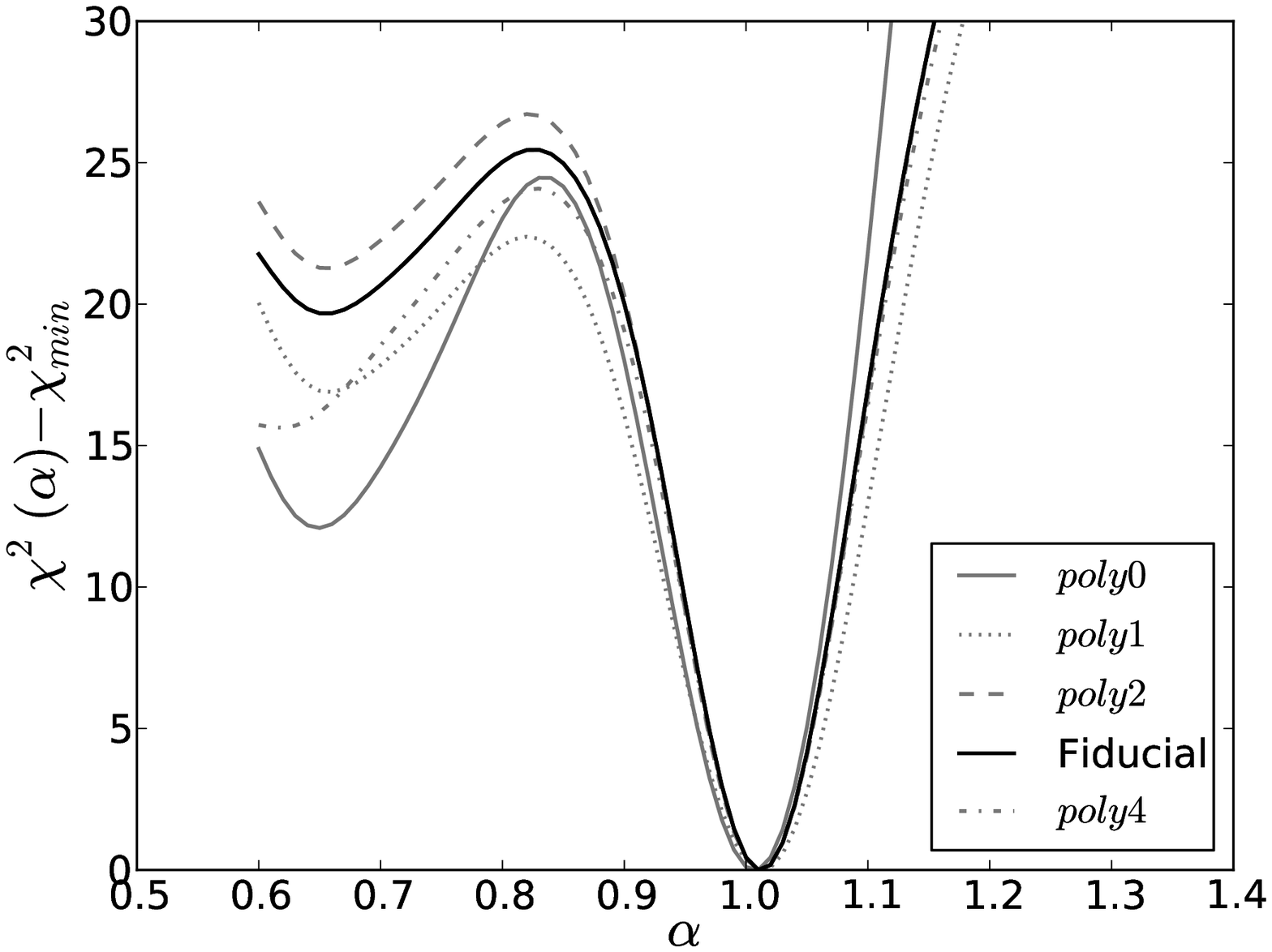, width=0.4\linewidth, clip=} \\
\end{tabular} 
\caption{Significance of the BAO in the DR7 data. (top) $\Delta\chi^2 =
\chi^2_{\rm{BAO}} - \chi^2_{\rm{no \; BAO}}$ versus $\alpha$ for different
$A(r)$ before reconstruction (left) and after reconstruction (right). The
different forms of $A(r)$ are represented by different line styles
as indicated in the legend. For our fiducial form (solid black line),
$\Delta\chi^2$ reaches a minimum of $\sim -11$ before reconstruction
and $\sim -18$ after reconstruction. Hence, a model containing BAO is a
better fit to the data than a model without BAO at more than 3$\sigma$
significance ($\Delta\chi^2=-9$) before reconstruction and at more than
4$\sigma$ significance ($\Delta\chi^2=-16$) after reconstruction. (bottom)
$\Delta \chi^2 = \chi^2(\alpha) - \chi^2_{min}$ versus $\alpha$ for
different $A(r)$ before reconstruction (left) and after reconstruction
(right). For our fiducial form, the curve is parabolic around the minimum
that corresponds to the best-fit value of $\alpha$. Before reconstruction,
the $\chi^2$ difference between the minimum and where the curve starts
plateauing at small $\alpha$ is $\sim 10-15$. This difference becomes
even more pronounced after reconstruction, measuring a $\Delta\chi^2
\sim 25$. Hence, the measured acoustic scale is favoured at slightly
more than 3$\sigma$ ($\Delta \chi^2 = 9$) before reconstruction and at
5$\sigma$ ($\Delta \chi^2 = 25$) post-reconstruction. Both the top and
bottom panels show an increase in significance of the BAO detection after
reconstruction. Also, one can see that in general (and especially before
reconstruction), the fits with higher order $A(r)$ terms (i.e. $poly2$
and fiducial) have more prominent $\Delta\chi^2$ minima in both the
top and bottom panels. This indicates that we obtain more robust BAO
detections when fitting with non-trivial $A(r)$ terms. However, fits
with $poly4$ appear to perform worse than the lower order fits before
reconstruction. This indicates that we are likely beginning to afford
the model too much flexibility.
\label{fig:sigdr7}}
\end{figure*}

We attempt to address these 2 questions in Figure \ref{fig:sigdr7}. In
the top panels we have plotted $\Delta\chi^2 = \chi^2_{\rm{BAO}} -
\chi^2_{\rm{no \; BAO}}$ at various values of $\alpha$ for different
$A(r)$. As in Figure \ref{fig:hibao}, the $\chi^2_{\rm{BAO}}$ values are
obtained through fits using a model containing BAO and the $\chi^2_{\rm{no
\; BAO}}$ are obtained through fits using a model without BAO. The left
panel shows the results before reconstruction and the right panel shows
the results after reconstruction. These plots answer the first question of
whether we have detected the BAO assuming that we are fairly confident in
our cosmology. With this assumption, we know that $\alpha$ must be close
to 1 and hence we can restrict our attention to this region. One can
see that the $\Delta\chi^2$ values are all negative around $\alpha=1$
and reach a minimum of $\Delta\chi^2 \sim -11$ before reconstruction
and $\sim -18$ after reconstruction for the fiducial $A(r)$ fits (solid
black line). This indicates that a model containing BAO is favoured over a
model without BAO at more than 3$\sigma$ ($\Delta\chi^2=-9$) confidence
before reconstruction and more than 4$\sigma$ ($\Delta\chi^2=-16$)
confidence after reconstruction. Note that these $\Delta\chi^2$ values
are comparable to the median values we measured for the LasDamas mocks
($-10$ and $-16$ respectively). While it is true that these $\chi^2$
values are not directly comparable due to the volume difference between
DR7 in a LasDamas cosmology and a WMAP7 cosmology, this discrepancy
is small. Hence, DR7 should be a fairly typical sample consistent with
cosmic variance.

The bottom panels show $\Delta \chi^2 = \chi^2(\alpha) - \chi^2_{min}$
versus $\alpha$ for various $A(r)$. Here, $\chi^2_{min}$ is the value
of $\chi^2$ that corresponds to the best-fit value of $\alpha$. These
plots are similar to the middle columns of Figures \ref{fig:exfigs}
and \ref{fig:exfigs_rec}. They answer the second question of whether the
BAO scale we measure is significantly favoured over other values. Again,
the left column corresponds to the results before reconstruction and the
right column corresponds to the results after reconstruction. For the
fiducial $A(r)$ fit before reconstruction, the curve is parabolic around
the best-fit $\alpha$ indicating the Gaussian nature of $\alpha$. The
corresponding $\chi^2_{min}$ lies at $\Delta\chi^2 \sim 10-15$ below where
the curve starts plateauing. Recall that the plateau is due to the fitter
having an easier time hiding the acoustic peak in the errors at large $r$
and is not actually physical. Post-reconstruction, the parabola becomes
tighter around the best-fit $\alpha$ and the $\chi^2$ difference between
the minimum and the plateau grows to 25. This indicates that while the
measured acoustic scale is favoured at slightly more than 3$\sigma$
($\Delta \chi^2 = 9$) before reconstruction, after reconstruction,
it becomes favoured at 5$\sigma$ ($\Delta \chi^2 = 25$). Hence,
reconstruction increases the BAO detection significance in both of the
tests considered here.

Another point to note is that in general, the $\Delta\chi^2$ minima in
both of the above mentioned cases is more prominent when one fits with
a higher degree $A(r)$ (i.e. $poly2$ or fiducial). This is especially
true before reconstruction when the acoustic scale is more difficult to
measure due to non-linear effects. Hence, we see that a more robust BAO
detection is achieved when fitting with a non-zero $A(r)$. However,
we also see that before reconstruction, $poly4$ (dash-dotted line)
performs worse than the lower order fits. This is because when we give
the model too much freedom, it acquires more flexibility to hide the
acoustic peak in the errors at large $r$ while using the $A(r)$ nuisance
terms to compensate for the shape of the acoustic peak in the data.

\section{Conclusions}\label{sec:theend}

We develop a series of tools and methods to study the baryon acoustic
oscillations in the SDSS DR7 LRG sample. These allow us to carefully
treat the uncertainties and covariances involved in measuring the
acoustic scale. Such tools include reconstruction, an algorithm for
estimating a reliable covariance matrix and a robust fitting model for the
correlation function. In this study, we demonstrate the first application
of reconstruction to a galaxy redshift survey and more careful treatments
of the covariance matrix and fitting model. This paper is the second
in a series of three papers. Paper I discusses the details of the DR7
LRG sample and our reconstruction technique. Paper III discusses the
cosmological implications of our measured acoustic scale.

Through our analysis of 160 SDSS DR7 mock catalogues from the LasDamas
simulations, we find that the covariance matrix derived directly from
the mocks is very noisy. We present a new method for obtaining a smooth
approximation to the mock covariance matrix using the analytic Gaussian
covariance matrix. This process introduces appropriate modifications
to the Gaussian covariance matrix using a maximum likelihood fit to the
mock covariances. We show that the modified Gaussian covariance matrix
obtained this way is a good fit to the mock covariances and produces
consistent measurements of the acoustic scale.

Some of the mocks have weak acoustic signals and hence the acoustic scale
can be poorly determined in these. In order to identify these poorly
constrained mocks, we find that looking at the probability distribution of
the shift in the acoustic scale, $\alpha$, can be a good gauge. For mocks
that have distributions with a larger standard deviation, the constraint
on $\alpha$ is poorer and vice versa. We impose a cutoff at a standard
deviation of 7\% in our mocks and find that in redshift space, 8 mocks
lie above this cutoff and in real space, 5 do. After reconstruction,
no poorly constrained mocks remain in redshift or real space.

We find that in redshift space, we obtain consistent measurements of
$\alpha$ when the fiducial model parameters (template cosmology, $\snl$,
degree of $A(r)$ and fitting range) are slightly tweaked. This implies
that the values of $\alpha$ we measure are robust against small changes
in model parameters. Hence, our fiducial model should return reliable
measurements of the acoustic scale. However, we note that in order to
afford the model enough flexibility, $A(r)$ should be non-zero as in the
fiducial form, Equation (\ref{eqn:aform}). This is because the $A(r)$
term is required to marginalize out all the broadband contributions
not accounted for by the template such as scale-dependent bias and
redshift-space distortions (or residual redshift-space distortions in
the post-reconstruction case). This term also accounts for any errors
in our choice of model cosmology. We find that if we use a template
cosmology that does not match the simulations to perform the fit, a low
order $A(r)$ does not recover the correct acoustic scale as well. One
must also be careful not to use an $A(r)$ term that is very high order
as the model will begin fitting the noise in the data.

Using our fitting scheme on our mock correlation functions, we
consistently measure the acoustic scale to $\sim3.3\%$ accuracy
in redshift space before reconstruction and $\sim2.1\%$ after
reconstruction. The fit to the average redshift-space mock correlation
function before reconstruction gives an $\alpha$ that is already very
close to 1. Hence, we do not expect reconstruction to shift the acoustic
scale much closer to its predicted linear theory position. However,
the decrease in best-fit $\snl$ from $8.1\hMpc$ before reconstruction to
$4.4\hMpc$ after reconstruction shows that reconstruction is effective
at removing the smearing of the acoustic peak caused by non-linear
structure growth.

We demonstrated the detectability of the acoustic signature in redshift
space both before and after reconstruction. In both cases we fit each of
the mocks using a model containing BAO and a model without BAO and find
that $\Delta\chi^2 = \chi^2_{\rm BAO} - \chi^2_{\mathrm{no \; BAO}}$
is negative on average. This indicates that the mock data prefers a
model containing BAO over a model without BAO. Hence, we conclude that
we have a robust detection of the acoustic signal in our mocks. We note
that $\Delta\chi^2$ is even more negative after reconstruction, again
revealing the importance of the procedure. In addition, when we fit
using $poly0$, we still obtain negative average values of $\Delta\chi^2$
implying that even with this simple model we can robustly detect the
BAO in our mocks. Similar results are obtained in real space before and
after reconstruction.

We then apply our covariance matrix and fitting techniques to
the correlation function calculated from the DR7 data in the WMAP7
cosmology. We again vary the various parameters of the fit and recover
consistent values of $\alpha$. From the probability distribution
of $\alpha$ we measure a mean $\alpha = 1.013 \pm 0.035$ before
reconstruction which gives $D_v(z=0.35)/r_s = 8.89 \pm 0.31$. After
reconstruction we measure $\alpha = 1.012 \pm 0.019$ which gives
$D_v(z=0.35)/r_s = 8.88 \pm 0.17$. We see that the error on $\alpha$
has decreased by a factor of 1.8. Such a decrease is equivalent to
what we would expect if we increase the survey volume by a factor of
3. This again demonstrates the power of reconstruction in removing the
uncertainties introduced by non-linear structure growth.

Finally we assess the significance of our DR7 BAO measurement using 2
different tests. The first measures how confident we are that our data
contains a BAO signature and the second measures how confident we are
that our measurement of the BAO scale is correct. We find that before
reconstruction, our data favours a model containing BAO at more than
3$\sigma$ over a model without BAO and the acoustic scale we measure is
preferred at more than 3$\sigma$. After reconstruction, these confidence
levels become even more pronounced. The data favours a model containing
BAO at more than 4$\sigma$ and the measured acoustic scale is preferred
at 5$\sigma$. Hence, we conclude that our DR7 BAO measurement is robust.

The methods developed in this paper and its companions should be
applicable to future data sets with higher precision requirements such
as the Baryon Oscillation Spectroscopic Survey (BOSS) in SDSS-III. BOSS
aims to measure the acoustic scale to $\sim1\%$ precision at $z=0.35$
and $z=0.6$ which will grant us even more precise measurements of the
properties of dark energy.

\section{acknowledgments}
Funding for the Sloan Digital Sky Survey (SDSS) and SDSS-II has
been provided by the Alfred P. Sloan Foundation, the Participating
Institutions, the National Science Foundation, the U.S. Department of
Energy, the National Aeronautics and Space Administration, the Japanese
Monbukagakusho, the Max Planck Society and the Higher Education Funding
Council for England. The SDSS website is \texttt{http://www.sdss.org/.}

The SDSS is managed by the Astrophysical Research Consortium (ARC) for
the Participating Institutions. The Participating Institutions are the
American Museum of Natural History, Astrophysical Institute Potsdam,
University of Basel, University of Chicago, Drexel University, Fermilab,
the Institute for Advanced Study, the Japan Participation Group, the
Johns Hopkins University, the Joint Institute for Nuclear Astrophysics,
the Kavli Institute for Particle Astrophysics and Cosmology, the Korean
Scientist Group, the Chinese Academy of Sciences (LAMOST), Los Alamos
National Laboratory, the Max-Planck-Institute for Astronomy (MPIA), New
Mexico State University, Ohio State University, University of Pittsburgh,
University of Portsmouth, Princeton University, the United States Naval
Observatory and the University of Washington.

We thank the LasDamas collaboration for making their galaxy mock catalogs
public. We thank Cameron McBride for assistance in using the LasDamas
mocks and comments on earlier versions of this work. We thank Martin
White for useful conversations on reconstruction. XX thanks Hee-Jong
Seo for her insightful comments. XX, DJE, and KTM were supported by NSF
grant AST-0707725 and NASA grant NNX07AH11G. NP and AJC are partially
supported by NASA grant NNX11AF43G. This work was supported in part
by the facilities and staff of the Yale University Faculty of Arts and
Sciences High Performance Computing Center.



\clearpage

\newcommand{\tableskip}{\\[-8pt]}
\newcommand{\singleline}{\tableskip\hline\tableskip}
\newcommand{\doubleline}{\tableskip\hline\tableskip}
\newlength{\tablespread}\setlength{\tablespread}{30pt}
\newcommand{\dje}{\hspace{\tablespread}}
\def\arraystretch{1.1}


\end{document}